\documentclass[aps,prl,reprint,showpacs,showkeys,nofootinbib,
superscriptaddress,preprintnumbers, longbibliography]{revtex4-1}
\usepackage{amsmath,amssymb,bm,mathrsfs}
\usepackage{srcltx}
\usepackage{dsfont}
\usepackage{epsfig}
\usepackage{slashed}
\usepackage{bbold}
\usepackage{psfrag}
\usepackage{xcolor}
\PassOptionsToPackage{caption=false}{subfig}
\usepackage{subfig}
\usepackage{xfrac}
\usepackage{multirow}
\usepackage{booktabs}
\usepackage{hyperref}
\hypersetup{colorlinks=true,linkcolor=blue,citecolor=blue,urlcolor=violet}
\usepackage{relsize}
\usepackage{siunitx}
\usepackage{tensor}

\newcommand{\be}{\begin{equation}}
\newcommand{\ee}{\end{equation}}

\newcommand{\beq}{\begin{equation}}
\newcommand{\eeq}{\end{equation}}
\newcommand{\ba}{\begin{array}}
\newcommand{\ea}{\end{array}}
\newcommand{\bea}{\begin{eqnarray}}
\newcommand{\eea}{\end{eqnarray} }
\newcommand{\bal}{\begin{align}}
\newcommand{\eal}{\end{align}}
\newcommand{\bi}{\begin{itemize}}
\newcommand{\ei}{\end{itemize}}
\newcommand{\ben}{\begin{enumerate}}
\newcommand{\een}{\end{enumerate}}
\newcommand{\bc}{\begin{center}}
\newcommand{\ec}{\end{center}}
\newcommand{\bt}{\begin{table}}
\newcommand{\et}{\end{table}}
\newcommand{\btb}{\begin{tabular}}
\newcommand{\etb}{\end{tabular}}

\newcommand{\vecb}[1]{\boldsymbol{#1}}
\newcommand{\la}{\left\langle}
\newcommand{\ra}{\right\rangle}
\newcommand{\thb}[1]{\la #1\ra}

\DeclareSIUnit\electronvolt{e\kern-.05em V}
\DeclareSIUnit\parsec{pc}
\hypersetup{colorlinks,citecolor= nicegreen,linkcolor= nicered}
\definecolor{nicered}{rgb}{0.7,0.1,0.1}
\definecolor{nicegreen}{rgb}{0.1,0.5,0.1}

\def\({\left(}
\def\){\right)}

\usepackage{soul}

\begin{document}

\title{Anticipating a New-Physics Signal in Upcoming 21-cm Power Spectrum Observations}

\author{Rennan Barkana}
\email{barkana@tau.ac.il}
\affiliation{School of Physics and Astronomy, Tel-Aviv University, Tel-Aviv, 69978, Israel}
\affiliation{Institute for Advanced Study, 1 Einstein Drive, Princeton, New Jersey 08540, USA}
\affiliation{Department of Astronomy and Astrophysics, University of California, Santa Cruz, CA 95064, USA}

\author{Anastasia Fialkov}
\email{anastasia.fialkov@gmail.com}
\affiliation{Institute of Astronomy, University of Cambridge, Madingley Road, Cambridge, CB3 0HA, UK}
\affiliation{Kavli Institute for Cosmology, Madingley Road, Cambridge CB3 0HA, UK}
	
\author{Hongwan Liu}
\email{hongwanl@princeton.edu}
\affiliation{Center for Cosmology and Particle Physics, Department of Physics, New York University, New York, NY 10003, U.S.A.}
\affiliation{Department of Physics, Princeton University, Princeton, New Jersey, 08544, U.S.A.}

\author{Nadav Joseph Outmezguine}
\email{njo@berkeley.edu}
\affiliation{Berkeley Center for Theoretical Physics, University of California, Berkeley, CA 94720, U.S.A.}
\affiliation{Theory Group, Lawrence Berkeley National Laboratory, Berkeley, CA 94720, U.S.A.}

\begin{abstract}
\noindent Dark matter-baryon interactions can cool the baryonic fluid, which has been shown to modify the cosmological 21-cm global signal. We show that in a two-component dark sector with an interacting millicharged component, dark matter-baryon scattering can produce a 21-cm power spectrum signal with acoustic oscillations. The signal can be up to three orders of magnitude larger than expected in $\Lambda$CDM cosmology, given realistic astrophysical models. This model provides a new-physics target for near-future experiments such as HERA or NenuFAR, which can potentially discover or strongly constrain the dark matter explanation of the putative EDGES anomaly. 
\end{abstract}

\maketitle

In recent years, rapid progress has been made toward turning 21-cm cosmology into reality, opening a valuable window into the Universe at $6 \lesssim z \lesssim 30$. During this epoch, the intergalactic medium (IGM) reached its lowest temperature in cosmological history, before heating up due to star formation, which began in dark matter (DM) halos of sufficient mass. As a result, 21-cm measurements probe the previously unknown period of cosmic dawn and the first stars~\cite{Madau:1996cs,Loeb:2003ya,Tseliakhovich:2010bj,2006PhR...433..181F, Barkana:2016nyr}, and are particularly sensitive to new-physics processes that impact the thermal state of the IGM or the process of star formation during this epoch~\cite{Sitwell:2013fpa,Sekiguchi:2014wfa,Nebrin:2018vqt,Munoz:2019hjh,Jones:2021mrs,Hotinli:2021vxg,Flitter:2022pzf,1408.1109,1603.06795,1803.03629,1803.09739,1803.09390,1803.09398,1803.11169,Barkana:2018lgd,Tashiro:2014tsa,Munoz:2015bca,Munoz:2018pzp,Barkana:2018qrx,Fialkov:2018xre,Berlin:2018sjs,Munoz:2018jwq,Kovetz:2018zan,Liu:2019knx,Creque-Sarbinowski:2019mcm,Aboubrahim:2021ohe,Adshead:2022ovo}.

The observable in 21-cm cosmology is the brightness temperature of radiation with wavelength \SI{21}{\centi\meter}, $T_{21}$, absorbed or emitted by the hyperfine states of neutral hydrogen atoms, and observed at a redshifted wavelength.  Measurements of the sky-averaged $T_{21}$ (i.e., the global 21-cm signal) have generated significant excitement in recent years. The EDGES collaboration reported a detection of the global signal~\cite{Bowman:2018yin}, finding a large absorption trough of $ -0.5^{+0.2}_{-0.5}~\SI{}{\kelvin}$ at $z\sim 17$ at 99\% confidence. This result is in significant tension with expectations from $\Lambda$CDM cosmology~\cite{Barkana:2018lgd}. More recently, however, the SARAS experiment found that the central value of the EDGES absorption profile is inconsistent with their measurements at 95\% confidence~\cite{Singh:2021mxo}. Near-future global signal experiments such as PRIZM~\cite{2019JAI.....850004P}, SCI-HI~\cite{Voytek:2013nua},  REACH~\cite{deLeraAcedo2022}, and MIST~\cite{Liu:2019yfw}, as well as future results from EDGES and SARAS will help to clarify the situation soon. 

Meanwhile, $T_{21}$ power spectrum measurements have been improving steadily. Over the last decade, experiments such as GMRT~\cite{Paciga:2013fj}, MWA~\cite{Dillon:2013rfa,Beardsley:2016njr,Li:2019kqp,Barry:2019qxp,Trott:2020szf,Yoshiura:2021yfx}, LOFAR~\cite{Patil:2017zqk,Mertens:2020llj}, PAPER~\cite{Kolopanis:2019vbl}, LEDA~\cite{Garsden:2021kdo} and HERA~\cite{HERA:2021bsv,HERA:2022wmy} have set increasingly strong upper limits on the power spectrum for comoving wavenumbers \SIrange{0.03}{3}{\per\mega\parsec}, in a broad redshift range of $6 \lesssim z \lesssim 17$; current limits are potentially one order of magnitude away from optimistic $\Lambda$CDM expectations~\cite{Reis:2021nqf}.

Inspired by the EDGES result, recent effort has been directed toward finding models that enhance the global signal. This can be accomplished by either increasing the brightness of the background radiation~\cite{Feng:2018rje,Bowman:2018yin,Ewall-Wice:2018bzf,Pospelov:2018kdh,Fialkov:2019vnb,Reis:2020arr}, or by reducing the baryon temperature~\cite{1408.1109,1603.06795,1803.03629,1803.09739,1803.09390,1803.09398,1803.11169}, typically through DM-baryon scattering~\cite{Barkana:2018lgd,Tashiro:2014tsa,Munoz:2015bca,Munoz:2018jwq,Munoz:2018pzp,Fialkov:2018xre,Berlin:2018sjs,Barkana:2018qrx,Kovetz:2018zan,Liu:2019knx,Creque-Sarbinowski:2019mcm,Aboubrahim:2021ohe,Adshead:2022ovo}. In this \textit{Letter}, we revisit the two-fluid dark sector---comprising a dominant cold DM (CDM) and a subdominant millicharged DM (mDM)---first presented in Ref.~\cite{Liu:2019knx}. This model generates a large global signal absorption trough by cooling baryons efficiently, without introducing significant drag on the baryonic fluid in the early Universe. Here, we show that the same model also leads to an enhanced power spectrum that can be several orders of magnitude larger than $\Lambda$CDM expectations. A strong correlation between baryon temperature and the baryon-DM bulk relative velocity naturally imprints large acoustic oscillations on the signal~\cite{Barkana:2018lgd,Munoz:2015bca,Fialkov:2018xre}, with a much larger amplitude than is possible through the effect on galaxy formation in standard astrophysical models~\cite{Tseliakhovich:2010bj,Dalal:2010yt,McQuinn:2012rt,Visbal:2012aw,Munoz:2018jwq,Munoz:2019rhi}. An oscillation signature is possible in principle within the simpler, non-interacting millicharged DM model~\cite{Munoz:2018pzp,Munoz:2018jwq}, but it is erased by drag at early times throughout the small parameter space of this model that remains consistent with observational constraints, particularly the constraints from the cosmic microwave background (CMB)~\cite{PhysRevLett.122.041301,Boddy:2018wzy,Kovetz:2018zan}. The predicted power spectrum in the interacting millicharged DM model provides a near-future, new-physics target for 21-cm power spectrum experiments as their sensitivity improves.

\noindent {\bf 21-cm cosmology:} 
As background radiation photons pass through a region of neutral hydrogen, they interact with the hydrogen hyperfine states. Consequently, absorption, spontaneous emission and stimulated emission of 21-cm photons change the background radiation intensity at that wavelength. The photon intensity then redshifts; the temperature contrast between the transmitted radiation and the background radiation, observed today at a wavelength $21 (1+z)$ \SI{}{\centi\meter}, is referred to as the \textit{21-cm brightness temperature} and  denoted by $T_{21}(z)$. We assume throughout this \textit{Letter} that the background radiation temperature is the CMB temperature, $T_\gamma(z) = 2.725 (1+z) \, \SI{}{\kelvin}$. $T_{21}$ is determined by the spin temperature $T_\text{S}$ of neutral hydrogen gas, via~\cite{Madau:1996cs,2006PhR...433..181F,Barkana:2016nyr}
\begin{alignat}{1}\label{eq:def_T21}
	T_{21}(z) \simeq \frac{1}{1+z} \left[
	T_\text{S}(z) - T_\gamma(z) \right] \left[ 1 - e^{-\tau(z)} \right] \,, 
\end{alignat}
where $\tau(z)$ is the effective optical depth of photons with wavelength 21 cm at redshift $z$.

$T_\text{S}$ is determined by the interaction of neutral hydrogen atoms with: \textit{1)}  the CMB photons at temperature $T_\gamma$; \textit{2)} other hydrogen atoms in the gas with temperature $T_\text{b}$; and \textit{3)} Lyman-$\alpha$ (Ly$\alpha$) radiation, which influences $T_\text{S}$ through the Wouthuysen-Field (WF) effect~\cite{1952AJ.....57R..31W,1958PIRE...46..240F}. The first process drives $T_\text{S} \to T_\gamma$, while the other two processes pull $T_\text{S} \to T_{\rm b}$ instead; as a result, $T_\text{S}$ takes a value between $T_\gamma$ and $T_\text{b}$. The relative strength of these processes can be parameterized by a single coupling coefficient $x_\text{tot,eff}$, encapsulating the collisional and Ly$\alpha$ interactions~\cite{Barkana:2016nyr},
\begin{alignat}{1}
    T_\text{S}^{-1}(\vecb{x}, z) = \frac{T_\gamma^{-1}(z) + x_\text{tot,eff}(\vecb{x}, z) T_\text{b}^{-1}(\vecb{x}, z)}{1 + x_\text{tot,eff}(\vecb{x}, z)} \,,
    \label{eq:spin_temp_main}
\end{alignat}
where we have  also introduced the spatial ($\vecb{x}$) dependence of $T_{\rm S}$, resulting from inhomogeneities in $T_{\rm b}$ and $x_\text{tot,eff}$. Further details on the value of $x_\text{tot,eff}$ and other details of the calculation of $T_{21}$ are discussed in the Supplemental Material.

There are two main types of 21-cm observables. The first is the global or sky-averaged brightness temperature, which we denote by $\thb{T_{21}}(z)$. After the first stars formed and started emitting significant Ly$\alpha$ radiation at $z \lesssim 25$, but prior to substantial X-ray heating of the intergalactic medium (IGM) at $z \lesssim 15$, $T_\text{b} \leq T_\text{S} \leq T_\gamma$, leading to a global signal that is in absorption, i.e., $\thb{T_{21}} \leq 0$. A value of $\thb{T_{21}}(z = 17)$ no lower than approximately $ -\SI{150}{\milli\kelvin}$ is expected within $\Lambda$CDM cosmology~\cite{Reis:2021nqf}. 

The second observable is the power spectrum of spatial fluctuations in the brightness temperature, $P_{T_{21}}(k, z)$, which is a function of comoving wavenumber $k$ and redshift $z$. The dimensionless power spectrum $P_{T_{21}}/\thb{T_{21}}^2$ is the Fourier transform of the two-point correlation function (2PCF) of fluctuations, $\xi_{T_{21}}(\vecb{x}) = \langle \delta_{T_{21}} (\vecb{x}_1) \delta_{T_{21}} (\vecb{x}_2) \rangle_{\vecb{x}}$, where $\langle \cdots \rangle_{\vecb{x}}$ denotes a spatial average over all pairs of points $\vecb{x}_1$ and $\vecb{x}_2$ such that $\vecb{x} = \vecb{x}_1 - \vecb{x}_2$, and where $\delta_{T_{21}} \equiv T_{21} /\langle T_{21} \rangle-1$. For our isotropic Universe,  $\xi_{T_{21}}$ is a function only of $x \equiv |\vecb{x}|$ and $P_{T_{21}}$ can be written as
\begin{equation}\label{eq:xi_to_P}
	\frac{P_{T_{21}}(k,z)}{ \thb{T_{21}}^{2}(z)}= 4 \pi \int_{0}^{\infty} d x \, x^{2} \xi_{T_{21}}(x, z) \frac{\sin(k x)}{kx} \,.
\end{equation}
$P_{T_{21}}$ is often equivalently expressed in terms of the power per $\log k$, $\Delta_{21}^2(k, z) \equiv k^3 P_{T_{21}}(k,z) / (2\pi^2)$. \vspace{0.4 cm}

\noindent{\bf Two-fluid dark sector:}
The two-fluid interacting millicharged DM model,  first proposed in Ref.~\cite{Liu:2019knx}, is capable of cooling baryons during the cosmic dark ages sufficiently to produce a global signal $\langle T_{21} \rangle (z = 17)$ consistent with the EDGES result, while avoiding stringent CMB constraints on momentum transfer between baryons and dark matter~\cite{Dvorkin:2013cea,Xu:2018efh,Gluscevic:2017ywp,Boddy:2018wzy,Kovetz:2018zan}. To accomplish this, the dark sector comprises two components: a millicharged component (mDM) with mass $m_\text{m}$ and electric charge $Q$ that makes up $f_\text{m} \lesssim 4 \times 10^{-3}$ of the total DM energy density, and a cold component (CDM) which accounts for the remainder. A light mediator between mDM and CDM allows for energy transfer between these two fluids. 

Prior to recombination, the mDM electric charge and mDM-CDM couplings are set such that the mDM fluid is tightly coupled to the baryons. After recombination, the baryonic fluid becomes mostly neutral; as a result, the mDM fluid decouples from the baryons, and instead becomes coupled to the CDM fluid, which has a temperature below the baryon temperature at all times. 
The mDM-baryon interactions now transfer heat from the baryons to the entire CDM-mDM fluid, cooling $T_\text{b}$ well below the $\Lambda$CDM expectation. The tight mDM-CDM coupling allows for heat flowing from the baryonic fluid to be shared among all  dark-sector particles, greatly enhancing the available heat capacity for cooling. In this \textit{Letter}, we focus on $m_\text{m} \gtrsim \SI{1}{\giga\eV}$, which are sufficiently massive to avoid CMB $N_\text{eff}$ limits on a combination of millicharged particles and light dark photons~\cite{Adshead:2022ovo}.

As in the $\Lambda$CDM paradigm, baryon acoustic oscillations set up a local bulk relative velocity between the baryons (together with the mDM) and CDM, $\vecb{v}_\text{bC}(\vecb{x})$, with a root-mean-square velocity of $v_\text{rms} \equiv \langle v_\text{bC}^2 (\vecb{x}) \rangle^{1/2} \approx \SI{29}{\kilo\meter\per\second}$ at $z = 1010$~\cite{Tseliakhovich:2010bj}. Unlike the non-interacting millicharged DM model~\cite{Munoz:2018pzp}, the mDM-CDM coupling in the interacting model restores the mDM-baryon velocity difference after cosmic recombination. 
Since the mDM-baryon interaction responsible for heat transfer from the baryons weakens rapidly with velocity, patches with large initial $v_\text{bC}$ remain hotter than the rapidly-cooling patches with vanishing $v_\text{bC}$~\cite{Barkana:2018lgd,Munoz:2015bca}. This is an important point: the relative bulk motion at $z\sim1010$ results in different initial conditions for the baryon temperature evolution at each spatial location. This spatial variation in $T_{\rm b}$ leads to spatial variation in the brightness temperature $T_{21}$ through Eqs.~\eqref{eq:def_T21} and \eqref{eq:spin_temp_main}. In particular, the correlation with $v_\text{bC}$ imprints the acoustic oscillations in the $v_\text{bC}$ power spectrum onto the 21-cm power spectrum~\cite{Tseliakhovich:2010bj,Barkana:2018lgd}.

Following Ref.~\cite{Liu:2019knx}, we compute $T_\text{b}$ as a function of dark matter parameters $Q$ and $m_\text{m}$, fixing the CDM mass at $m_\text{C} = \SI{100}{\mega\eV}$ and also $f_\text{m} = 10^{-4}$. We choose the maximal coupling between mDM and CDM, permitting both tight coupling between mDM and baryons  before recombination, and a sufficiently small drag on the baryonic fluid to avoid CMB power spectrum constraints (\cite{Boddy:2018wzy,Kovetz:2018zan}, also see Appendix C of Ref.~\cite{Liu:2019knx}). We integrate the differential equations governing the properties of the mDM, CDM and baryon fluids starting from photon decoupling at $z = 1010$, for various initial bulk velocities $v_{\rm bC}$, ultimately obtaining $T_\text{b}(Q, m_\text{m}; v_\text{bC};z)$. 
\vspace{0.4 cm}

\noindent \textbf{Astrophysics Modeling:}
To relate $T_\text{b}(Q, m_\text{m}; v_\text{bC}, z)$ from our model to a value of $T_{21}(Q, m_\text{m}; v_\text{bC}, z)$ through Eqs.~\eqref{eq:def_T21} and~\eqref{eq:spin_temp_main}, we need an astrophysical model for the Ly$\alpha$ and X-ray radiation fields. Ly$\alpha$ photons determine the coupling of $T_\text{S}$ to $T_\text{b}$ in Eq.~\eqref{eq:spin_temp_main}, while both Ly$\alpha$ and X-ray photons lead to IGM heating, partially counteracting cooling by mDM. To determine both of these effects, we rely on a large-scale, semi-numerical 21-cm code based on Refs.~\cite{Visbal:2012aw,Fialkov:2013uwm,Reis:2021nqf}. In this \textit{Letter}, we aim to highlight the discovery potential of the two-fluid dark sector model in the 21-cm power spectrum; we therefore account for the minimal effect of realistic astrophysical models. This contrasts with the typically-adopted simplistic approach that derives the maximum possible signal by assuming no astrophysical heating together with full Ly$\alpha$ coupling (i.e., $x_\text{tot,eff} \rightarrow \infty$); in practice, this limit is not possible, since strong Ly$\alpha$ coupling brings along with it significant heating, thus narrowing the range of possible 21-cm signals \cite{Reis:2021nqf}. We use an ensemble of 140 realistic astrophysical models from the semi-numerical simulations, chosen to minimize astrophysical heating and maximize Ly$\alpha$ coupling. As shown below, we must use an ensemble since the astrophysical model that gives the maximum absorption of $T_{21}$ depends on the DM parameters and on redshift. Further details on the simulations and the astrophysical parameters are discussed in the Supplemental Material. 

As we focus on models with subdominant X-ray heating, the dominant process counteracting dark cooling is Ly$\alpha$ heating. Ly$\alpha$ heating results from  scattering of Ly$\alpha$ photons, either directly from atomic recoil~\cite{Chen:2003gc,Chuzhoy:2005wv,2006MNRAS.372.1093F}, or by mediating heat transfer from the CMB to the baryons~\cite{Venumadhav:2018uwn}. From each simulation, we obtain the spatially averaged baryon temperature ${T_\text{b}^A}(z)$, with $A$ indexing the 140 astrophysical models. To account for astrophysical heating in our calculations we define the excess heating as $\delta T_\text{b}^A(z) \equiv {T_\text{b}^A}(z) - T_\text{b}^0(z)$, where $T_\text{b}^0(z)$ is the $\Lambda$CDM prediction for the baryon temperature in the absence of heating. The final baryon temperature is then
\begin{equation}\label{eq:T_b_Astro}
	T_\text{b}^A(Q, m_\text{m} ; v_\text{bC}; z) = T_\text{b}(Q, m_\text{m} ; v_\text{bC} ; z) + \delta T_\text{b}^A (z)
\end{equation}
for each astrophysical model. The second ingredient we obtain from each simulation is the average effective coupling $x_\text{tot,eff}^A(T_\text{b};z)$ from the spatially averaged Ly$\alpha$ radiation field. $T^A_{21}(Q, m_\text{m}; v_\text{bC}; z)$ is then determined by substituting $x_\text{tot,eff}^A(T_\text{b};z)$ into Eq.~\eqref{eq:spin_temp_main} together with Eq.~\eqref{eq:T_b_Astro}. 

Note that we have ignored spatial variations in $\delta T_\text{b}^A$, as well as in the Ly$\alpha$ radiation field, using only their mean values, and thus leaving the dependence of $T_\text{b}$ on $v_{\rm bC}$ as the only source of fluctuations. While a fully self-consistent treatment including astrophysical heating and dark cooling would account for all spatial variations concurrently, the simplified prescription presented here is computationally much more feasible, and is a reasonable approximation within the parameter range of interest, for which the effect of velocity fluctuations strongly dominate. 
\vspace{0.4 cm}

\noindent{\bf Bulk relative velocity:} 
In the two-fluid interacting dark-sector model, the fluctuations of $T_\text{21}$ are set primarily by the dependence of the baryon temperature on $v_\text{bC}$. This correlates the spatial variations of $T_\text{21}$ (at any redshift), to those of $v_{\rm bC}$ at $z\sim1010$, assuming that any other sources of such fluctuations are subdominant. 

Since the drag between the baryons and CDM is small, $\vecb{v}_\text{bC}$ at recombination is as it is in the $\Lambda$CDM paradigm: a Gaussian random field with a power spectrum $P_v(k)$ defined through $\langle \tilde{v}_\text{bC}(\vecb{k}) \tilde{v}_\text{bC}(\vecb{k}') \rangle = (2\pi)^3 \delta^{(3)}(\vecb{k} + \vecb{k}') P_v(k)$, where $\hat{k} \tilde{v}_\text{bC}(\vecb{k}) \equiv \tilde{\vecb{v}}_\text{bC}(\vecb{k})$, the Fourier transform of $\vecb{v}_\text{bC}(\vecb{x})$. $P_v(k)$ exhibits the characteristic acoustic oscillations in $k$~\cite{Tseliakhovich:2010bj}.

Under our assumption that the spatial fluctuations of $T_{21}$ are dominated by its dependence on $v_\text{bC}$, the global signal is evaluated as
\begin{alignat}{1}
    \langle T_{21} \rangle = \left( \frac{3}{2\pi v_\text{rms}^2} \right)^{3/2} \int d^3 \vecb{v}_\text{bC} \, T_{21} (v_\text{bC}) \exp \left( - \frac{3 v_\text{bC}^2}{2 v_\text{rms}^2} \right) \,.
    \label{eq:global_signal}
\end{alignat}
Similarly, the $T_{21}$ 2PCF is given by
\begin{multline}
    \xi_{T_{21}}(\vecb{x}) = \int d^3 \vecb{v}_{\text{bC},a} \int d^3 \vecb{v}_{\text{bC},b} \\
    \times \mathcal{P}(\vecb{v}_{\text{bC},a}, \vecb{v}_{\text{bC},b}; \vecb{x}) \delta_{T_{21}} (v_{\text{bC},a}) \delta_{T_{21}} (v_{\text{bC},b}) \,,
    \label{eq:6D_gaussian}
\end{multline}
where $\mathcal{P}\left(\vecb{v}_{\text{bC},a}, \vecb{v}_{\text{bC},b} ; \vecb{x}\right)$ is the joint PDF of 3D bulk relative velocities at points $a$ and $b$ separated by vector $\vecb{x}$.  Since $v_\text{bC}$ is a Gaussian random field, $\mathcal{P}$ is completely specified by $P_v(k)$.  In  the Supplemental Material, we specify the exact structure of $\mathcal{P}$, and explain the details of our computation of $\xi_{T_{21}}$. In particular, we improve on previous results \cite{Dalal:2010yt,Ali-Haimoud:2013hpa} by reducing the 6D integral in Eq.~\eqref{eq:6D_gaussian} to a 3D integral instead of a 4D one, making the integral much easier to evaluate numerically. The power spectrum is then calculated through Eq.~\eqref{eq:xi_to_P}.\vspace{0.4 cm}

\begin{figure*}
\centering
\includegraphics[width=0.49\textwidth]{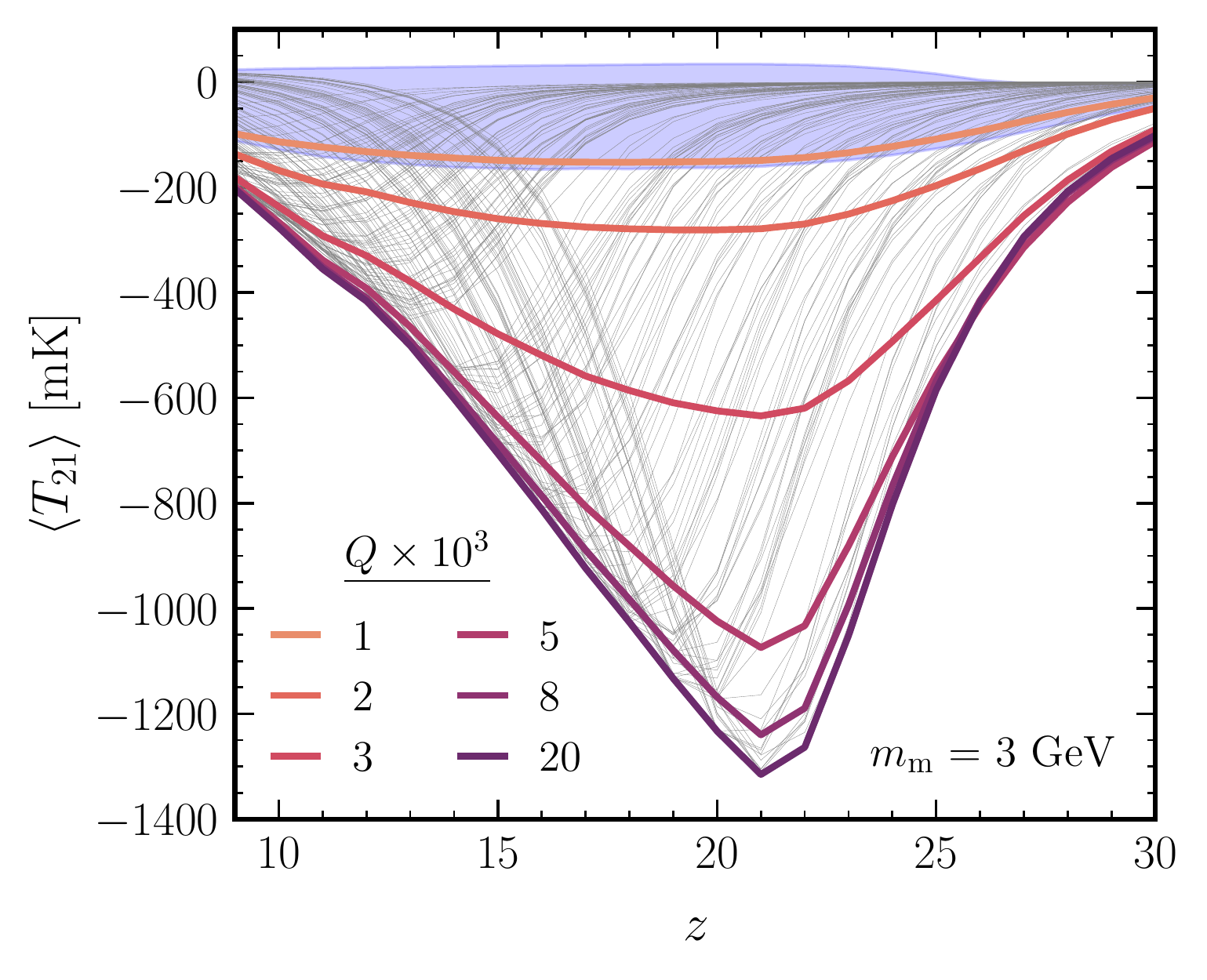}\;\;\;\;\includegraphics[width=0.49\textwidth]{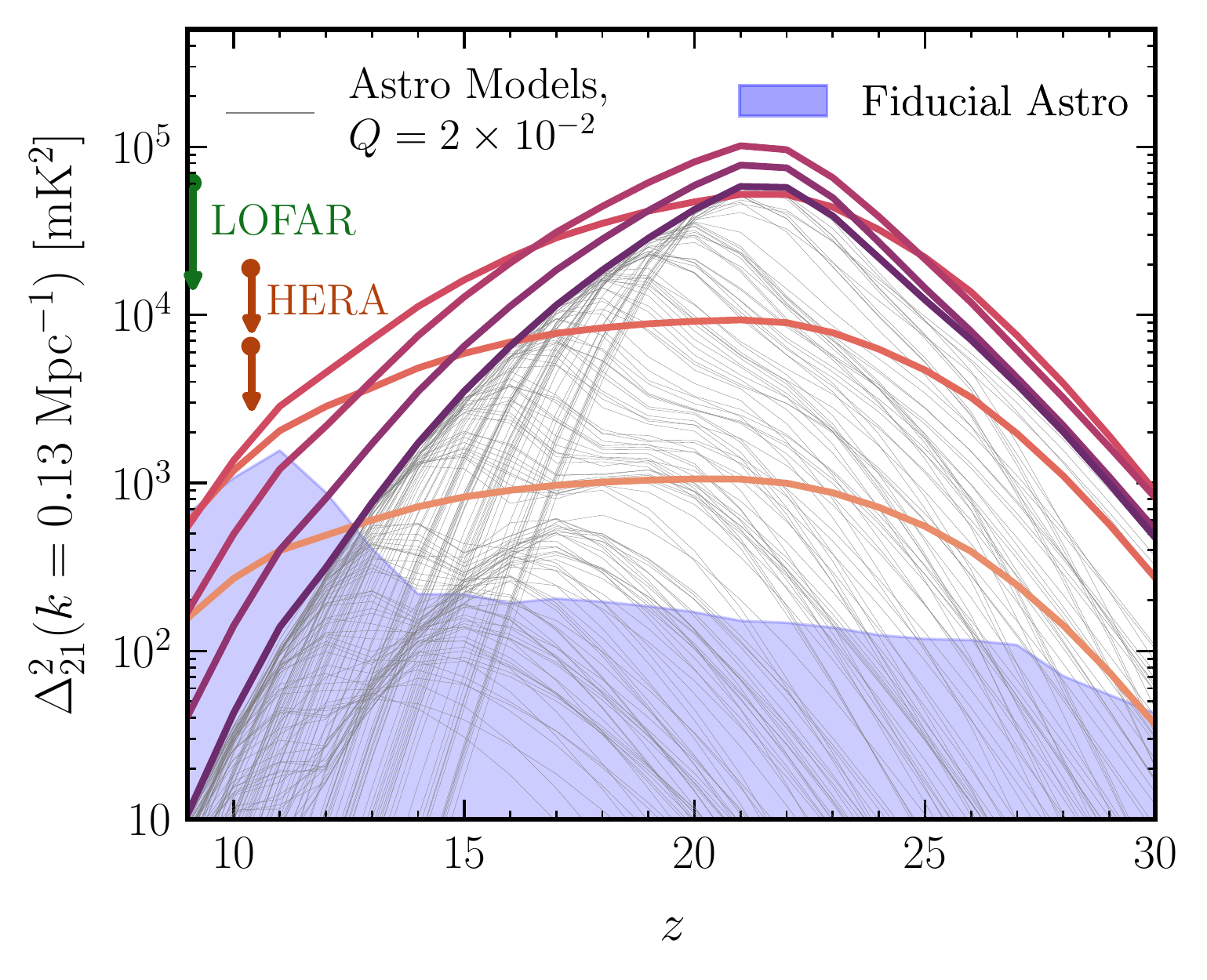}
\caption{
{\it Left:} The minimum envelope (i.e., maximum absorption) of the predicted global signal $\thb{T_{21}}$ for $m_\text{m} = \SI{3}{\giga\eV}$ across all astrophysical models, for various values of the charge (colored lines). {\it Right:} The maximum envelope of the power spectrum $\Delta_{21}^2$ at $k = \SI{0.13}{\per\mega\parsec}$ for the same values of the charge (colored lines). Existing upper limits over a range of redshifts reported by LOFAR~\cite{Patil:2017zqk,Mertens:2020llj} (green), and HERA~\cite{HERA:2021bsv,HERA:2022wmy} (orange) are shown as arrows. In both panels, the signals for all astrophysical models for $Q = 2 \times 10^{-2}$ are shown in gray; the fiducial range of both quantities expected in standard $\Lambda$CDM cosmology is shaded in purple~\cite{Reis:2021nqf}. 
}
\label{fig:global_and_PS}
\end{figure*}

\noindent \textbf{Results:}
Figure~\ref{fig:global_and_PS} shows the predicted $\thb{T_{21}}$ and $\Delta^2_{21}$ for comoving wavenumber $k = \SI{0.13}{\per\mega\parsec}$ for the two-fluid interacting millicharged DM model with $m_\text{m} = \SI{3}{\giga\eV}$, for a range of $Q$ values that are viable given current constraints. To indicate the most easily observable models, we plot lines in color representing the minimum $\langle T_{21} \rangle$ and maximum $\Delta_{21}^2$ envelopes, obtained by varying over all 140 astrophysical models. To give a sense of the variability in these models, we show in gray lines the 140 different models considered for the fixed value $Q = 2 \times 10^{-2}$. The full fiducial range of values that are possible in standard $\Lambda$CDM cosmology are shaded in purple. 

The global signal shown on the left has the characteristic absorption profile, corresponding to Ly$\alpha$ emission driving $T_\text{S} \to T_\text{b}$ before heating of the IGM causes $T_\text{b}$ to increase; by adding X-ray heating it is possible to vary the shape further. As pointed out in Ref.~\cite{Liu:2019knx}, the global signal from our model can attain the central value of the EDGES absorption profile at $z = 17$, producing signals as large as $\langle T_{21} \rangle (z = 21) \sim -\SI{1250}{\milli\kelvin}$. 

The power spectrum shown on the right is predicted to reach $\Delta_{21}^2(z = 21) \sim \SI{e5}{\milli\kelvin\squared}$ for $Q = 5 \times 10^{-3}$, several orders of magnitude larger than the conventional $\Lambda$CDM expectation~\cite{Reis:2021nqf} indicated by the purple band. For $9 \leq z \leq 30$, 21-cm power spectrum experiments such as MWA~\cite{Dillon:2013rfa,Yoshiura:2021yfx}, LOFAR~\cite{Patil:2017zqk,Mertens:2020llj}, and HERA~\cite{HERA:2021bsv,HERA:2022wmy} have already reported upper limits; we limit ourselves to $z \geq 9$ to avoid uncertainties due to reionization, even though stringent upper limits at $z = 7.9$ have recently been reported by HERA~\cite{HERA:2022wmy}. 

\begin{figure*}
\centering
\includegraphics[width=0.49\textwidth]{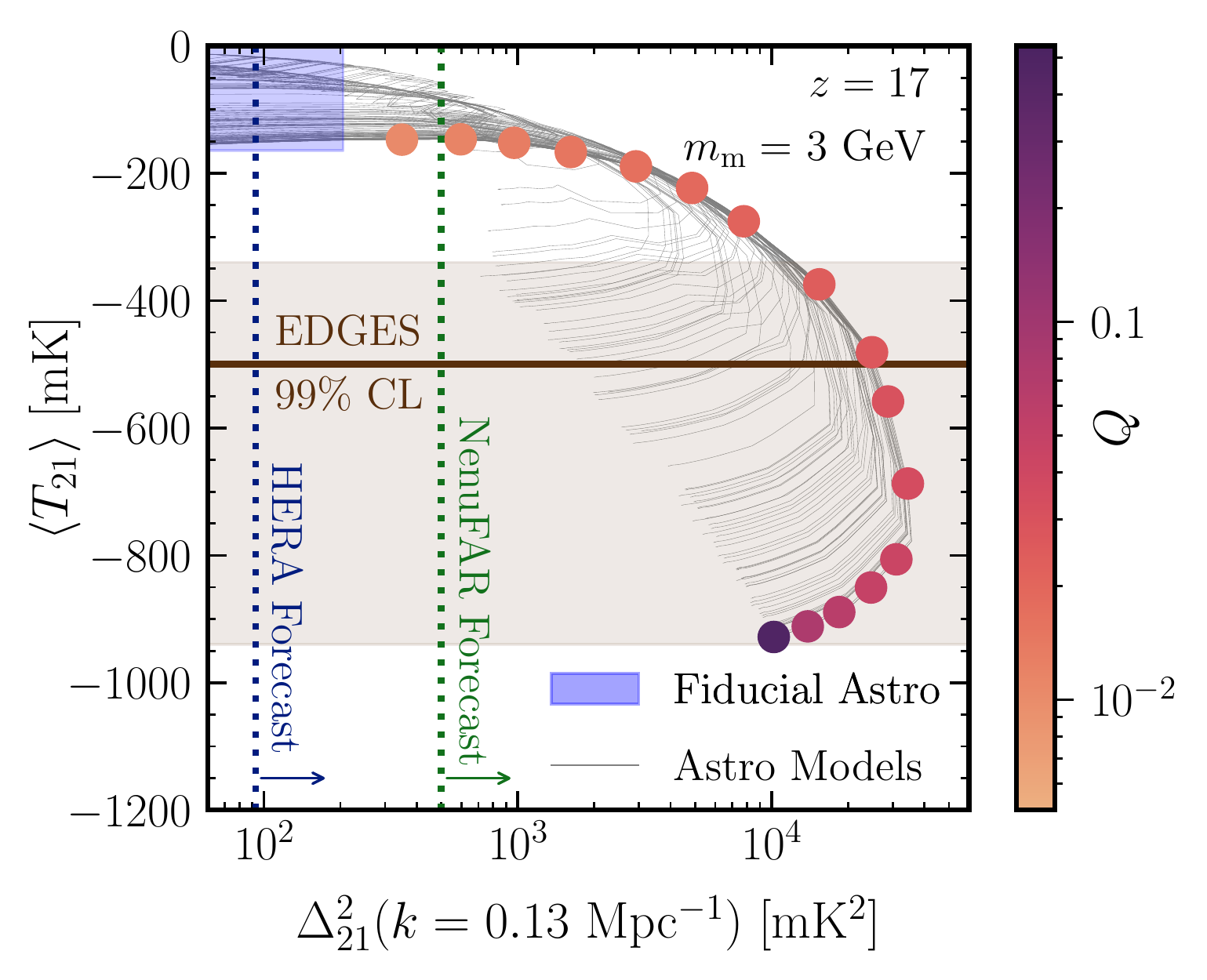}\;\;\;\;\includegraphics[width=0.49\textwidth]{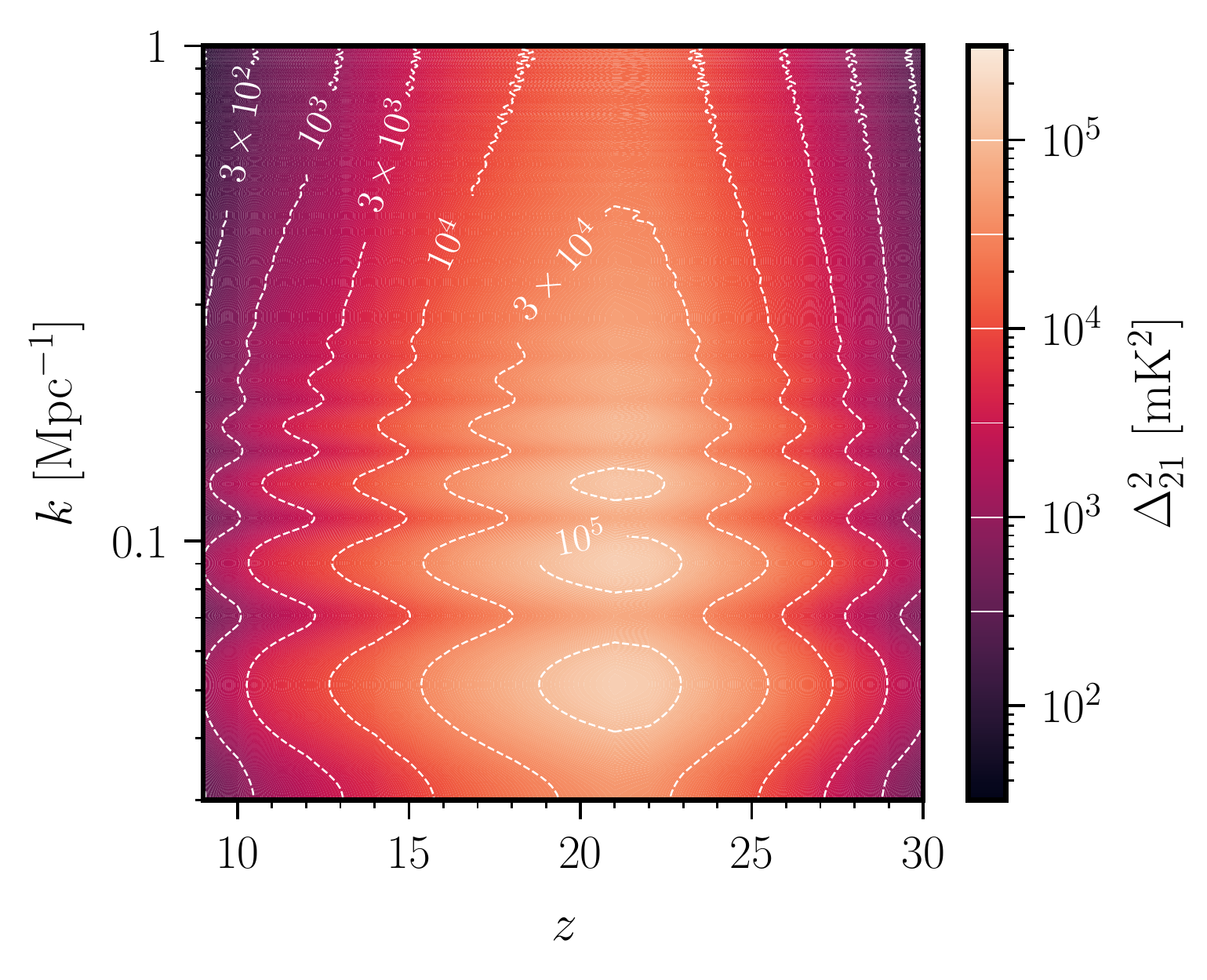}
\caption{
{\it Left:} Scatter plot of the global signal and the power spectrum at $k=0.13\,\rm Mpc^{-1}$ at $z = 17$, for $m_\text{m} = \SI{3}{\giga\eV}$. Each gray line is a fixed astrophysical model with varying charge $Q$, while the color of each large dot represents a value of $Q$ and indicates the maximum fluctuation and absorption signal for that value. The EDGES central value of $\langle T_{21} \rangle = -\SI{500}{\milli\kelvin}$ with its error range is shown for reference (brown, horizontal), although the EDGES profile is ruled out at 95\% confidence by SARAS. The projected sensitivities of NenuFAR~\cite{Mertens:2021apf} (green, dotted) and HERA~\cite{Munoz:2018jwq} (blue, dotted) with $10^3$ hours of observation are shown. 
{\it Right:} Contours of $\Delta_{T_{21}}^2(k, z)$, maximized over both astrophysics and particle physics parameters.
}
\label{fig:global_vs_PS_and_max_PS}
\end{figure*}

From comparing the $Q=2\times10^{-2}$ envelopes of both panels of Fig.~\ref{fig:global_and_PS}, it is evident that models leading to a maximal global signal absorption feature do not correspond to the largest $\Delta_{21}^2$ values. In the left panel of Fig.~\ref{fig:global_vs_PS_and_max_PS}, we show a scatter plot in the space of $\Delta_{21}^2(k = \SI{0.13}{\per\mega\parsec})$ and $\langle T_{21} \rangle$ at $z=17$, for $m_\text{m}=\SI{3}{\giga\eV}$ and the experimentally allowed range of $Q$~\cite{Liu:2019knx}. Each gray line is a fixed astrophysical model, with varying charge $Q$. The colored dots indicated the maximal $\Delta_{21}^2$ and minimal $\thb{T_{21}}$ across all astrophysical models, for a fixed $Q$. We see that increasing $Q$ first leads to gradually increasing values of $\Delta_{21}^2$, before $\langle T_{21} \rangle$ starts decreasing significantly. This behavior stems from the $\sigma \propto v_\text{rel}^{-4}$ relative velocity dependence of the Rutherford scattering cross section between mDM and baryons---cooling is most efficient for regions where $v_\text{bC} = 0$, while for higher velocities, less efficient cooling or even heating can take place~\cite{Munoz:2015bca}. This leads to enhanced fluctuations in $T_\text{b}$ and hence $T_{21}$. As $Q$ is increased further, these fluctuations are diminished, since large regions of the IGM---with most values of $v_\text{bC}$---experience strong and early cooling; this behavior is reflected in the right panel of Fig.~\ref{fig:global_and_PS}, where $\Delta_{21}^2$ begins to decrease for $Q \gtrsim 5 \times 10^{-3}$. 

In the same figure, we show the forecast sensitivity of NenuFAR~\cite{Mertens:2021apf} and HERA~\cite{Munoz:2018jwq} after $10^3$ hours of observation, for $k = \SI{0.13}{\per\mega\parsec}$. We note that the Square Kilometre Array (SKA) is projected to do even better, reaching a sensitivity of around $1~$mK$^2$~\cite{SKA}. For many choices of $(Q, m_\text{m})$, we see that 21-cm power spectrum experiments are sensitive to these models, even though the global signal only has a value of $\langle T_{21} \rangle \approx -\SI{200}{\milli\kelvin}$, only slightly larger than the $\Lambda$CDM expectation shown in Fig.~\ref{fig:global_and_PS}~\cite{Reis:2021nqf}. At present, measurements by the SARAS experiment have excluded the central value of the EDGES absorption profile at the 95\% confidence level, but with the exclusion likely depending on the signal shape and not just on the amplitude.

Finally, in the right panel of Fig.~\ref{fig:global_vs_PS_and_max_PS}, we plot the maximum $\Delta_{21}^2$ across all astrophysical and new physics models parameters, for experimentally relevant values of $k$ and $z$. Across a broad range of $k$ and $z$, $\Delta_{21}^2 \gtrsim \SI{e3}{\milli\kelvin\squared}$, reaching values as large as $\Delta_{21}^2 \sim \SI{e5}{\milli\kelvin\squared}$, exceeding the fiducial expectation from $\Lambda$CDM cosmology by several orders of magnitude. Furthermore, large acoustic oscillations are imprinted in the $T_{21}$ signal through the correlation between $T_\text{b}$ and $v_\text{bC}$. This is currently the only viable new-physics model with large acoustic oscillations in the 21-cm power spectrum, and provides a new-physics target for experiments like NenuFAR and HERA. \vspace{0.4 cm}

\noindent{\bf Conclusion:}
We study the 21-cm power spectrum of the two-fluid interacting millicharged DM model first studied in Ref.~\cite{Liu:2019knx}, finding that the model can produce a power spectrum that will be readily detectable in near-future runs of 21-cm power spectrum experiments such as NenuFAR and HERA, and eventually the SKA. Large power spectra are possible even in models in which the predicted global signal is close to the standard $\Lambda$CDM range. This model is currently the only viable model which produces a large 21-cm power spectrum with acoustic oscillation features, demonstrating the power of such experiments in searching for new physics. Our results should provide a useful new-physics benchmark for upcoming 21-cm power spectrum experimental results. \vspace{0.4 cm}

\noindent{\bf Acknowledgments:} The authors would like to thank Yacine Ali-Ha\"{i}moud, Tomer Volansky and Omer Katz for useful discussions RB acknowledges the support of the Israel Science Foundation (grant No. 2359/20), The Ambrose Monell Foundation and the Institute for Advanced Study as well as the Vera Rubin Presidential Chair in Astronomy at UCSC and the Packard Foundation. AF was supported by the Royal Society University Research Fellowship. HL is supported by NSF
grant PHY-1915409, the DOE under Award Number DESC0007968 and the Simons Foundation. The work of NJO was supported in part by the Zuckerman STEM Leadership Program and by the National Science Foundation (NSF) under the grant No.~PHY-1915314. This work was performed in part at the Aspen Center for Physics, which is supported by NSF grant PHY-1607611. This work was also performed using the Princeton Research Computing resources at Princeton University which is a consortium of groups including the Princeton Institute for Computational Science and Engineering
and the Princeton University Office of Information Technology’s Research Computing department. This research made use of the \texttt{IPython}~\cite{PER-GRA:2007}, \texttt{Jupyter}~\cite{Kluyver2016JupyterN}, \texttt{matplotlib}~\cite{Hunter:2007}, \texttt{NumPy}~\cite{harris2020array}, \texttt{pyfftlog}~\cite{TALMAN197835,Hamilton:1999uv,dieter_werthmuller_2020_3830476}, \texttt{seaborn}~\cite{seaborn}, \texttt{pandas}~\cite{pandas:2010}, \texttt{SciPy}~\cite{2020SciPy-NMeth}, and \texttt{tqdm}~\cite{da2019tqdm} software packages.

\bibliography{21cmbib.bib}

\begin{thebibliography}{100}%
\makeatletter
\providecommand \@ifxundefined [1]{%
 \@ifx{#1\undefined}
}%
\providecommand \@ifnum [1]{%
 \ifnum #1\expandafter \@firstoftwo
 \else \expandafter \@secondoftwo
 \fi
}%
\providecommand \@ifx [1]{%
 \ifx #1\expandafter \@firstoftwo
 \else \expandafter \@secondoftwo
 \fi
}%
\providecommand \natexlab [1]{#1}%
\providecommand \enquote  [1]{``#1''}%
\providecommand \bibnamefont  [1]{#1}%
\providecommand \bibfnamefont [1]{#1}%
\providecommand \citenamefont [1]{#1}%
\providecommand \href@noop [0]{\@secondoftwo}%
\providecommand \href [0]{\begingroup \@sanitize@url \@href}%
\providecommand \@href[1]{\@@startlink{#1}\@@href}%
\providecommand \@@href[1]{\endgroup#1\@@endlink}%
\providecommand \@sanitize@url [0]{\catcode `\\12\catcode `\$12\catcode
  `\&12\catcode `\#12\catcode `\^12\catcode `\_12\catcode `\%12\relax}%
\providecommand \@@startlink[1]{}%
\providecommand \@@endlink[0]{}%
\providecommand \url  [0]{\begingroup\@sanitize@url \@url }%
\providecommand \@url [1]{\endgroup\@href {#1}{\urlprefix }}%
\providecommand \urlprefix  [0]{URL }%
\providecommand \Eprint [0]{\href }%
\providecommand \doibase [0]{http://dx.doi.org/}%
\providecommand \selectlanguage [0]{\@gobble}%
\providecommand \bibinfo  [0]{\@secondoftwo}%
\providecommand \bibfield  [0]{\@secondoftwo}%
\providecommand \translation [1]{[#1]}%
\providecommand \BibitemOpen [0]{}%
\providecommand \bibitemStop [0]{}%
\providecommand \bibitemNoStop [0]{.\EOS\space}%
\providecommand \EOS [0]{\spacefactor3000\relax}%
\providecommand \BibitemShut  [1]{\csname bibitem#1\endcsname}%
\let\auto@bib@innerbib\@empty
\bibitem [{\citenamefont {Madau}\ \emph {et~al.}(1997)\citenamefont {Madau},
  \citenamefont {Meiksin},\ and\ \citenamefont {Rees}}]{Madau:1996cs}%
  \BibitemOpen
  \bibfield  {author} {\bibinfo {author} {\bibfnamefont {Piero}\ \bibnamefont
  {Madau}}, \bibinfo {author} {\bibfnamefont {Avery}\ \bibnamefont {Meiksin}},
  \ and\ \bibinfo {author} {\bibfnamefont {Martin~J.}\ \bibnamefont {Rees}},\
  }\bibfield  {title} {\enquote {\bibinfo {title} {{21-cm tomography of the
  intergalactic medium at high redshift}},}\ }\href {\doibase 10.1086/303549}
  {\bibfield  {journal} {\bibinfo  {journal} {Astrophys. J.}\ }\textbf
  {\bibinfo {volume} {475}},\ \bibinfo {pages} {429} (\bibinfo {year}
  {1997})},\ \Eprint {http://arxiv.org/abs/astro-ph/9608010}
  {arXiv:astro-ph/9608010} \BibitemShut {NoStop}%
\bibitem [{\citenamefont {Loeb}\ and\ \citenamefont
  {Zaldarriaga}(2004)}]{Loeb:2003ya}%
  \BibitemOpen
  \bibfield  {author} {\bibinfo {author} {\bibfnamefont {Abraham}\ \bibnamefont
  {Loeb}}\ and\ \bibinfo {author} {\bibfnamefont {Matias}\ \bibnamefont
  {Zaldarriaga}},\ }\bibfield  {title} {\enquote {\bibinfo {title} {{Measuring
  the small - scale power spectrum of cosmic density fluctuations through 21 cm
  tomography prior to the epoch of structure formation}},}\ }\href {\doibase
  10.1103/PhysRevLett.92.211301} {\bibfield  {journal} {\bibinfo  {journal}
  {Phys. Rev. Lett.}\ }\textbf {\bibinfo {volume} {92}},\ \bibinfo {pages}
  {211301} (\bibinfo {year} {2004})},\ \Eprint
  {http://arxiv.org/abs/astro-ph/0312134} {arXiv:astro-ph/0312134} \BibitemShut
  {NoStop}%
\bibitem [{\citenamefont {Tseliakhovich}\ and\ \citenamefont
  {Hirata}(2010)}]{Tseliakhovich:2010bj}%
  \BibitemOpen
  \bibfield  {author} {\bibinfo {author} {\bibfnamefont {Dmitriy}\ \bibnamefont
  {Tseliakhovich}}\ and\ \bibinfo {author} {\bibfnamefont {Christopher}\
  \bibnamefont {Hirata}},\ }\bibfield  {title} {\enquote {\bibinfo {title}
  {{Relative Velocity of Dark Matter and Baryonic Fluids and the Formation of
  the First Structures}},}\ }\href {\doibase 10.1103/PhysRevD.82.083520}
  {\bibfield  {journal} {\bibinfo  {journal} {Phys. Rev.}\ }\textbf {\bibinfo
  {volume} {D82}},\ \bibinfo {pages} {083520} (\bibinfo {year} {2010})},\
  \Eprint {http://arxiv.org/abs/1005.2416} {arXiv:1005.2416 [astro-ph.CO]}
  \BibitemShut {NoStop}%
\bibitem [{\citenamefont {{Furlanetto}}\ \emph {et~al.}(2006)\citenamefont
  {{Furlanetto}}, \citenamefont {{Oh}},\ and\ \citenamefont
  {{Briggs}}}]{2006PhR...433..181F}%
  \BibitemOpen
  \bibfield  {author} {\bibinfo {author} {\bibfnamefont {Steven~R.}\
  \bibnamefont {{Furlanetto}}}, \bibinfo {author} {\bibfnamefont {S.~Peng}\
  \bibnamefont {{Oh}}}, \ and\ \bibinfo {author} {\bibfnamefont {Frank~H.}\
  \bibnamefont {{Briggs}}},\ }\bibfield  {title} {\enquote {\bibinfo {title}
  {{Cosmology at low frequencies: The 21 cm transition and the high-redshift
  Universe}},}\ }\href {\doibase 10.1016/j.physrep.2006.08.002} {\bibfield
  {journal} {\bibinfo  {journal} {Physics Reports}\ }\textbf {\bibinfo {volume}
  {433}},\ \bibinfo {pages} {181--301} (\bibinfo {year} {2006})},\ \Eprint
  {http://arxiv.org/abs/astro-ph/0608032} {arXiv:astro-ph/0608032 [astro-ph]}
  \BibitemShut {NoStop}%
\bibitem [{\citenamefont {Barkana}(2016)}]{Barkana:2016nyr}%
  \BibitemOpen
  \bibfield  {author} {\bibinfo {author} {\bibfnamefont {Rennan}\ \bibnamefont
  {Barkana}},\ }\bibfield  {title} {\enquote {\bibinfo {title} {{The Rise of
  the First Stars: Supersonic Streaming, Radiative Feedback, and 21-cm
  Cosmology}},}\ }\href {\doibase 10.1016/j.physrep.2016.06.006} {\bibfield
  {journal} {\bibinfo  {journal} {Phys. Rept.}\ }\textbf {\bibinfo {volume}
  {645}},\ \bibinfo {pages} {1--59} (\bibinfo {year} {2016})},\ \Eprint
  {http://arxiv.org/abs/1605.04357} {arXiv:1605.04357 [astro-ph.CO]}
  \BibitemShut {NoStop}%
\bibitem [{\citenamefont {Sitwell}\ \emph {et~al.}(2014)\citenamefont
  {Sitwell}, \citenamefont {Mesinger}, \citenamefont {Ma},\ and\ \citenamefont
  {Sigurdson}}]{Sitwell:2013fpa}%
  \BibitemOpen
  \bibfield  {author} {\bibinfo {author} {\bibfnamefont {Michael}\ \bibnamefont
  {Sitwell}}, \bibinfo {author} {\bibfnamefont {Andrei}\ \bibnamefont
  {Mesinger}}, \bibinfo {author} {\bibfnamefont {Yin-Zhe}\ \bibnamefont {Ma}},
  \ and\ \bibinfo {author} {\bibfnamefont {Kris}\ \bibnamefont {Sigurdson}},\
  }\bibfield  {title} {\enquote {\bibinfo {title} {{The Imprint of Warm Dark
  Matter on the Cosmological 21-cm Signal}},}\ }\href {\doibase
  10.1093/mnras/stt2392} {\bibfield  {journal} {\bibinfo  {journal} {Mon. Not.
  Roy. Astron. Soc.}\ }\textbf {\bibinfo {volume} {438}},\ \bibinfo {pages}
  {2664--2671} (\bibinfo {year} {2014})},\ \Eprint
  {http://arxiv.org/abs/1310.0029} {arXiv:1310.0029 [astro-ph.CO]} \BibitemShut
  {NoStop}%
\bibitem [{\citenamefont {Sekiguchi}\ and\ \citenamefont
  {Tashiro}(2014)}]{Sekiguchi:2014wfa}%
  \BibitemOpen
  \bibfield  {author} {\bibinfo {author} {\bibfnamefont {Toyokazu}\
  \bibnamefont {Sekiguchi}}\ and\ \bibinfo {author} {\bibfnamefont {Hiroyuki}\
  \bibnamefont {Tashiro}},\ }\bibfield  {title} {\enquote {\bibinfo {title}
  {{Constraining warm dark matter with 21 cm line fluctuations due to
  minihalos}},}\ }\href {\doibase 10.1088/1475-7516/2014/08/007} {\bibfield
  {journal} {\bibinfo  {journal} {JCAP}\ }\textbf {\bibinfo {volume} {08}},\
  \bibinfo {pages} {007} (\bibinfo {year} {2014})},\ \Eprint
  {http://arxiv.org/abs/1401.5563} {arXiv:1401.5563 [astro-ph.CO]} \BibitemShut
  {NoStop}%
\bibitem [{\citenamefont {Nebrin}\ \emph {et~al.}(2019)\citenamefont {Nebrin},
  \citenamefont {Ghara},\ and\ \citenamefont {Mellema}}]{Nebrin:2018vqt}%
  \BibitemOpen
  \bibfield  {author} {\bibinfo {author} {\bibfnamefont {Olof}\ \bibnamefont
  {Nebrin}}, \bibinfo {author} {\bibfnamefont {Raghunath}\ \bibnamefont
  {Ghara}}, \ and\ \bibinfo {author} {\bibfnamefont {Garrelt}\ \bibnamefont
  {Mellema}},\ }\bibfield  {title} {\enquote {\bibinfo {title} {{Fuzzy Dark
  Matter at Cosmic Dawn: New 21-cm Constraints}},}\ }\href {\doibase
  10.1088/1475-7516/2019/04/051} {\bibfield  {journal} {\bibinfo  {journal}
  {JCAP}\ }\textbf {\bibinfo {volume} {04}},\ \bibinfo {pages} {051} (\bibinfo
  {year} {2019})},\ \Eprint {http://arxiv.org/abs/1812.09760} {arXiv:1812.09760
  [astro-ph.CO]} \BibitemShut {NoStop}%
\bibitem [{\citenamefont {Mu\~noz}\ \emph {et~al.}(2020)\citenamefont
  {Mu\~noz}, \citenamefont {Dvorkin},\ and\ \citenamefont
  {Cyr-Racine}}]{Munoz:2019hjh}%
  \BibitemOpen
  \bibfield  {author} {\bibinfo {author} {\bibfnamefont {Julian~B.}\
  \bibnamefont {Mu\~noz}}, \bibinfo {author} {\bibfnamefont {Cora}\
  \bibnamefont {Dvorkin}}, \ and\ \bibinfo {author} {\bibfnamefont
  {Francis-Yan}\ \bibnamefont {Cyr-Racine}},\ }\bibfield  {title} {\enquote
  {\bibinfo {title} {{Probing the Small-Scale Matter Power Spectrum with
  Large-Scale 21-cm Data}},}\ }\href {\doibase 10.1103/PhysRevD.101.063526}
  {\bibfield  {journal} {\bibinfo  {journal} {Phys. Rev. D}\ }\textbf {\bibinfo
  {volume} {101}},\ \bibinfo {pages} {063526} (\bibinfo {year} {2020})},\
  \Eprint {http://arxiv.org/abs/1911.11144} {arXiv:1911.11144 [astro-ph.CO]}
  \BibitemShut {NoStop}%
\bibitem [{\citenamefont {Jones}\ \emph {et~al.}(2021)\citenamefont {Jones},
  \citenamefont {Palatnick}, \citenamefont {Chen}, \citenamefont {Beane},\ and\
  \citenamefont {Lidz}}]{Jones:2021mrs}%
  \BibitemOpen
  \bibfield  {author} {\bibinfo {author} {\bibfnamefont {Dana}\ \bibnamefont
  {Jones}}, \bibinfo {author} {\bibfnamefont {Skyler}\ \bibnamefont
  {Palatnick}}, \bibinfo {author} {\bibfnamefont {Richard}\ \bibnamefont
  {Chen}}, \bibinfo {author} {\bibfnamefont {Angus}\ \bibnamefont {Beane}}, \
  and\ \bibinfo {author} {\bibfnamefont {Adam}\ \bibnamefont {Lidz}},\
  }\bibfield  {title} {\enquote {\bibinfo {title} {{Fuzzy Dark Matter and the
  21 cm Power Spectrum}},}\ }\href {\doibase 10.3847/1538-4357/abf0a9}
  {\bibfield  {journal} {\bibinfo  {journal} {Astrophys. J.}\ }\textbf
  {\bibinfo {volume} {913}},\ \bibinfo {pages} {7} (\bibinfo {year} {2021})},\
  \Eprint {http://arxiv.org/abs/2101.07177} {arXiv:2101.07177 [astro-ph.CO]}
  \BibitemShut {NoStop}%
\bibitem [{\citenamefont {Hotinli}\ \emph {et~al.}(2021)\citenamefont
  {Hotinli}, \citenamefont {Marsh},\ and\ \citenamefont
  {Kamionkowski}}]{Hotinli:2021vxg}%
  \BibitemOpen
  \bibfield  {author} {\bibinfo {author} {\bibfnamefont {Selim~C.}\
  \bibnamefont {Hotinli}}, \bibinfo {author} {\bibfnamefont {David J.~E.}\
  \bibnamefont {Marsh}}, \ and\ \bibinfo {author} {\bibfnamefont {Marc}\
  \bibnamefont {Kamionkowski}},\ }\bibfield  {title} {\enquote {\bibinfo
  {title} {{Probing ultra-light axions with the 21-cm Signal during Cosmic
  Dawn}},}\ }\href@noop {} {\  (\bibinfo {year} {2021})},\ \Eprint
  {http://arxiv.org/abs/2112.06943} {arXiv:2112.06943 [astro-ph.CO]}
  \BibitemShut {NoStop}%
\bibitem [{\citenamefont {Flitter}\ and\ \citenamefont
  {Kovetz}(2022)}]{Flitter:2022pzf}%
  \BibitemOpen
  \bibfield  {author} {\bibinfo {author} {\bibfnamefont {Jordan}\ \bibnamefont
  {Flitter}}\ and\ \bibinfo {author} {\bibfnamefont {Ely~D.}\ \bibnamefont
  {Kovetz}},\ }\bibfield  {title} {\enquote {\bibinfo {title} {{Closing the
  window on fuzzy dark matter with the 21cm signal}},}\ }\href@noop {} {\
  (\bibinfo {year} {2022})},\ \Eprint {http://arxiv.org/abs/2207.05083}
  {arXiv:2207.05083 [astro-ph.CO]} \BibitemShut {NoStop}%
\bibitem [{\citenamefont {Evoli}\ \emph {et~al.}(2014)\citenamefont {Evoli},
  \citenamefont {Mesinger},\ and\ \citenamefont {Ferrara}}]{1408.1109}%
  \BibitemOpen
  \bibfield  {author} {\bibinfo {author} {\bibfnamefont {Carmelo}\ \bibnamefont
  {Evoli}}, \bibinfo {author} {\bibfnamefont {Andrei}\ \bibnamefont
  {Mesinger}}, \ and\ \bibinfo {author} {\bibfnamefont {Andrea}\ \bibnamefont
  {Ferrara}},\ }\bibfield  {title} {\enquote {\bibinfo {title} {{Unveiling the
  nature of dark matter with high redshift 21 cm line experiments}},}\ }\href
  {\doibase 10.1088/1475-7516/2014/11/024} {\bibfield  {journal} {\bibinfo
  {journal} {JCAP}\ }\textbf {\bibinfo {volume} {11}},\ \bibinfo {pages} {024}
  (\bibinfo {year} {2014})},\ \Eprint {http://arxiv.org/abs/1408.1109}
  {arXiv:1408.1109 [astro-ph.HE]} \BibitemShut {NoStop}%
\bibitem [{\citenamefont {Lopez-Honorez}\ \emph {et~al.}(2016)\citenamefont
  {Lopez-Honorez}, \citenamefont {Mena}, \citenamefont {Molin\'e},
  \citenamefont {Palomares-Ruiz},\ and\ \citenamefont {Vincent}}]{1603.06795}%
  \BibitemOpen
  \bibfield  {author} {\bibinfo {author} {\bibfnamefont {Laura}\ \bibnamefont
  {Lopez-Honorez}}, \bibinfo {author} {\bibfnamefont {Olga}\ \bibnamefont
  {Mena}}, \bibinfo {author} {\bibfnamefont {\'Angeles}\ \bibnamefont
  {Molin\'e}}, \bibinfo {author} {\bibfnamefont {Sergio}\ \bibnamefont
  {Palomares-Ruiz}}, \ and\ \bibinfo {author} {\bibfnamefont {Aaron~C.}\
  \bibnamefont {Vincent}},\ }\bibfield  {title} {\enquote {\bibinfo {title}
  {{The 21 cm signal and the interplay between dark matter annihilations and
  astrophysical processes}},}\ }\href {\doibase 10.1088/1475-7516/2016/08/004}
  {\bibfield  {journal} {\bibinfo  {journal} {JCAP}\ }\textbf {\bibinfo
  {volume} {08}},\ \bibinfo {pages} {004} (\bibinfo {year} {2016})},\ \Eprint
  {http://arxiv.org/abs/1603.06795} {arXiv:1603.06795 [astro-ph.CO]}
  \BibitemShut {NoStop}%
\bibitem [{\citenamefont {D'Amico}\ \emph {et~al.}(2018)\citenamefont
  {D'Amico}, \citenamefont {Panci},\ and\ \citenamefont
  {Strumia}}]{1803.03629}%
  \BibitemOpen
  \bibfield  {author} {\bibinfo {author} {\bibfnamefont {Guido}\ \bibnamefont
  {D'Amico}}, \bibinfo {author} {\bibfnamefont {Paolo}\ \bibnamefont {Panci}},
  \ and\ \bibinfo {author} {\bibfnamefont {Alessandro}\ \bibnamefont
  {Strumia}},\ }\bibfield  {title} {\enquote {\bibinfo {title} {{Bounds on Dark
  Matter annihilations from 21 cm data}},}\ }\href {\doibase
  10.1103/PhysRevLett.121.011103} {\bibfield  {journal} {\bibinfo  {journal}
  {Phys. Rev. Lett.}\ }\textbf {\bibinfo {volume} {121}},\ \bibinfo {pages}
  {011103} (\bibinfo {year} {2018})},\ \Eprint
  {http://arxiv.org/abs/1803.03629} {arXiv:1803.03629 [astro-ph.CO]}
  \BibitemShut {NoStop}%
\bibitem [{\citenamefont {Liu}\ and\ \citenamefont
  {Slatyer}(2018)}]{1803.09739}%
  \BibitemOpen
  \bibfield  {author} {\bibinfo {author} {\bibfnamefont {Hongwan}\ \bibnamefont
  {Liu}}\ and\ \bibinfo {author} {\bibfnamefont {Tracy~R.}\ \bibnamefont
  {Slatyer}},\ }\bibfield  {title} {\enquote {\bibinfo {title} {{Implications
  of a 21-cm signal for dark matter annihilation and decay}},}\ }\href
  {\doibase 10.1103/PhysRevD.98.023501} {\bibfield  {journal} {\bibinfo
  {journal} {Phys. Rev. D}\ }\textbf {\bibinfo {volume} {98}},\ \bibinfo
  {pages} {023501} (\bibinfo {year} {2018})},\ \Eprint
  {http://arxiv.org/abs/1803.09739} {arXiv:1803.09739 [astro-ph.CO]}
  \BibitemShut {NoStop}%
\bibitem [{\citenamefont {Clark}\ \emph {et~al.}(2018)\citenamefont {Clark},
  \citenamefont {Dutta}, \citenamefont {Gao}, \citenamefont {Ma},\ and\
  \citenamefont {Strigari}}]{1803.09390}%
  \BibitemOpen
  \bibfield  {author} {\bibinfo {author} {\bibfnamefont {Steven}\ \bibnamefont
  {Clark}}, \bibinfo {author} {\bibfnamefont {Bhaskar}\ \bibnamefont {Dutta}},
  \bibinfo {author} {\bibfnamefont {Yu}~\bibnamefont {Gao}}, \bibinfo {author}
  {\bibfnamefont {Yin-Zhe}\ \bibnamefont {Ma}}, \ and\ \bibinfo {author}
  {\bibfnamefont {Louis~E.}\ \bibnamefont {Strigari}},\ }\bibfield  {title}
  {\enquote {\bibinfo {title} {{21 cm limits on decaying dark matter and
  primordial black holes}},}\ }\href {\doibase 10.1103/PhysRevD.98.043006}
  {\bibfield  {journal} {\bibinfo  {journal} {Phys. Rev. D}\ }\textbf {\bibinfo
  {volume} {98}},\ \bibinfo {pages} {043006} (\bibinfo {year} {2018})},\
  \Eprint {http://arxiv.org/abs/1803.09390} {arXiv:1803.09390 [astro-ph.HE]}
  \BibitemShut {NoStop}%
\bibitem [{\citenamefont {Cheung}\ \emph {et~al.}(2019)\citenamefont {Cheung},
  \citenamefont {Kuo}, \citenamefont {Ng},\ and\ \citenamefont
  {Tsai}}]{1803.09398}%
  \BibitemOpen
  \bibfield  {author} {\bibinfo {author} {\bibfnamefont {Kingman}\ \bibnamefont
  {Cheung}}, \bibinfo {author} {\bibfnamefont {Jui-Lin}\ \bibnamefont {Kuo}},
  \bibinfo {author} {\bibfnamefont {Kin-Wang}\ \bibnamefont {Ng}}, \ and\
  \bibinfo {author} {\bibfnamefont {Yue-Lin~Sming}\ \bibnamefont {Tsai}},\
  }\bibfield  {title} {\enquote {\bibinfo {title} {{The impact of EDGES 21-cm
  data on dark matter interactions}},}\ }\href {\doibase
  10.1016/j.physletb.2018.11.058} {\bibfield  {journal} {\bibinfo  {journal}
  {Phys. Lett. B}\ }\textbf {\bibinfo {volume} {789}},\ \bibinfo {pages}
  {137--144} (\bibinfo {year} {2019})},\ \Eprint
  {http://arxiv.org/abs/1803.09398} {arXiv:1803.09398 [astro-ph.CO]}
  \BibitemShut {NoStop}%
\bibitem [{\citenamefont {Mitridate}\ and\ \citenamefont
  {Podo}(2018)}]{1803.11169}%
  \BibitemOpen
  \bibfield  {author} {\bibinfo {author} {\bibfnamefont {Andrea}\ \bibnamefont
  {Mitridate}}\ and\ \bibinfo {author} {\bibfnamefont {Alessandro}\
  \bibnamefont {Podo}},\ }\bibfield  {title} {\enquote {\bibinfo {title}
  {{Bounds on Dark Matter decay from 21 cm line}},}\ }\href {\doibase
  10.1088/1475-7516/2018/05/069} {\bibfield  {journal} {\bibinfo  {journal}
  {JCAP}\ }\textbf {\bibinfo {volume} {05}},\ \bibinfo {pages} {069} (\bibinfo
  {year} {2018})},\ \Eprint {http://arxiv.org/abs/1803.11169} {arXiv:1803.11169
  [hep-ph]} \BibitemShut {NoStop}%
\bibitem [{\citenamefont {Barkana}(2018)}]{Barkana:2018lgd}%
  \BibitemOpen
  \bibfield  {author} {\bibinfo {author} {\bibfnamefont {Rennan}\ \bibnamefont
  {Barkana}},\ }\bibfield  {title} {\enquote {\bibinfo {title} {{Possible
  Interaction Between Baryons and Dark-Matter Particles Revealed by the First
  Stars}},}\ }\href {\doibase 10.1038/nature25791} {\bibfield  {journal}
  {\bibinfo  {journal} {Nature}\ }\textbf {\bibinfo {volume} {555}},\ \bibinfo
  {pages} {71--74} (\bibinfo {year} {2018})},\ \Eprint
  {http://arxiv.org/abs/1803.06698} {arXiv:1803.06698 [astro-ph.CO]}
  \BibitemShut {NoStop}%
\bibitem [{\citenamefont {Tashiro}\ \emph {et~al.}(2014)\citenamefont
  {Tashiro}, \citenamefont {Kadota},\ and\ \citenamefont
  {Silk}}]{Tashiro:2014tsa}%
  \BibitemOpen
  \bibfield  {author} {\bibinfo {author} {\bibfnamefont {Hiroyuki}\
  \bibnamefont {Tashiro}}, \bibinfo {author} {\bibfnamefont {Kenji}\
  \bibnamefont {Kadota}}, \ and\ \bibinfo {author} {\bibfnamefont {Joseph}\
  \bibnamefont {Silk}},\ }\bibfield  {title} {\enquote {\bibinfo {title}
  {{Effects of Dark Matter-Baryon Scattering on Redshifted 21 Cm Signals}},}\
  }\href {\doibase 10.1103/PhysRevD.90.083522} {\bibfield  {journal} {\bibinfo
  {journal} {Phys. Rev. D}\ }\textbf {\bibinfo {volume} {90}},\ \bibinfo
  {pages} {083522} (\bibinfo {year} {2014})},\ \Eprint
  {http://arxiv.org/abs/1408.2571} {arXiv:1408.2571 [astro-ph.CO]} \BibitemShut
  {NoStop}%
\bibitem [{\citenamefont {Muñoz}\ \emph {et~al.}(2015)\citenamefont {Muñoz},
  \citenamefont {Kovetz},\ and\ \citenamefont {Ali-Haïmoud}}]{Munoz:2015bca}%
  \BibitemOpen
  \bibfield  {author} {\bibinfo {author} {\bibfnamefont {Julian~B.}\
  \bibnamefont {Muñoz}}, \bibinfo {author} {\bibfnamefont {Ely~D.}\
  \bibnamefont {Kovetz}}, \ and\ \bibinfo {author} {\bibfnamefont {Yacine}\
  \bibnamefont {Ali-Haïmoud}},\ }\bibfield  {title} {\enquote {\bibinfo
  {title} {{Heating of Baryons due to Scattering with Dark Matter During the
  Dark Ages}},}\ }\href {\doibase 10.1103/PhysRevD.92.083528} {\bibfield
  {journal} {\bibinfo  {journal} {Phys. Rev.}\ }\textbf {\bibinfo {volume}
  {D92}},\ \bibinfo {pages} {083528} (\bibinfo {year} {2015})},\ \Eprint
  {http://arxiv.org/abs/1509.00029} {arXiv:1509.00029 [astro-ph.CO]}
  \BibitemShut {NoStop}%
\bibitem [{\citenamefont {Muñoz}\ and\ \citenamefont
  {Loeb}(2018)}]{Munoz:2018pzp}%
  \BibitemOpen
  \bibfield  {author} {\bibinfo {author} {\bibfnamefont {Julian~B.}\
  \bibnamefont {Muñoz}}\ and\ \bibinfo {author} {\bibfnamefont {Abraham}\
  \bibnamefont {Loeb}},\ }\bibfield  {title} {\enquote {\bibinfo {title} {{A
  small amount of mini-charged dark matter could cool the baryons in the early
  Universe}},}\ }\href {\doibase 10.1038/s41586-018-0151-x} {\bibfield
  {journal} {\bibinfo  {journal} {Nature}\ }\textbf {\bibinfo {volume} {557}},\
  \bibinfo {pages} {684} (\bibinfo {year} {2018})},\ \Eprint
  {http://arxiv.org/abs/1802.10094} {arXiv:1802.10094 [astro-ph.CO]}
  \BibitemShut {NoStop}%
\bibitem [{\citenamefont {Barkana}\ \emph {et~al.}(2018)\citenamefont
  {Barkana}, \citenamefont {Outmezguine}, \citenamefont {Redigolo},\ and\
  \citenamefont {Volansky}}]{Barkana:2018qrx}%
  \BibitemOpen
  \bibfield  {author} {\bibinfo {author} {\bibfnamefont {Rennan}\ \bibnamefont
  {Barkana}}, \bibinfo {author} {\bibfnamefont {Nadav~Joseph}\ \bibnamefont
  {Outmezguine}}, \bibinfo {author} {\bibfnamefont {Diego}\ \bibnamefont
  {Redigolo}}, \ and\ \bibinfo {author} {\bibfnamefont {Tomer}\ \bibnamefont
  {Volansky}},\ }\bibfield  {title} {\enquote {\bibinfo {title} {{Strong
  Constraints on Light Dark Matter Interpretation of the Edges Signal}},}\
  }\href {\doibase 10.1103/PhysRevD.98.103005} {\bibfield  {journal} {\bibinfo
  {journal} {Phys. Rev. D}\ }\textbf {\bibinfo {volume} {98}},\ \bibinfo
  {pages} {103005} (\bibinfo {year} {2018})},\ \Eprint
  {http://arxiv.org/abs/1803.03091} {arXiv:1803.03091 [hep-ph]} \BibitemShut
  {NoStop}%
\bibitem [{\citenamefont {Fialkov}\ \emph {et~al.}(2018)\citenamefont
  {Fialkov}, \citenamefont {Barkana},\ and\ \citenamefont
  {Cohen}}]{Fialkov:2018xre}%
  \BibitemOpen
  \bibfield  {author} {\bibinfo {author} {\bibfnamefont {Anastasia}\
  \bibnamefont {Fialkov}}, \bibinfo {author} {\bibfnamefont {Rennan}\
  \bibnamefont {Barkana}}, \ and\ \bibinfo {author} {\bibfnamefont {Aviad}\
  \bibnamefont {Cohen}},\ }\bibfield  {title} {\enquote {\bibinfo {title}
  {{Constraining Baryon--Dark Matter Scattering with the Cosmic Dawn 21-Cm
  Signal}},}\ }\href {\doibase 10.1103/PhysRevLett.121.011101} {\bibfield
  {journal} {\bibinfo  {journal} {Phys. Rev. Lett.}\ }\textbf {\bibinfo
  {volume} {121}},\ \bibinfo {pages} {011101} (\bibinfo {year} {2018})},\
  \Eprint {http://arxiv.org/abs/1802.10577} {arXiv:1802.10577 [astro-ph.CO]}
  \BibitemShut {NoStop}%
\bibitem [{\citenamefont {Berlin}\ \emph {et~al.}(2018)\citenamefont {Berlin},
  \citenamefont {Hooper}, \citenamefont {Krnjaic},\ and\ \citenamefont
  {McDermott}}]{Berlin:2018sjs}%
  \BibitemOpen
  \bibfield  {author} {\bibinfo {author} {\bibfnamefont {Asher}\ \bibnamefont
  {Berlin}}, \bibinfo {author} {\bibfnamefont {Dan}\ \bibnamefont {Hooper}},
  \bibinfo {author} {\bibfnamefont {Gordan}\ \bibnamefont {Krnjaic}}, \ and\
  \bibinfo {author} {\bibfnamefont {Samuel~D.}\ \bibnamefont {McDermott}},\
  }\bibfield  {title} {\enquote {\bibinfo {title} {{Severely Constraining Dark
  Matter Interpretations of the 21-Cm Anomaly}},}\ }\href {\doibase
  10.1103/PhysRevLett.121.011102} {\bibfield  {journal} {\bibinfo  {journal}
  {Phys. Rev. Lett.}\ }\textbf {\bibinfo {volume} {121}},\ \bibinfo {pages}
  {011102} (\bibinfo {year} {2018})},\ \Eprint
  {http://arxiv.org/abs/1803.02804} {arXiv:1803.02804 [hep-ph]} \BibitemShut
  {NoStop}%
\bibitem [{\citenamefont {Muñoz}\ \emph {et~al.}(2018)\citenamefont {Muñoz},
  \citenamefont {Dvorkin},\ and\ \citenamefont {Loeb}}]{Munoz:2018jwq}%
  \BibitemOpen
  \bibfield  {author} {\bibinfo {author} {\bibfnamefont {Julian~B.}\
  \bibnamefont {Muñoz}}, \bibinfo {author} {\bibfnamefont {Cora}\ \bibnamefont
  {Dvorkin}}, \ and\ \bibinfo {author} {\bibfnamefont {Abraham}\ \bibnamefont
  {Loeb}},\ }\bibfield  {title} {\enquote {\bibinfo {title} {{21-cm
  Fluctuations from Charged Dark Matter}},}\ }\href {\doibase
  10.1103/PhysRevLett.121.121301} {\bibfield  {journal} {\bibinfo  {journal}
  {Phys. Rev. Lett.}\ }\textbf {\bibinfo {volume} {121}},\ \bibinfo {pages}
  {121301} (\bibinfo {year} {2018})},\ \Eprint
  {http://arxiv.org/abs/1804.01092} {arXiv:1804.01092 [astro-ph.CO]}
  \BibitemShut {NoStop}%
\bibitem [{\citenamefont {Kovetz}\ \emph {et~al.}(2018)\citenamefont {Kovetz},
  \citenamefont {Poulin}, \citenamefont {Gluscevic}, \citenamefont {Boddy},
  \citenamefont {Barkana},\ and\ \citenamefont
  {Kamionkowski}}]{Kovetz:2018zan}%
  \BibitemOpen
  \bibfield  {author} {\bibinfo {author} {\bibfnamefont {Ely~D.}\ \bibnamefont
  {Kovetz}}, \bibinfo {author} {\bibfnamefont {Vivian}\ \bibnamefont {Poulin}},
  \bibinfo {author} {\bibfnamefont {Vera}\ \bibnamefont {Gluscevic}}, \bibinfo
  {author} {\bibfnamefont {Kimberly~K.}\ \bibnamefont {Boddy}}, \bibinfo
  {author} {\bibfnamefont {Rennan}\ \bibnamefont {Barkana}}, \ and\ \bibinfo
  {author} {\bibfnamefont {Marc}\ \bibnamefont {Kamionkowski}},\ }\bibfield
  {title} {\enquote {\bibinfo {title} {{Tighter limits on dark matter
  explanations of the anomalous EDGES 21 cm signal}},}\ }\href {\doibase
  10.1103/PhysRevD.98.103529} {\bibfield  {journal} {\bibinfo  {journal} {Phys.
  Rev.}\ }\textbf {\bibinfo {volume} {D98}},\ \bibinfo {pages} {103529}
  (\bibinfo {year} {2018})},\ \Eprint {http://arxiv.org/abs/1807.11482}
  {arXiv:1807.11482 [astro-ph.CO]} \BibitemShut {NoStop}%
\bibitem [{\citenamefont {Liu}\ \emph {et~al.}(2019{\natexlab{a}})\citenamefont
  {Liu}, \citenamefont {Outmezguine}, \citenamefont {Redigolo},\ and\
  \citenamefont {Volansky}}]{Liu:2019knx}%
  \BibitemOpen
  \bibfield  {author} {\bibinfo {author} {\bibfnamefont {Hongwan}\ \bibnamefont
  {Liu}}, \bibinfo {author} {\bibfnamefont {Nadav~Joseph}\ \bibnamefont
  {Outmezguine}}, \bibinfo {author} {\bibfnamefont {Diego}\ \bibnamefont
  {Redigolo}}, \ and\ \bibinfo {author} {\bibfnamefont {Tomer}\ \bibnamefont
  {Volansky}},\ }\bibfield  {title} {\enquote {\bibinfo {title} {{Reviving
  Millicharged Dark Matter for 21-Cm Cosmology}},}\ }\href {\doibase
  10.1103/PhysRevD.100.123011} {\bibfield  {journal} {\bibinfo  {journal}
  {Phys. Rev.}\ }\textbf {\bibinfo {volume} {D100}},\ \bibinfo {pages} {123011}
  (\bibinfo {year} {2019}{\natexlab{a}})},\ \Eprint
  {http://arxiv.org/abs/1908.06986} {arXiv:1908.06986 [hep-ph]} \BibitemShut
  {NoStop}%
\bibitem [{\citenamefont {Creque-Sarbinowski}\ \emph
  {et~al.}(2019)\citenamefont {Creque-Sarbinowski}, \citenamefont {Ji},
  \citenamefont {Kovetz},\ and\ \citenamefont
  {Kamionkowski}}]{Creque-Sarbinowski:2019mcm}%
  \BibitemOpen
  \bibfield  {author} {\bibinfo {author} {\bibfnamefont {Cyril}\ \bibnamefont
  {Creque-Sarbinowski}}, \bibinfo {author} {\bibfnamefont {Lingyuan}\
  \bibnamefont {Ji}}, \bibinfo {author} {\bibfnamefont {Ely~D.}\ \bibnamefont
  {Kovetz}}, \ and\ \bibinfo {author} {\bibfnamefont {Marc}\ \bibnamefont
  {Kamionkowski}},\ }\bibfield  {title} {\enquote {\bibinfo {title} {{Direct
  millicharged dark matter cannot explain the EDGES signal}},}\ }\href
  {\doibase 10.1103/PhysRevD.100.023528} {\bibfield  {journal} {\bibinfo
  {journal} {Phys. Rev. D}\ }\textbf {\bibinfo {volume} {100}},\ \bibinfo
  {pages} {023528} (\bibinfo {year} {2019})},\ \Eprint
  {http://arxiv.org/abs/1903.09154} {arXiv:1903.09154 [astro-ph.CO]}
  \BibitemShut {NoStop}%
\bibitem [{\citenamefont {Aboubrahim}\ \emph {et~al.}(2021)\citenamefont
  {Aboubrahim}, \citenamefont {Nath},\ and\ \citenamefont
  {Wang}}]{Aboubrahim:2021ohe}%
  \BibitemOpen
  \bibfield  {author} {\bibinfo {author} {\bibfnamefont {Amin}\ \bibnamefont
  {Aboubrahim}}, \bibinfo {author} {\bibfnamefont {Pran}\ \bibnamefont {Nath}},
  \ and\ \bibinfo {author} {\bibfnamefont {Zhu-Yao}\ \bibnamefont {Wang}},\
  }\bibfield  {title} {\enquote {\bibinfo {title} {{A Cosmologically Consistent
  Millicharged Dark Matter Solution to the Edges Anomaly of Possible String
  Theory Origin}},}\ }\href {\doibase 10.1007/JHEP12(2021)148} {\bibfield
  {journal} {\bibinfo  {journal} {JHEP}\ }\textbf {\bibinfo {volume} {12}},\
  \bibinfo {pages} {148} (\bibinfo {year} {2021})},\ \Eprint
  {http://arxiv.org/abs/2108.05819} {arXiv:2108.05819 [hep-ph]} \BibitemShut
  {NoStop}%
\bibitem [{\citenamefont {Adshead}\ \emph {et~al.}(2022)\citenamefont
  {Adshead}, \citenamefont {Ralegankar},\ and\ \citenamefont
  {Shelton}}]{Adshead:2022ovo}%
  \BibitemOpen
  \bibfield  {author} {\bibinfo {author} {\bibfnamefont {Peter}\ \bibnamefont
  {Adshead}}, \bibinfo {author} {\bibfnamefont {Pranjal}\ \bibnamefont
  {Ralegankar}}, \ and\ \bibinfo {author} {\bibfnamefont {Jessie}\ \bibnamefont
  {Shelton}},\ }\bibfield  {title} {\enquote {\bibinfo {title} {{Dark radiation
  constraints on portal interactions with hidden sectors}},}\ }\href@noop {} {\
   (\bibinfo {year} {2022})},\ \Eprint {http://arxiv.org/abs/2206.13530}
  {arXiv:2206.13530 [hep-ph]} \BibitemShut {NoStop}%
\bibitem [{\citenamefont {Bowman}\ \emph {et~al.}(2018)\citenamefont {Bowman},
  \citenamefont {Rogers}, \citenamefont {Monsalve}, \citenamefont {Mozdzen},\
  and\ \citenamefont {Mahesh}}]{Bowman:2018yin}%
  \BibitemOpen
  \bibfield  {author} {\bibinfo {author} {\bibfnamefont {Judd~D.}\ \bibnamefont
  {Bowman}}, \bibinfo {author} {\bibfnamefont {Alan E.~E.}\ \bibnamefont
  {Rogers}}, \bibinfo {author} {\bibfnamefont {Raul~A.}\ \bibnamefont
  {Monsalve}}, \bibinfo {author} {\bibfnamefont {Thomas~J.}\ \bibnamefont
  {Mozdzen}}, \ and\ \bibinfo {author} {\bibfnamefont {Nivedita}\ \bibnamefont
  {Mahesh}},\ }\bibfield  {title} {\enquote {\bibinfo {title} {{An Absorption
  Profile Centred at 78 Megahertz in the Sky-Averaged Spectrum}},}\ }\href
  {\doibase 10.1038/nature25792} {\bibfield  {journal} {\bibinfo  {journal}
  {Nature}\ }\textbf {\bibinfo {volume} {555}},\ \bibinfo {pages} {67--70}
  (\bibinfo {year} {2018})},\ \Eprint {http://arxiv.org/abs/1810.05912}
  {arXiv:1810.05912 [astro-ph.CO]} \BibitemShut {NoStop}%
\bibitem [{\citenamefont {Singh}\ \emph {et~al.}(2021)\citenamefont {Singh},
  \citenamefont {T.}, \citenamefont {Subrahmanyan}, \citenamefont {Shankar},
  \citenamefont {Girish}, \citenamefont {Raghunathan}, \citenamefont
  {Somashekar}, \citenamefont {Srivani},\ and\ \citenamefont
  {Rao}}]{Singh:2021mxo}%
  \BibitemOpen
  \bibfield  {author} {\bibinfo {author} {\bibfnamefont {Saurabh}\ \bibnamefont
  {Singh}}, \bibinfo {author} {\bibfnamefont {Jishnu~Nambissan}\ \bibnamefont
  {T.}}, \bibinfo {author} {\bibfnamefont {Ravi}\ \bibnamefont {Subrahmanyan}},
  \bibinfo {author} {\bibfnamefont {N.~Udaya}\ \bibnamefont {Shankar}},
  \bibinfo {author} {\bibfnamefont {B.~S.}\ \bibnamefont {Girish}}, \bibinfo
  {author} {\bibfnamefont {A.}~\bibnamefont {Raghunathan}}, \bibinfo {author}
  {\bibfnamefont {R.}~\bibnamefont {Somashekar}}, \bibinfo {author}
  {\bibfnamefont {K.~S.}\ \bibnamefont {Srivani}}, \ and\ \bibinfo {author}
  {\bibfnamefont {Mayuri~Sathyanarayana}\ \bibnamefont {Rao}},\ }\bibfield
  {title} {\enquote {\bibinfo {title} {{On the detection of a cosmic dawn
  signal in the radio background}},}\ }\href@noop {} {\  (\bibinfo {year}
  {2021})},\ \Eprint {http://arxiv.org/abs/2112.06778} {arXiv:2112.06778
  [astro-ph.CO]} \BibitemShut {NoStop}%
\bibitem [{\citenamefont {{Philip}}\ \emph {et~al.}(2019)\citenamefont
  {{Philip}}, \citenamefont {{Abdurashidova}}, \citenamefont {{Chiang}},
  \citenamefont {{Ghazi}}, \citenamefont {{Gumba}}, \citenamefont
  {{Heilgendorff}}, \citenamefont {{J{\'a}uregui-Garc{\'\i}a}}, \citenamefont
  {{Malepe}}, \citenamefont {{Nunhokee}}, \citenamefont {{Peterson}},
  \citenamefont {{Sievers}}, \citenamefont {{Simes}},\ and\ \citenamefont
  {{Spann}}}]{2019JAI.....850004P}%
  \BibitemOpen
  \bibfield  {author} {\bibinfo {author} {\bibfnamefont {L.}~\bibnamefont
  {{Philip}}}, \bibinfo {author} {\bibfnamefont {Z.}~\bibnamefont
  {{Abdurashidova}}}, \bibinfo {author} {\bibfnamefont {H.~C.}\ \bibnamefont
  {{Chiang}}}, \bibinfo {author} {\bibfnamefont {N.}~\bibnamefont {{Ghazi}}},
  \bibinfo {author} {\bibfnamefont {A.}~\bibnamefont {{Gumba}}}, \bibinfo
  {author} {\bibfnamefont {H.~M.}\ \bibnamefont {{Heilgendorff}}}, \bibinfo
  {author} {\bibfnamefont {J.~M.}\ \bibnamefont {{J{\'a}uregui-Garc{\'\i}a}}},
  \bibinfo {author} {\bibfnamefont {K.}~\bibnamefont {{Malepe}}}, \bibinfo
  {author} {\bibfnamefont {C.~D.}\ \bibnamefont {{Nunhokee}}}, \bibinfo
  {author} {\bibfnamefont {J.}~\bibnamefont {{Peterson}}}, \bibinfo {author}
  {\bibfnamefont {J.~L.}\ \bibnamefont {{Sievers}}}, \bibinfo {author}
  {\bibfnamefont {V.}~\bibnamefont {{Simes}}}, \ and\ \bibinfo {author}
  {\bibfnamefont {R.}~\bibnamefont {{Spann}}},\ }\bibfield  {title} {\enquote
  {\bibinfo {title} {{Probing Radio Intensity at High-Z from Marion: 2017
  Instrument}},}\ }\href {\doibase 10.1142/S2251171719500041} {\bibfield
  {journal} {\bibinfo  {journal} {Journal of Astronomical Instrumentation}\
  }\textbf {\bibinfo {volume} {8}},\ \bibinfo {eid} {1950004} (\bibinfo {year}
  {2019})},\ \Eprint {http://arxiv.org/abs/1806.09531} {arXiv:1806.09531
  [astro-ph.IM]} \BibitemShut {NoStop}%
\bibitem [{\citenamefont {Voytek}\ \emph {et~al.}(2014)\citenamefont {Voytek},
  \citenamefont {Natarajan}, \citenamefont {J\'auregui~Garc\'\i{}a},
  \citenamefont {Peterson},\ and\ \citenamefont
  {L\'opez-Cruz}}]{Voytek:2013nua}%
  \BibitemOpen
  \bibfield  {author} {\bibinfo {author} {\bibfnamefont {Tabitha~C.}\
  \bibnamefont {Voytek}}, \bibinfo {author} {\bibfnamefont {Aravind}\
  \bibnamefont {Natarajan}}, \bibinfo {author} {\bibfnamefont {Jos\'e~Miguel}\
  \bibnamefont {J\'auregui~Garc\'\i{}a}}, \bibinfo {author} {\bibfnamefont
  {Jeffrey~B.}\ \bibnamefont {Peterson}}, \ and\ \bibinfo {author}
  {\bibfnamefont {Omar}\ \bibnamefont {L\'opez-Cruz}},\ }\bibfield  {title}
  {\enquote {\bibinfo {title} {{Probing the Dark Ages at $z \sim$ 20: The
  SCI-HI 21 cm All-sky Spectrum Experiment}},}\ }\href {\doibase
  10.1088/2041-8205/782/1/L9} {\bibfield  {journal} {\bibinfo  {journal}
  {Astrophys. J. Lett.}\ }\textbf {\bibinfo {volume} {782}},\ \bibinfo {pages}
  {L9} (\bibinfo {year} {2014})},\ \Eprint {http://arxiv.org/abs/1311.0014}
  {arXiv:1311.0014 [astro-ph.CO]} \BibitemShut {NoStop}%
\bibitem [{\citenamefont {de~Lera~Acedo}\ \emph {et~al.}(2022)\citenamefont
  {de~Lera~Acedo}, \citenamefont {de~Villiers}, \citenamefont {Razavi-Ghods},
  \citenamefont {Handley}, \citenamefont {Fialkov}, \citenamefont {Magro},
  \citenamefont {Anstey}, \citenamefont {Bevins}, \citenamefont {Chiello},
  \citenamefont {Cumner}, \citenamefont {Josaitis}, \citenamefont {Roque},
  \citenamefont {Sims}, \citenamefont {Scheutwinkel}, \citenamefont
  {Alexander}, \citenamefont {Bernardi}, \citenamefont {Carey}, \citenamefont
  {Cavillot}, \citenamefont {Croukamp}, \citenamefont {Ely}, \citenamefont
  {Gessey-Jones}, \citenamefont {Gueuning}, \citenamefont {Hills},
  \citenamefont {Kulkarni}, \citenamefont {Maiolino}, \citenamefont {Meerburg},
  \citenamefont {Mittal}, \citenamefont {Pritchard}, \citenamefont {Puchwein},
  \citenamefont {Saxena}, \citenamefont {Shen}, \citenamefont {Smirnov},
  \citenamefont {Spinelli},\ and\ \citenamefont
  {Zarb-Adami}}]{deLeraAcedo2022}%
  \BibitemOpen
  \bibfield  {author} {\bibinfo {author} {\bibfnamefont {E.}~\bibnamefont
  {de~Lera~Acedo}}, \bibinfo {author} {\bibfnamefont {D.~I.~L.}\ \bibnamefont
  {de~Villiers}}, \bibinfo {author} {\bibfnamefont {N.}~\bibnamefont
  {Razavi-Ghods}}, \bibinfo {author} {\bibfnamefont {W.}~\bibnamefont
  {Handley}}, \bibinfo {author} {\bibfnamefont {A.}~\bibnamefont {Fialkov}},
  \bibinfo {author} {\bibfnamefont {A.}~\bibnamefont {Magro}}, \bibinfo
  {author} {\bibfnamefont {D.}~\bibnamefont {Anstey}}, \bibinfo {author}
  {\bibfnamefont {H.~T.~J.}\ \bibnamefont {Bevins}}, \bibinfo {author}
  {\bibfnamefont {R.}~\bibnamefont {Chiello}}, \bibinfo {author} {\bibfnamefont
  {J.}~\bibnamefont {Cumner}}, \bibinfo {author} {\bibfnamefont {A.~T.}\
  \bibnamefont {Josaitis}}, \bibinfo {author} {\bibfnamefont {I.~L.~V.}\
  \bibnamefont {Roque}}, \bibinfo {author} {\bibfnamefont {P.~H.}\ \bibnamefont
  {Sims}}, \bibinfo {author} {\bibfnamefont {K.~H.}\ \bibnamefont
  {Scheutwinkel}}, \bibinfo {author} {\bibfnamefont {P.}~\bibnamefont
  {Alexander}}, \bibinfo {author} {\bibfnamefont {G.}~\bibnamefont {Bernardi}},
  \bibinfo {author} {\bibfnamefont {S.}~\bibnamefont {Carey}}, \bibinfo
  {author} {\bibfnamefont {J.}~\bibnamefont {Cavillot}}, \bibinfo {author}
  {\bibfnamefont {W.}~\bibnamefont {Croukamp}}, \bibinfo {author}
  {\bibfnamefont {J.~A.}\ \bibnamefont {Ely}}, \bibinfo {author} {\bibfnamefont
  {T.}~\bibnamefont {Gessey-Jones}}, \bibinfo {author} {\bibfnamefont
  {Q.}~\bibnamefont {Gueuning}}, \bibinfo {author} {\bibfnamefont
  {R.}~\bibnamefont {Hills}}, \bibinfo {author} {\bibfnamefont
  {G.}~\bibnamefont {Kulkarni}}, \bibinfo {author} {\bibfnamefont
  {R.}~\bibnamefont {Maiolino}}, \bibinfo {author} {\bibfnamefont {P.~D.}\
  \bibnamefont {Meerburg}}, \bibinfo {author} {\bibfnamefont {S.}~\bibnamefont
  {Mittal}}, \bibinfo {author} {\bibfnamefont {J.~R.}\ \bibnamefont
  {Pritchard}}, \bibinfo {author} {\bibfnamefont {E.}~\bibnamefont {Puchwein}},
  \bibinfo {author} {\bibfnamefont {A.}~\bibnamefont {Saxena}}, \bibinfo
  {author} {\bibfnamefont {E.}~\bibnamefont {Shen}}, \bibinfo {author}
  {\bibfnamefont {O.}~\bibnamefont {Smirnov}}, \bibinfo {author} {\bibfnamefont
  {M.}~\bibnamefont {Spinelli}}, \ and\ \bibinfo {author} {\bibfnamefont
  {K.}~\bibnamefont {Zarb-Adami}},\ }\bibfield  {title} {\enquote {\bibinfo
  {title} {The {REACH} radiometer for detecting the 21-cm hydrogen signal from
  redshift z{\hspace{0.167em}}$\approx${\hspace{0.167em}}7.5{\textendash}28},}\
  }\href {\doibase 10.1038/s41550-022-01709-9} {\bibfield  {journal} {\bibinfo
  {journal} {Nature Astronomy}\ } (\bibinfo {year} {2022}),\
  10.1038/s41550-022-01709-9}\BibitemShut {NoStop}%
\bibitem [{\citenamefont {Liu}\ \emph {et~al.}(2019{\natexlab{b}})\citenamefont
  {Liu}, \citenamefont {Chiang}, \citenamefont {Crites}, \citenamefont
  {Sievers},\ and\ \citenamefont {Hlo\v{z}ek}}]{Liu:2019yfw}%
  \BibitemOpen
  \bibfield  {author} {\bibinfo {author} {\bibfnamefont {Adrian}\ \bibnamefont
  {Liu}}, \bibinfo {author} {\bibfnamefont {H.~Cynthia}\ \bibnamefont
  {Chiang}}, \bibinfo {author} {\bibfnamefont {Abigail}\ \bibnamefont
  {Crites}}, \bibinfo {author} {\bibfnamefont {Jonathan}\ \bibnamefont
  {Sievers}}, \ and\ \bibinfo {author} {\bibfnamefont {Ren\'ee}\ \bibnamefont
  {Hlo\v{z}ek}},\ }\bibfield  {title} {\enquote {\bibinfo {title}
  {{High-redshift 21cm Cosmology in Canada}},}\ }\href {\doibase
  10.5281/zenodo.3756080} {\  (\bibinfo {year} {2019}{\natexlab{b}}),\
  10.5281/zenodo.3756080},\ \Eprint {http://arxiv.org/abs/1910.03153}
  {arXiv:1910.03153 [astro-ph.CO]} \BibitemShut {NoStop}%
\bibitem [{\citenamefont {Paciga}\ \emph {et~al.}(2013)\citenamefont {Paciga}
  \emph {et~al.}}]{Paciga:2013fj}%
  \BibitemOpen
  \bibfield  {author} {\bibinfo {author} {\bibfnamefont {Gregory}\ \bibnamefont
  {Paciga}} \emph {et~al.},\ }\bibfield  {title} {\enquote {\bibinfo {title}
  {{A refined foreground-corrected limit on the HI power spectrum at z=8.6 from
  the GMRT Epoch of Reionization Experiment}},}\ }\href {\doibase
  10.1093/mnras/stt753} {\bibfield  {journal} {\bibinfo  {journal} {Mon. Not.
  Roy. Astron. Soc.}\ }\textbf {\bibinfo {volume} {433}},\ \bibinfo {pages}
  {639} (\bibinfo {year} {2013})},\ \Eprint {http://arxiv.org/abs/1301.5906}
  {arXiv:1301.5906 [astro-ph.CO]} \BibitemShut {NoStop}%
\bibitem [{\citenamefont {Dillon}\ \emph {et~al.}(2014)\citenamefont {Dillon}
  \emph {et~al.}}]{Dillon:2013rfa}%
  \BibitemOpen
  \bibfield  {author} {\bibinfo {author} {\bibfnamefont {Joshua~S.}\
  \bibnamefont {Dillon}} \emph {et~al.},\ }\bibfield  {title} {\enquote
  {\bibinfo {title} {{Overcoming real-world obstacles in 21 cm power spectrum
  estimation: A method demonstration and results from early Murchison Widefield
  Array data}},}\ }\href {\doibase 10.1103/PhysRevD.89.023002} {\bibfield
  {journal} {\bibinfo  {journal} {Phys. Rev. D}\ }\textbf {\bibinfo {volume}
  {89}},\ \bibinfo {pages} {023002} (\bibinfo {year} {2014})},\ \Eprint
  {http://arxiv.org/abs/1304.4229} {arXiv:1304.4229 [astro-ph.CO]} \BibitemShut
  {NoStop}%
\bibitem [{\citenamefont {Beardsley}\ \emph {et~al.}(2016)\citenamefont
  {Beardsley} \emph {et~al.}}]{Beardsley:2016njr}%
  \BibitemOpen
  \bibfield  {author} {\bibinfo {author} {\bibfnamefont {A.~P.}\ \bibnamefont
  {Beardsley}} \emph {et~al.},\ }\bibfield  {title} {\enquote {\bibinfo {title}
  {{First Season MWA EoR Power Spectrum Results at Redshift 7}},}\ }\href
  {\doibase 10.3847/1538-4357/833/1/102} {\bibfield  {journal} {\bibinfo
  {journal} {Astrophys. J.}\ }\textbf {\bibinfo {volume} {833}},\ \bibinfo
  {pages} {102} (\bibinfo {year} {2016})},\ \Eprint
  {http://arxiv.org/abs/1608.06281} {arXiv:1608.06281 [astro-ph.IM]}
  \BibitemShut {NoStop}%
\bibitem [{\citenamefont {Li}\ \emph {et~al.}(2019)\citenamefont {Li} \emph
  {et~al.}}]{Li:2019kqp}%
  \BibitemOpen
  \bibfield  {author} {\bibinfo {author} {\bibfnamefont {W.}~\bibnamefont {Li}}
  \emph {et~al.},\ }\bibfield  {title} {\enquote {\bibinfo {title} {{First
  Season MWA Phase II EoR Power Spectrum Results at Redshift 7}},}\ }\href
  {\doibase 10.3847/1538-4357/ab55e4} {\bibfield  {journal} {\bibinfo
  {journal} {Astrophys. J.}\ }\textbf {\bibinfo {volume} {887}},\ \bibinfo
  {pages} {141} (\bibinfo {year} {2019})},\ \Eprint
  {http://arxiv.org/abs/1911.10216} {arXiv:1911.10216 [astro-ph.CO]}
  \BibitemShut {NoStop}%
\bibitem [{\citenamefont {Barry}\ \emph {et~al.}(2019)\citenamefont {Barry}
  \emph {et~al.}}]{Barry:2019qxp}%
  \BibitemOpen
  \bibfield  {author} {\bibinfo {author} {\bibfnamefont {N.}~\bibnamefont
  {Barry}} \emph {et~al.},\ }\bibfield  {title} {\enquote {\bibinfo {title}
  {{Improving the Epoch of Reionization Power Spectrum Results from Murchison
  Widefield Array Season 1 Observations}},}\ }\href {\doibase
  10.3847/1538-4357/ab40a8} {\  (\bibinfo {year} {2019}),\
  10.3847/1538-4357/ab40a8},\ \Eprint {http://arxiv.org/abs/1909.00561}
  {arXiv:1909.00561 [astro-ph.IM]} \BibitemShut {NoStop}%
\bibitem [{\citenamefont {Trott}\ \emph {et~al.}(2020)\citenamefont {Trott}
  \emph {et~al.}}]{Trott:2020szf}%
  \BibitemOpen
  \bibfield  {author} {\bibinfo {author} {\bibfnamefont {Cathryn~M.}\
  \bibnamefont {Trott}} \emph {et~al.},\ }\bibfield  {title} {\enquote
  {\bibinfo {title} {{Deep multiredshift limits on Epoch of Reionization 21 cm
  power spectra from four seasons of Murchison Widefield Array
  observations}},}\ }\href {\doibase 10.1093/mnras/staa414} {\bibfield
  {journal} {\bibinfo  {journal} {Mon. Not. Roy. Astron. Soc.}\ }\textbf
  {\bibinfo {volume} {493}},\ \bibinfo {pages} {4711--4727} (\bibinfo {year}
  {2020})},\ \Eprint {http://arxiv.org/abs/2002.02575} {arXiv:2002.02575
  [astro-ph.CO]} \BibitemShut {NoStop}%
\bibitem [{\citenamefont {Yoshiura}\ \emph {et~al.}(2021)\citenamefont
  {Yoshiura} \emph {et~al.}}]{Yoshiura:2021yfx}%
  \BibitemOpen
  \bibfield  {author} {\bibinfo {author} {\bibfnamefont {S.}~\bibnamefont
  {Yoshiura}} \emph {et~al.},\ }\bibfield  {title} {\enquote {\bibinfo {title}
  {{A new MWA limit on the 21~cm power spectrum at redshifts
  \ensuremath{\sim}13\textendash{}17}},}\ }\href {\doibase
  10.1093/mnras/stab1560} {\bibfield  {journal} {\bibinfo  {journal} {Mon. Not.
  Roy. Astron. Soc.}\ }\textbf {\bibinfo {volume} {505}},\ \bibinfo {pages}
  {4775--4790} (\bibinfo {year} {2021})},\ \Eprint
  {http://arxiv.org/abs/2105.12888} {arXiv:2105.12888 [astro-ph.CO]}
  \BibitemShut {NoStop}%
\bibitem [{\citenamefont {Patil}\ \emph {et~al.}(2017)\citenamefont {Patil}
  \emph {et~al.}}]{Patil:2017zqk}%
  \BibitemOpen
  \bibfield  {author} {\bibinfo {author} {\bibfnamefont {A.~H.}\ \bibnamefont
  {Patil}} \emph {et~al.},\ }\bibfield  {title} {\enquote {\bibinfo {title}
  {{Upper limits on the 21-cm Epoch of Reionization power spectrum from one
  night with LOFAR}},}\ }\href {\doibase 10.3847/1538-4357/aa63e7} {\bibfield
  {journal} {\bibinfo  {journal} {Astrophys. J.}\ }\textbf {\bibinfo {volume}
  {838}},\ \bibinfo {pages} {65} (\bibinfo {year} {2017})},\ \Eprint
  {http://arxiv.org/abs/1702.08679} {arXiv:1702.08679 [astro-ph.CO]}
  \BibitemShut {NoStop}%
\bibitem [{\citenamefont {Mertens}\ \emph {et~al.}(2020)\citenamefont {Mertens}
  \emph {et~al.}}]{Mertens:2020llj}%
  \BibitemOpen
  \bibfield  {author} {\bibinfo {author} {\bibfnamefont {F.~G.}\ \bibnamefont
  {Mertens}} \emph {et~al.},\ }\bibfield  {title} {\enquote {\bibinfo {title}
  {{Improved upper limits on the 21-cm signal power spectrum of neutral
  hydrogen at $\boldsymbol{z \approx 9.1}$ from LOFAR}},}\ }\href {\doibase
  10.1093/mnras/staa327} {\bibfield  {journal} {\bibinfo  {journal} {Mon. Not.
  Roy. Astron. Soc.}\ }\textbf {\bibinfo {volume} {493}},\ \bibinfo {pages}
  {1662--1685} (\bibinfo {year} {2020})},\ \Eprint
  {http://arxiv.org/abs/2002.07196} {arXiv:2002.07196 [astro-ph.CO]}
  \BibitemShut {NoStop}%
\bibitem [{\citenamefont {Kolopanis}\ \emph {et~al.}(2019)\citenamefont
  {Kolopanis} \emph {et~al.}}]{Kolopanis:2019vbl}%
  \BibitemOpen
  \bibfield  {author} {\bibinfo {author} {\bibfnamefont {Matthew}\ \bibnamefont
  {Kolopanis}} \emph {et~al.},\ }\bibfield  {title} {\enquote {\bibinfo {title}
  {{A simplified, lossless re-analysis of PAPER-64}},}\ }\href {\doibase
  10.3847/1538-4357/ab3e3a} {\  (\bibinfo {year} {2019}),\
  10.3847/1538-4357/ab3e3a},\ \Eprint {http://arxiv.org/abs/1909.02085}
  {arXiv:1909.02085 [astro-ph.CO]} \BibitemShut {NoStop}%
\bibitem [{\citenamefont {Garsden}\ \emph {et~al.}(2021)\citenamefont
  {Garsden}, \citenamefont {Greenhill}, \citenamefont {Bernardi}, \citenamefont
  {Fialkov}, \citenamefont {Price}, \citenamefont {Mitchell}, \citenamefont
  {Dowell}, \citenamefont {Spinelli},\ and\ \citenamefont
  {Schinzel}}]{Garsden:2021kdo}%
  \BibitemOpen
  \bibfield  {author} {\bibinfo {author} {\bibfnamefont {Hugh}\ \bibnamefont
  {Garsden}}, \bibinfo {author} {\bibfnamefont {Lincoln}\ \bibnamefont
  {Greenhill}}, \bibinfo {author} {\bibfnamefont {Gianni}\ \bibnamefont
  {Bernardi}}, \bibinfo {author} {\bibfnamefont {Anastasia}\ \bibnamefont
  {Fialkov}}, \bibinfo {author} {\bibfnamefont {Daniel~C.}\ \bibnamefont
  {Price}}, \bibinfo {author} {\bibfnamefont {Daniel}\ \bibnamefont
  {Mitchell}}, \bibinfo {author} {\bibfnamefont {Jayce}\ \bibnamefont
  {Dowell}}, \bibinfo {author} {\bibfnamefont {Marta}\ \bibnamefont
  {Spinelli}}, \ and\ \bibinfo {author} {\bibfnamefont {Frank~K.}\ \bibnamefont
  {Schinzel}},\ }\bibfield  {title} {\enquote {\bibinfo {title} {{A 21-cm power
  spectrum at 48~MHz, using the Owens Valley Long Wavelength Array}},}\ }\href
  {\doibase 10.1093/mnras/stab1671} {\bibfield  {journal} {\bibinfo  {journal}
  {Mon. Not. Roy. Astron. Soc.}\ }\textbf {\bibinfo {volume} {506}},\ \bibinfo
  {pages} {5802--5817} (\bibinfo {year} {2021})},\ \Eprint
  {http://arxiv.org/abs/2102.09596} {arXiv:2102.09596 [astro-ph.CO]}
  \BibitemShut {NoStop}%
\bibitem [{\citenamefont {Abdurashidova}\ \emph
  {et~al.}(2022{\natexlab{a}})\citenamefont {Abdurashidova} \emph
  {et~al.}}]{HERA:2021bsv}%
  \BibitemOpen
  \bibfield  {author} {\bibinfo {author} {\bibfnamefont {Zara}\ \bibnamefont
  {Abdurashidova}} \emph {et~al.} (\bibinfo {collaboration} {HERA}),\
  }\bibfield  {title} {\enquote {\bibinfo {title} {{First Results from HERA
  Phase I: Upper Limits on the Epoch of Reionization 21 cm Power Spectrum}},}\
  }\href {\doibase 10.3847/1538-4357/ac1c78} {\bibfield  {journal} {\bibinfo
  {journal} {Astrophys. J.}\ }\textbf {\bibinfo {volume} {925}},\ \bibinfo
  {pages} {221} (\bibinfo {year} {2022}{\natexlab{a}})},\ \Eprint
  {http://arxiv.org/abs/2108.02263} {arXiv:2108.02263 [astro-ph.CO]}
  \BibitemShut {NoStop}%
\bibitem [{\citenamefont {Abdurashidova}\ \emph
  {et~al.}(2022{\natexlab{b}})\citenamefont {Abdurashidova} \emph
  {et~al.}}]{HERA:2022wmy}%
  \BibitemOpen
  \bibfield  {author} {\bibinfo {author} {\bibfnamefont {Zara}\ \bibnamefont
  {Abdurashidova}} \emph {et~al.} (\bibinfo {collaboration} {HERA}),\
  }\bibfield  {title} {\enquote {\bibinfo {title} {{Improved Constraints on the
  21 cm EoR Power Spectrum and the X-Ray Heating of the IGM with HERA Phase I
  Observations}},}\ }\href@noop {} {\  (\bibinfo {year}
  {2022}{\natexlab{b}})},\ \Eprint {http://arxiv.org/abs/2210.04912}
  {arXiv:2210.04912 [astro-ph.CO]} \BibitemShut {NoStop}%
\bibitem [{\citenamefont {Reis}\ \emph {et~al.}(2021)\citenamefont {Reis},
  \citenamefont {Fialkov},\ and\ \citenamefont {Barkana}}]{Reis:2021nqf}%
  \BibitemOpen
  \bibfield  {author} {\bibinfo {author} {\bibfnamefont {Itamar}\ \bibnamefont
  {Reis}}, \bibinfo {author} {\bibfnamefont {Anastasia}\ \bibnamefont
  {Fialkov}}, \ and\ \bibinfo {author} {\bibfnamefont {Rennan}\ \bibnamefont
  {Barkana}},\ }\bibfield  {title} {\enquote {\bibinfo {title} {{The subtlety
  of Ly-a photons: changing the expected range of the 21-cm signal}},}\
  }\href@noop {} {\  (\bibinfo {year} {2021})},\ \Eprint
  {http://arxiv.org/abs/2101.01777} {arXiv:2101.01777 [astro-ph.CO]}
  \BibitemShut {NoStop}%
\bibitem [{\citenamefont {Feng}\ and\ \citenamefont
  {Holder}(2018)}]{Feng:2018rje}%
  \BibitemOpen
  \bibfield  {author} {\bibinfo {author} {\bibfnamefont {Chang}\ \bibnamefont
  {Feng}}\ and\ \bibinfo {author} {\bibfnamefont {Gilbert}\ \bibnamefont
  {Holder}},\ }\bibfield  {title} {\enquote {\bibinfo {title} {{Enhanced global
  signal of neutral hydrogen due to excess radiation at cosmic dawn}},}\ }\href
  {\doibase 10.3847/2041-8213/aac0fe} {\bibfield  {journal} {\bibinfo
  {journal} {Astrophys. J. Lett.}\ }\textbf {\bibinfo {volume} {858}},\
  \bibinfo {pages} {L17} (\bibinfo {year} {2018})},\ \Eprint
  {http://arxiv.org/abs/1802.07432} {arXiv:1802.07432 [astro-ph.CO]}
  \BibitemShut {NoStop}%
\bibitem [{\citenamefont {Ewall-Wice}\ \emph {et~al.}(2018)\citenamefont
  {Ewall-Wice}, \citenamefont {Chang}, \citenamefont {Lazio}, \citenamefont
  {Dore}, \citenamefont {Seiffert},\ and\ \citenamefont
  {Monsalve}}]{Ewall-Wice:2018bzf}%
  \BibitemOpen
  \bibfield  {author} {\bibinfo {author} {\bibfnamefont {A.}~\bibnamefont
  {Ewall-Wice}}, \bibinfo {author} {\bibfnamefont {T.-C.}\ \bibnamefont
  {Chang}}, \bibinfo {author} {\bibfnamefont {J.}~\bibnamefont {Lazio}},
  \bibinfo {author} {\bibfnamefont {O.}~\bibnamefont {Dore}}, \bibinfo {author}
  {\bibfnamefont {M.}~\bibnamefont {Seiffert}}, \ and\ \bibinfo {author}
  {\bibfnamefont {R.A.}\ \bibnamefont {Monsalve}},\ }\bibfield  {title}
  {\enquote {\bibinfo {title} {{Modeling the Radio Background from the First
  Black Holes at Cosmic Dawn: Implications for the 21 cm Absorption
  Amplitude}},}\ }\href {\doibase 10.3847/1538-4357/aae51d} {\bibfield
  {journal} {\bibinfo  {journal} {Astrophys. J.}\ }\textbf {\bibinfo {volume}
  {868}},\ \bibinfo {pages} {63} (\bibinfo {year} {2018})},\ \Eprint
  {http://arxiv.org/abs/1803.01815} {arXiv:1803.01815 [astro-ph.CO]}
  \BibitemShut {NoStop}%
\bibitem [{\citenamefont {Pospelov}\ \emph {et~al.}(2018)\citenamefont
  {Pospelov}, \citenamefont {Pradler}, \citenamefont {Ruderman},\ and\
  \citenamefont {Urbano}}]{Pospelov:2018kdh}%
  \BibitemOpen
  \bibfield  {author} {\bibinfo {author} {\bibfnamefont {Maxim}\ \bibnamefont
  {Pospelov}}, \bibinfo {author} {\bibfnamefont {Josef}\ \bibnamefont
  {Pradler}}, \bibinfo {author} {\bibfnamefont {Joshua~T.}\ \bibnamefont
  {Ruderman}}, \ and\ \bibinfo {author} {\bibfnamefont {Alfredo}\ \bibnamefont
  {Urbano}},\ }\bibfield  {title} {\enquote {\bibinfo {title} {{Room for New
  Physics in the Rayleigh-Jeans Tail of the Cosmic Microwave Background}},}\
  }\href {\doibase 10.1103/PhysRevLett.121.031103} {\bibfield  {journal}
  {\bibinfo  {journal} {Phys. Rev. Lett.}\ }\textbf {\bibinfo {volume} {121}},\
  \bibinfo {pages} {031103} (\bibinfo {year} {2018})},\ \Eprint
  {http://arxiv.org/abs/1803.07048} {arXiv:1803.07048 [hep-ph]} \BibitemShut
  {NoStop}%
\bibitem [{\citenamefont {Fialkov}\ and\ \citenamefont
  {Barkana}(2019)}]{Fialkov:2019vnb}%
  \BibitemOpen
  \bibfield  {author} {\bibinfo {author} {\bibfnamefont {Anastasia}\
  \bibnamefont {Fialkov}}\ and\ \bibinfo {author} {\bibfnamefont {Rennan}\
  \bibnamefont {Barkana}},\ }\bibfield  {title} {\enquote {\bibinfo {title}
  {{Signature of Excess Radio Background in the 21-cm Global Signal and Power
  Spectrum}},}\ }\href {\doibase 10.1093/mnras/stz873} {\bibfield  {journal}
  {\bibinfo  {journal} {Mon. Not. Roy. Astron. Soc.}\ }\textbf {\bibinfo
  {volume} {486}},\ \bibinfo {pages} {1763--1773} (\bibinfo {year} {2019})},\
  \Eprint {http://arxiv.org/abs/1902.02438} {arXiv:1902.02438 [astro-ph.CO]}
  \BibitemShut {NoStop}%
\bibitem [{\citenamefont {Reis}\ \emph {et~al.}(2020)\citenamefont {Reis},
  \citenamefont {Fialkov},\ and\ \citenamefont {Barkana}}]{Reis:2020arr}%
  \BibitemOpen
  \bibfield  {author} {\bibinfo {author} {\bibfnamefont {Itamar}\ \bibnamefont
  {Reis}}, \bibinfo {author} {\bibfnamefont {Anastasia}\ \bibnamefont
  {Fialkov}}, \ and\ \bibinfo {author} {\bibfnamefont {Rennan}\ \bibnamefont
  {Barkana}},\ }\bibfield  {title} {\enquote {\bibinfo {title} {{High-redshift
  radio galaxies: a potential new source of 21-cm fluctuations}},}\ }\href
  {\doibase 10.1093/mnras/staa3091} {\bibfield  {journal} {\bibinfo  {journal}
  {Mon. Not. Roy. Astron. Soc.}\ }\textbf {\bibinfo {volume} {499}},\ \bibinfo
  {pages} {5993--6008} (\bibinfo {year} {2020})},\ \Eprint
  {http://arxiv.org/abs/2008.04315} {arXiv:2008.04315 [astro-ph.CO]}
  \BibitemShut {NoStop}%
\bibitem [{\citenamefont {Dalal}\ \emph {et~al.}(2010)\citenamefont {Dalal},
  \citenamefont {Pen},\ and\ \citenamefont {Seljak}}]{Dalal:2010yt}%
  \BibitemOpen
  \bibfield  {author} {\bibinfo {author} {\bibfnamefont {Neal}\ \bibnamefont
  {Dalal}}, \bibinfo {author} {\bibfnamefont {Ue-Li}\ \bibnamefont {Pen}}, \
  and\ \bibinfo {author} {\bibfnamefont {Uros}\ \bibnamefont {Seljak}},\
  }\bibfield  {title} {\enquote {\bibinfo {title} {{Large-Scale Bao Signatures
  of the Smallest Galaxies}},}\ }\href {\doibase 10.1088/1475-7516/2010/11/007}
  {\bibfield  {journal} {\bibinfo  {journal} {JCAP}\ }\textbf {\bibinfo
  {volume} {1011}},\ \bibinfo {pages} {007} (\bibinfo {year} {2010})},\ \Eprint
  {http://arxiv.org/abs/1009.4704} {arXiv:1009.4704 [astro-ph.CO]} \BibitemShut
  {NoStop}%
\bibitem [{\citenamefont {McQuinn}\ and\ \citenamefont
  {O'Leary}(2012)}]{McQuinn:2012rt}%
  \BibitemOpen
  \bibfield  {author} {\bibinfo {author} {\bibfnamefont {Matthew}\ \bibnamefont
  {McQuinn}}\ and\ \bibinfo {author} {\bibfnamefont {Ryan~M.}\ \bibnamefont
  {O'Leary}},\ }\bibfield  {title} {\enquote {\bibinfo {title} {{The impact of
  the supersonic baryon-dark matter velocity difference on the
  z\textasciitilde{}20 21cm background}},}\ }\href {\doibase
  10.1088/0004-637X/760/1/3} {\bibfield  {journal} {\bibinfo  {journal}
  {Astrophys. J.}\ }\textbf {\bibinfo {volume} {760}},\ \bibinfo {pages} {3}
  (\bibinfo {year} {2012})},\ \Eprint {http://arxiv.org/abs/1204.1345}
  {arXiv:1204.1345 [astro-ph.CO]} \BibitemShut {NoStop}%
\bibitem [{\citenamefont {Visbal}\ \emph {et~al.}(2012)\citenamefont {Visbal},
  \citenamefont {Barkana}, \citenamefont {Fialkov}, \citenamefont
  {Tseliakhovich},\ and\ \citenamefont {Hirata}}]{Visbal:2012aw}%
  \BibitemOpen
  \bibfield  {author} {\bibinfo {author} {\bibfnamefont {Eli}\ \bibnamefont
  {Visbal}}, \bibinfo {author} {\bibfnamefont {Rennan}\ \bibnamefont
  {Barkana}}, \bibinfo {author} {\bibfnamefont {Anastasia}\ \bibnamefont
  {Fialkov}}, \bibinfo {author} {\bibfnamefont {Dmitriy}\ \bibnamefont
  {Tseliakhovich}}, \ and\ \bibinfo {author} {\bibfnamefont {Christopher}\
  \bibnamefont {Hirata}},\ }\bibfield  {title} {\enquote {\bibinfo {title}
  {{The signature of the first stars in atomic hydrogen at redshift 20}},}\
  }\href {\doibase 10.1038/nature11177} {\bibfield  {journal} {\bibinfo
  {journal} {Nature}\ }\textbf {\bibinfo {volume} {487}},\ \bibinfo {pages}
  {70} (\bibinfo {year} {2012})},\ \Eprint {http://arxiv.org/abs/1201.1005}
  {arXiv:1201.1005 [astro-ph.CO]} \BibitemShut {NoStop}%
\bibitem [{\citenamefont {Mu\~noz}(2019)}]{Munoz:2019rhi}%
  \BibitemOpen
  \bibfield  {author} {\bibinfo {author} {\bibfnamefont {Julian~B.}\
  \bibnamefont {Mu\~noz}},\ }\bibfield  {title} {\enquote {\bibinfo {title}
  {{Robust Velocity-induced Acoustic Oscillations at Cosmic Dawn}},}\ }\href
  {\doibase 10.1103/PhysRevD.100.063538} {\bibfield  {journal} {\bibinfo
  {journal} {Phys. Rev. D}\ }\textbf {\bibinfo {volume} {100}},\ \bibinfo
  {pages} {063538} (\bibinfo {year} {2019})},\ \Eprint
  {http://arxiv.org/abs/1904.07881} {arXiv:1904.07881 [astro-ph.CO]}
  \BibitemShut {NoStop}%
\bibitem [{\citenamefont {de~Putter}\ \emph {et~al.}(2019)\citenamefont
  {de~Putter}, \citenamefont {Dor\'e}, \citenamefont {Gleyzes}, \citenamefont
  {Green},\ and\ \citenamefont {Meyers}}]{PhysRevLett.122.041301}%
  \BibitemOpen
  \bibfield  {author} {\bibinfo {author} {\bibfnamefont {Roland}\ \bibnamefont
  {de~Putter}}, \bibinfo {author} {\bibfnamefont {Olivier}\ \bibnamefont
  {Dor\'e}}, \bibinfo {author} {\bibfnamefont {J\'er\^ome}\ \bibnamefont
  {Gleyzes}}, \bibinfo {author} {\bibfnamefont {Daniel}\ \bibnamefont {Green}},
  \ and\ \bibinfo {author} {\bibfnamefont {Joel}\ \bibnamefont {Meyers}},\
  }\bibfield  {title} {\enquote {\bibinfo {title} {Dark matter interactions,
  helium, and the cosmic microwave background},}\ }\href {\doibase
  10.1103/PhysRevLett.122.041301} {\bibfield  {journal} {\bibinfo  {journal}
  {Phys. Rev. Lett.}\ }\textbf {\bibinfo {volume} {122}},\ \bibinfo {pages}
  {041301} (\bibinfo {year} {2019})}\BibitemShut {NoStop}%
\bibitem [{\citenamefont {Boddy}\ \emph {et~al.}(2018)\citenamefont {Boddy},
  \citenamefont {Gluscevic}, \citenamefont {Poulin}, \citenamefont {Kovetz},
  \citenamefont {Kamionkowski},\ and\ \citenamefont {Barkana}}]{Boddy:2018wzy}%
  \BibitemOpen
  \bibfield  {author} {\bibinfo {author} {\bibfnamefont {Kimberly~K.}\
  \bibnamefont {Boddy}}, \bibinfo {author} {\bibfnamefont {Vera}\ \bibnamefont
  {Gluscevic}}, \bibinfo {author} {\bibfnamefont {Vivian}\ \bibnamefont
  {Poulin}}, \bibinfo {author} {\bibfnamefont {Ely~D.}\ \bibnamefont {Kovetz}},
  \bibinfo {author} {\bibfnamefont {Marc}\ \bibnamefont {Kamionkowski}}, \ and\
  \bibinfo {author} {\bibfnamefont {Rennan}\ \bibnamefont {Barkana}},\
  }\bibfield  {title} {\enquote {\bibinfo {title} {{Critical Assessment of CMB
  Limits on Dark Matter-Baryon Scattering: New Treatment of the Relative Bulk
  Velocity}},}\ }\href {\doibase 10.1103/PhysRevD.98.123506} {\bibfield
  {journal} {\bibinfo  {journal} {Phys. Rev.}\ }\textbf {\bibinfo {volume}
  {D98}},\ \bibinfo {pages} {123506} (\bibinfo {year} {2018})},\ \Eprint
  {http://arxiv.org/abs/1808.00001} {arXiv:1808.00001 [astro-ph.CO]}
  \BibitemShut {NoStop}%
\bibitem [{\citenamefont {{Wouthuysen}}(1952)}]{1952AJ.....57R..31W}%
  \BibitemOpen
  \bibfield  {author} {\bibinfo {author} {\bibfnamefont {S.~A.}\ \bibnamefont
  {{Wouthuysen}}},\ }\bibfield  {title} {\enquote {\bibinfo {title} {{On the
  excitation mechanism of the 21-cm (radio-frequency) interstellar hydrogen
  emission line.}}}\ }\href {\doibase 10.1086/106661} {\bibfield  {journal}
  {\bibinfo  {journal} {Astron. J.}\ }\textbf {\bibinfo {volume} {57}},\
  \bibinfo {pages} {31--32} (\bibinfo {year} {1952})}\BibitemShut {NoStop}%
\bibitem [{\citenamefont {{Field}}(1958)}]{1958PIRE...46..240F}%
  \BibitemOpen
  \bibfield  {author} {\bibinfo {author} {\bibfnamefont {George~B.}\
  \bibnamefont {{Field}}},\ }\bibfield  {title} {\enquote {\bibinfo {title}
  {{Excitation of the Hydrogen 21-cm Line}},}\ }\href {\doibase
  10.1109/JRPROC.1958.286741} {\bibfield  {journal} {\bibinfo  {journal}
  {Proceedings of the IRE}\ }\textbf {\bibinfo {volume} {46}},\ \bibinfo
  {pages} {240--250} (\bibinfo {year} {1958})}\BibitemShut {NoStop}%
\bibitem [{\citenamefont {Dvorkin}\ \emph {et~al.}(2014)\citenamefont
  {Dvorkin}, \citenamefont {Blum},\ and\ \citenamefont
  {Kamionkowski}}]{Dvorkin:2013cea}%
  \BibitemOpen
  \bibfield  {author} {\bibinfo {author} {\bibfnamefont {Cora}\ \bibnamefont
  {Dvorkin}}, \bibinfo {author} {\bibfnamefont {Kfir}\ \bibnamefont {Blum}}, \
  and\ \bibinfo {author} {\bibfnamefont {Marc}\ \bibnamefont {Kamionkowski}},\
  }\bibfield  {title} {\enquote {\bibinfo {title} {{Constraining Dark
  Matter-Baryon Scattering with Linear Cosmology}},}\ }\href {\doibase
  10.1103/PhysRevD.89.023519} {\bibfield  {journal} {\bibinfo  {journal} {Phys.
  Rev. D}\ }\textbf {\bibinfo {volume} {89}},\ \bibinfo {pages} {023519}
  (\bibinfo {year} {2014})},\ \Eprint {http://arxiv.org/abs/1311.2937}
  {arXiv:1311.2937 [astro-ph.CO]} \BibitemShut {NoStop}%
\bibitem [{\citenamefont {Xu}\ \emph {et~al.}(2018)\citenamefont {Xu},
  \citenamefont {Dvorkin},\ and\ \citenamefont {Chael}}]{Xu:2018efh}%
  \BibitemOpen
  \bibfield  {author} {\bibinfo {author} {\bibfnamefont {Weishuang~Linda}\
  \bibnamefont {Xu}}, \bibinfo {author} {\bibfnamefont {Cora}\ \bibnamefont
  {Dvorkin}}, \ and\ \bibinfo {author} {\bibfnamefont {Andrew}\ \bibnamefont
  {Chael}},\ }\bibfield  {title} {\enquote {\bibinfo {title} {{Probing sub-GeV
  Dark Matter-Baryon Scattering with Cosmological Observables}},}\ }\href
  {\doibase 10.1103/PhysRevD.97.103530} {\bibfield  {journal} {\bibinfo
  {journal} {Phys. Rev. D}\ }\textbf {\bibinfo {volume} {97}},\ \bibinfo
  {pages} {103530} (\bibinfo {year} {2018})},\ \Eprint
  {http://arxiv.org/abs/1802.06788} {arXiv:1802.06788 [astro-ph.CO]}
  \BibitemShut {NoStop}%
\bibitem [{\citenamefont {Gluscevic}\ and\ \citenamefont
  {Boddy}(2018)}]{Gluscevic:2017ywp}%
  \BibitemOpen
  \bibfield  {author} {\bibinfo {author} {\bibfnamefont {Vera}\ \bibnamefont
  {Gluscevic}}\ and\ \bibinfo {author} {\bibfnamefont {Kimberly~K.}\
  \bibnamefont {Boddy}},\ }\bibfield  {title} {\enquote {\bibinfo {title}
  {{Constraints on Scattering of keV–TeV Dark Matter with Protons in the
  Early Universe}},}\ }\href {\doibase 10.1103/PhysRevLett.121.081301}
  {\bibfield  {journal} {\bibinfo  {journal} {Phys. Rev. Lett.}\ }\textbf
  {\bibinfo {volume} {121}},\ \bibinfo {pages} {081301} (\bibinfo {year}
  {2018})},\ \Eprint {http://arxiv.org/abs/1712.07133} {arXiv:1712.07133
  [astro-ph.CO]} \BibitemShut {NoStop}%
\bibitem [{\citenamefont {Fialkov}\ \emph
  {et~al.}(2014{\natexlab{a}})\citenamefont {Fialkov}, \citenamefont {Barkana},
  \citenamefont {Pinhas},\ and\ \citenamefont {Visbal}}]{Fialkov:2013uwm}%
  \BibitemOpen
  \bibfield  {author} {\bibinfo {author} {\bibfnamefont {Anastasia}\
  \bibnamefont {Fialkov}}, \bibinfo {author} {\bibfnamefont {Rennan}\
  \bibnamefont {Barkana}}, \bibinfo {author} {\bibfnamefont {Arazi}\
  \bibnamefont {Pinhas}}, \ and\ \bibinfo {author} {\bibfnamefont {Eli}\
  \bibnamefont {Visbal}},\ }\bibfield  {title} {\enquote {\bibinfo {title}
  {{Complete history of the observable 21-cm signal from the first stars during
  the pre-reionization era}},}\ }\href {\doibase 10.1093/mnrasl/slt135}
  {\bibfield  {journal} {\bibinfo  {journal} {Mon. Not. Roy. Astron. Soc.}\
  }\textbf {\bibinfo {volume} {437}},\ \bibinfo {pages} {36} (\bibinfo {year}
  {2014}{\natexlab{a}})},\ \Eprint {http://arxiv.org/abs/1306.2354}
  {arXiv:1306.2354 [astro-ph.CO]} \BibitemShut {NoStop}%
\bibitem [{\citenamefont {Chen}\ and\ \citenamefont
  {Miralda-Escude}(2004)}]{Chen:2003gc}%
  \BibitemOpen
  \bibfield  {author} {\bibinfo {author} {\bibfnamefont {Xue-Lei}\ \bibnamefont
  {Chen}}\ and\ \bibinfo {author} {\bibfnamefont {Jordi}\ \bibnamefont
  {Miralda-Escude}},\ }\bibfield  {title} {\enquote {\bibinfo {title} {{The
  spin - kinetic temperature coupling and the heating rate due to Lyman - alpha
  scattering before reionization: Predictions for 21cm emission and
  absorption}},}\ }\href {\doibase 10.1086/380829} {\bibfield  {journal}
  {\bibinfo  {journal} {Astrophys. J.}\ }\textbf {\bibinfo {volume} {602}},\
  \bibinfo {pages} {1--11} (\bibinfo {year} {2004})},\ \Eprint
  {http://arxiv.org/abs/astro-ph/0303395} {arXiv:astro-ph/0303395} \BibitemShut
  {NoStop}%
\bibitem [{\citenamefont {Chuzhoy}\ and\ \citenamefont
  {Shapiro}(2006)}]{Chuzhoy:2005wv}%
  \BibitemOpen
  \bibfield  {author} {\bibinfo {author} {\bibfnamefont {Leonid}\ \bibnamefont
  {Chuzhoy}}\ and\ \bibinfo {author} {\bibfnamefont {Paul~R.}\ \bibnamefont
  {Shapiro}},\ }\bibfield  {title} {\enquote {\bibinfo {title} {{Uv pumping of
  hyperfine transitions in the light elements, with application to 21-cm
  hydrogen and 92-cm deuterium lines from the early universe}},}\ }\href
  {\doibase 10.1086/507670} {\bibfield  {journal} {\bibinfo  {journal}
  {Astrophys. J.}\ }\textbf {\bibinfo {volume} {651}},\ \bibinfo {pages} {1--7}
  (\bibinfo {year} {2006})},\ \Eprint {http://arxiv.org/abs/astro-ph/0512206}
  {arXiv:astro-ph/0512206} \BibitemShut {NoStop}%
\bibitem [{\citenamefont {{Furlanetto}}\ and\ \citenamefont
  {{Pritchard}}(2006)}]{2006MNRAS.372.1093F}%
  \BibitemOpen
  \bibfield  {author} {\bibinfo {author} {\bibfnamefont {Steven~R.}\
  \bibnamefont {{Furlanetto}}}\ and\ \bibinfo {author} {\bibfnamefont
  {Jonathan~R.}\ \bibnamefont {{Pritchard}}},\ }\bibfield  {title} {\enquote
  {\bibinfo {title} {{The scattering of Lyman-series photons in the
  intergalactic medium}},}\ }\href {\doibase 10.1111/j.1365-2966.2006.10899.x}
  {\bibfield  {journal} {\bibinfo  {journal} {Mon. Not. Roy. Astron. Soc.}\
  }\textbf {\bibinfo {volume} {372}},\ \bibinfo {pages} {1093--1103} (\bibinfo
  {year} {2006})},\ \Eprint {http://arxiv.org/abs/astro-ph/0605680}
  {arXiv:astro-ph/0605680 [astro-ph]} \BibitemShut {NoStop}%
\bibitem [{\citenamefont {Venumadhav}\ \emph {et~al.}(2018)\citenamefont
  {Venumadhav}, \citenamefont {Dai}, \citenamefont {Kaurov},\ and\
  \citenamefont {Zaldarriaga}}]{Venumadhav:2018uwn}%
  \BibitemOpen
  \bibfield  {author} {\bibinfo {author} {\bibfnamefont {Tejaswi}\ \bibnamefont
  {Venumadhav}}, \bibinfo {author} {\bibfnamefont {Liang}\ \bibnamefont {Dai}},
  \bibinfo {author} {\bibfnamefont {Alexander}\ \bibnamefont {Kaurov}}, \ and\
  \bibinfo {author} {\bibfnamefont {Matias}\ \bibnamefont {Zaldarriaga}},\
  }\bibfield  {title} {\enquote {\bibinfo {title} {{Heating of the
  intergalactic medium by the cosmic microwave background during cosmic
  dawn}},}\ }\href {\doibase 10.1103/PhysRevD.98.103513} {\bibfield  {journal}
  {\bibinfo  {journal} {Phys. Rev. D}\ }\textbf {\bibinfo {volume} {98}},\
  \bibinfo {pages} {103513} (\bibinfo {year} {2018})},\ \Eprint
  {http://arxiv.org/abs/1804.02406} {arXiv:1804.02406 [astro-ph.CO]}
  \BibitemShut {NoStop}%
\bibitem [{\citenamefont {Ali-Haïmoud}\ \emph {et~al.}(2014)\citenamefont
  {Ali-Haïmoud}, \citenamefont {Meerburg},\ and\ \citenamefont
  {Yuan}}]{Ali-Haimoud:2013hpa}%
  \BibitemOpen
  \bibfield  {author} {\bibinfo {author} {\bibfnamefont {Yacine}\ \bibnamefont
  {Ali-Haïmoud}}, \bibinfo {author} {\bibfnamefont {P.~Daniel}\ \bibnamefont
  {Meerburg}}, \ and\ \bibinfo {author} {\bibfnamefont {Sihan}\ \bibnamefont
  {Yuan}},\ }\bibfield  {title} {\enquote {\bibinfo {title} {{New light on
  21 cm intensity fluctuations from the dark ages}},}\ }\href {\doibase
  10.1103/PhysRevD.89.083506} {\bibfield  {journal} {\bibinfo  {journal} {Phys.
  Rev.}\ }\textbf {\bibinfo {volume} {D89}},\ \bibinfo {pages} {083506}
  (\bibinfo {year} {2014})},\ \Eprint {http://arxiv.org/abs/1312.4948}
  {arXiv:1312.4948 [astro-ph.CO]} \BibitemShut {NoStop}%
\bibitem [{\citenamefont {Mertens}\ \emph {et~al.}(2021)\citenamefont
  {Mertens}, \citenamefont {Semelin},\ and\ \citenamefont
  {Koopmans}}]{Mertens:2021apf}%
  \BibitemOpen
  \bibfield  {author} {\bibinfo {author} {\bibfnamefont {F.~G.}\ \bibnamefont
  {Mertens}}, \bibinfo {author} {\bibfnamefont {B.}~\bibnamefont {Semelin}}, \
  and\ \bibinfo {author} {\bibfnamefont {L.~V.~E.}\ \bibnamefont {Koopmans}},\
  }\bibfield  {title} {\enquote {\bibinfo {title} {{Exploring the Cosmic Dawn
  with NenuFAR}},}\ }in\ \href@noop {} {\emph {\bibinfo {booktitle} {{Semaine
  de l'astrophysique fran\c{c}aise 2021}}}}\ (\bibinfo {year} {2021})\ \Eprint
  {http://arxiv.org/abs/2109.10055} {arXiv:2109.10055 [astro-ph.CO]}
  \BibitemShut {NoStop}%
\bibitem [{\citenamefont {{Koopmans}}\ \emph {et~al.}(2015)\citenamefont
  {{Koopmans}}, \citenamefont {{Pritchard}}, \citenamefont {{Mellema}},
  \citenamefont {{Aguirre}}, \citenamefont {{Ahn}}, \citenamefont {{Barkana}},
  \citenamefont {{van Bemmel}}, \citenamefont {{Bernardi}}, \citenamefont
  {{Bonaldi}}, \citenamefont {{Briggs}}, \citenamefont {{de Bruyn}},
  \citenamefont {{Chang}}, \citenamefont {{Chapman}}, \citenamefont {{Chen}},
  \citenamefont {{Ciardi}}, \citenamefont {{Dayal}}, \citenamefont {{Ferrara}},
  \citenamefont {{Fialkov}}, \citenamefont {{Fiore}}, \citenamefont {{Ichiki}},
  \citenamefont {{Illiev}}, \citenamefont {{Inoue}}, \citenamefont {{Jelic}},
  \citenamefont {{Jones}}, \citenamefont {{Lazio}}, \citenamefont {{Maio}},
  \citenamefont {{Majumdar}}, \citenamefont {{Mack}}, \citenamefont
  {{Mesinger}}, \citenamefont {{Morales}}, \citenamefont {{Parsons}},
  \citenamefont {{Pen}}, \citenamefont {{Santos}}, \citenamefont {{Schneider}},
  \citenamefont {{Semelin}}, \citenamefont {{de Souza}}, \citenamefont
  {{Subrahmanyan}}, \citenamefont {{Takeuchi}}, \citenamefont {{Vedantham}},
  \citenamefont {{Wagg}}, \citenamefont {{Webster}}, \citenamefont {{Wyithe}},
  \citenamefont {{Datta}},\ and\ \citenamefont {{Trott}}}]{SKA}%
  \BibitemOpen
  \bibfield  {author} {\bibinfo {author} {\bibfnamefont {L.}~\bibnamefont
  {{Koopmans}}}, \bibinfo {author} {\bibfnamefont {J.}~\bibnamefont
  {{Pritchard}}}, \bibinfo {author} {\bibfnamefont {G.}~\bibnamefont
  {{Mellema}}}, \bibinfo {author} {\bibfnamefont {J.}~\bibnamefont
  {{Aguirre}}}, \bibinfo {author} {\bibfnamefont {K.}~\bibnamefont {{Ahn}}},
  \bibinfo {author} {\bibfnamefont {R.}~\bibnamefont {{Barkana}}}, \bibinfo
  {author} {\bibfnamefont {I.}~\bibnamefont {{van Bemmel}}}, \bibinfo {author}
  {\bibfnamefont {G.}~\bibnamefont {{Bernardi}}}, \bibinfo {author}
  {\bibfnamefont {A.}~\bibnamefont {{Bonaldi}}}, \bibinfo {author}
  {\bibfnamefont {F.}~\bibnamefont {{Briggs}}}, \bibinfo {author}
  {\bibfnamefont {A.~G.}\ \bibnamefont {{de Bruyn}}}, \bibinfo {author}
  {\bibfnamefont {T.~C.}\ \bibnamefont {{Chang}}}, \bibinfo {author}
  {\bibfnamefont {E.}~\bibnamefont {{Chapman}}}, \bibinfo {author}
  {\bibfnamefont {X.}~\bibnamefont {{Chen}}}, \bibinfo {author} {\bibfnamefont
  {B.}~\bibnamefont {{Ciardi}}}, \bibinfo {author} {\bibfnamefont
  {P.}~\bibnamefont {{Dayal}}}, \bibinfo {author} {\bibfnamefont
  {A.}~\bibnamefont {{Ferrara}}}, \bibinfo {author} {\bibfnamefont
  {A.}~\bibnamefont {{Fialkov}}}, \bibinfo {author} {\bibfnamefont
  {F.}~\bibnamefont {{Fiore}}}, \bibinfo {author} {\bibfnamefont
  {K.}~\bibnamefont {{Ichiki}}}, \bibinfo {author} {\bibfnamefont {I.~T.}\
  \bibnamefont {{Illiev}}}, \bibinfo {author} {\bibfnamefont {S.}~\bibnamefont
  {{Inoue}}}, \bibinfo {author} {\bibfnamefont {V.}~\bibnamefont {{Jelic}}},
  \bibinfo {author} {\bibfnamefont {M.}~\bibnamefont {{Jones}}}, \bibinfo
  {author} {\bibfnamefont {J.}~\bibnamefont {{Lazio}}}, \bibinfo {author}
  {\bibfnamefont {U.}~\bibnamefont {{Maio}}}, \bibinfo {author} {\bibfnamefont
  {S.}~\bibnamefont {{Majumdar}}}, \bibinfo {author} {\bibfnamefont {K.~J.}\
  \bibnamefont {{Mack}}}, \bibinfo {author} {\bibfnamefont {A.}~\bibnamefont
  {{Mesinger}}}, \bibinfo {author} {\bibfnamefont {M.~F.}\ \bibnamefont
  {{Morales}}}, \bibinfo {author} {\bibfnamefont {A.}~\bibnamefont
  {{Parsons}}}, \bibinfo {author} {\bibfnamefont {U.~L.}\ \bibnamefont
  {{Pen}}}, \bibinfo {author} {\bibfnamefont {M.}~\bibnamefont {{Santos}}},
  \bibinfo {author} {\bibfnamefont {R.}~\bibnamefont {{Schneider}}}, \bibinfo
  {author} {\bibfnamefont {B.}~\bibnamefont {{Semelin}}}, \bibinfo {author}
  {\bibfnamefont {R.~S.}\ \bibnamefont {{de Souza}}}, \bibinfo {author}
  {\bibfnamefont {R.}~\bibnamefont {{Subrahmanyan}}}, \bibinfo {author}
  {\bibfnamefont {T.}~\bibnamefont {{Takeuchi}}}, \bibinfo {author}
  {\bibfnamefont {H.}~\bibnamefont {{Vedantham}}}, \bibinfo {author}
  {\bibfnamefont {J.}~\bibnamefont {{Wagg}}}, \bibinfo {author} {\bibfnamefont
  {R.}~\bibnamefont {{Webster}}}, \bibinfo {author} {\bibfnamefont
  {S.}~\bibnamefont {{Wyithe}}}, \bibinfo {author} {\bibfnamefont {K.~K.}\
  \bibnamefont {{Datta}}}, \ and\ \bibinfo {author} {\bibfnamefont
  {C.}~\bibnamefont {{Trott}}},\ }\bibfield  {title} {\enquote {\bibinfo
  {title} {{The Cosmic Dawn and Epoch of Reionisation with SKA}},}\ }in\ \href
  {\doibase 10.22323/1.215.0001} {\emph {\bibinfo {booktitle} {Advancing
  Astrophysics with the Square Kilometre Array (AASKA14)}}}\ (\bibinfo {year}
  {2015})\ p.~\bibinfo {pages} {1},\ \Eprint {http://arxiv.org/abs/1505.07568}
  {arXiv:1505.07568 [astro-ph.CO]} \BibitemShut {NoStop}%
\bibitem [{\citenamefont {{Perez}}\ and\ \citenamefont
  {{Granger}}(2007)}]{PER-GRA:2007}%
  \BibitemOpen
  \bibfield  {author} {\bibinfo {author} {\bibfnamefont {Fernando}\
  \bibnamefont {{Perez}}}\ and\ \bibinfo {author} {\bibfnamefont {Brian~E.}\
  \bibnamefont {{Granger}}},\ }\bibfield  {title} {\enquote {\bibinfo {title}
  {{IPython: A System for Interactive Scientific Computing}},}\ }\href
  {\doibase 10.1109/MCSE.2007.53} {\bibfield  {journal} {\bibinfo  {journal}
  {Computing in Science and Engineering}\ }\textbf {\bibinfo {volume} {9}},\
  \bibinfo {pages} {21--29} (\bibinfo {year} {2007})}\BibitemShut {NoStop}%
\bibitem [{\citenamefont {Kluyver}\ \emph {et~al.}(2016)\citenamefont {Kluyver}
  \emph {et~al.}}]{Kluyver2016JupyterN}%
  \BibitemOpen
  \bibfield  {author} {\bibinfo {author} {\bibfnamefont {Thomas}\ \bibnamefont
  {Kluyver}} \emph {et~al.},\ }\bibfield  {title} {\enquote {\bibinfo {title}
  {Jupyter notebooks - a publishing format for reproducible computational
  workflows},}\ }in\ \href@noop {} {\emph {\bibinfo {booktitle} {ELPUB}}}\
  (\bibinfo {year} {2016})\BibitemShut {NoStop}%
\bibitem [{\citenamefont {Hunter}(2007)}]{Hunter:2007}%
  \BibitemOpen
  \bibfield  {author} {\bibinfo {author} {\bibfnamefont {J.~D.}\ \bibnamefont
  {Hunter}},\ }\bibfield  {title} {\enquote {\bibinfo {title} {Matplotlib: A 2d
  graphics environment},}\ }\href@noop {} {\bibfield  {journal} {\bibinfo
  {journal} {Computing In Science \& Engineering}\ }\textbf {\bibinfo {volume}
  {9}},\ \bibinfo {pages} {90--95} (\bibinfo {year} {2007})}\BibitemShut
  {NoStop}%
\bibitem [{\citenamefont {Harris}\ \emph {et~al.}(2020)\citenamefont {Harris},
  \citenamefont {Millman}, \citenamefont {van~der Walt}, \citenamefont
  {Gommers}, \citenamefont {Virtanen}, \citenamefont {Cournapeau},
  \citenamefont {Wieser}, \citenamefont {Taylor}, \citenamefont {Berg},
  \citenamefont {Smith}, \citenamefont {Kern}, \citenamefont {Picus},
  \citenamefont {Hoyer}, \citenamefont {van Kerkwijk}, \citenamefont {Brett},
  \citenamefont {Haldane}, \citenamefont {del R{\'{i}}o}, \citenamefont
  {Wiebe}, \citenamefont {Peterson}, \citenamefont {G{\'{e}}rard-Marchant},
  \citenamefont {Sheppard}, \citenamefont {Reddy}, \citenamefont {Weckesser},
  \citenamefont {Abbasi}, \citenamefont {Gohlke},\ and\ \citenamefont
  {Oliphant}}]{harris2020array}%
  \BibitemOpen
  \bibfield  {author} {\bibinfo {author} {\bibfnamefont {Charles~R.}\
  \bibnamefont {Harris}}, \bibinfo {author} {\bibfnamefont {K.~Jarrod}\
  \bibnamefont {Millman}}, \bibinfo {author} {\bibfnamefont {St{\'{e}}fan~J.}\
  \bibnamefont {van~der Walt}}, \bibinfo {author} {\bibfnamefont {Ralf}\
  \bibnamefont {Gommers}}, \bibinfo {author} {\bibfnamefont {Pauli}\
  \bibnamefont {Virtanen}}, \bibinfo {author} {\bibfnamefont {David}\
  \bibnamefont {Cournapeau}}, \bibinfo {author} {\bibfnamefont {Eric}\
  \bibnamefont {Wieser}}, \bibinfo {author} {\bibfnamefont {Julian}\
  \bibnamefont {Taylor}}, \bibinfo {author} {\bibfnamefont {Sebastian}\
  \bibnamefont {Berg}}, \bibinfo {author} {\bibfnamefont {Nathaniel~J.}\
  \bibnamefont {Smith}}, \bibinfo {author} {\bibfnamefont {Robert}\
  \bibnamefont {Kern}}, \bibinfo {author} {\bibfnamefont {Matti}\ \bibnamefont
  {Picus}}, \bibinfo {author} {\bibfnamefont {Stephan}\ \bibnamefont {Hoyer}},
  \bibinfo {author} {\bibfnamefont {Marten~H.}\ \bibnamefont {van Kerkwijk}},
  \bibinfo {author} {\bibfnamefont {Matthew}\ \bibnamefont {Brett}}, \bibinfo
  {author} {\bibfnamefont {Allan}\ \bibnamefont {Haldane}}, \bibinfo {author}
  {\bibfnamefont {Jaime~Fern{\'{a}}ndez}\ \bibnamefont {del R{\'{i}}o}},
  \bibinfo {author} {\bibfnamefont {Mark}\ \bibnamefont {Wiebe}}, \bibinfo
  {author} {\bibfnamefont {Pearu}\ \bibnamefont {Peterson}}, \bibinfo {author}
  {\bibfnamefont {Pierre}\ \bibnamefont {G{\'{e}}rard-Marchant}}, \bibinfo
  {author} {\bibfnamefont {Kevin}\ \bibnamefont {Sheppard}}, \bibinfo {author}
  {\bibfnamefont {Tyler}\ \bibnamefont {Reddy}}, \bibinfo {author}
  {\bibfnamefont {Warren}\ \bibnamefont {Weckesser}}, \bibinfo {author}
  {\bibfnamefont {Hameer}\ \bibnamefont {Abbasi}}, \bibinfo {author}
  {\bibfnamefont {Christoph}\ \bibnamefont {Gohlke}}, \ and\ \bibinfo {author}
  {\bibfnamefont {Travis~E.}\ \bibnamefont {Oliphant}},\ }\bibfield  {title}
  {\enquote {\bibinfo {title} {Array programming with {NumPy}},}\ }\href
  {\doibase 10.1038/s41586-020-2649-2} {\bibfield  {journal} {\bibinfo
  {journal} {Nature}\ }\textbf {\bibinfo {volume} {585}},\ \bibinfo {pages}
  {357--362} (\bibinfo {year} {2020})}\BibitemShut {NoStop}%
\bibitem [{\citenamefont {Talman}(1978)}]{TALMAN197835}%
  \BibitemOpen
  \bibfield  {author} {\bibinfo {author} {\bibfnamefont {James~D}\ \bibnamefont
  {Talman}},\ }\bibfield  {title} {\enquote {\bibinfo {title} {Numerical
  fourier and bessel transforms in logarithmic variables},}\ }\href {\doibase
  https://doi.org/10.1016/0021-9991(78)90107-9} {\bibfield  {journal} {\bibinfo
   {journal} {Journal of Computational Physics}\ }\textbf {\bibinfo {volume}
  {29}},\ \bibinfo {pages} {35--48} (\bibinfo {year} {1978})}\BibitemShut
  {NoStop}%
\bibitem [{\citenamefont {Hamilton}(2000)}]{Hamilton:1999uv}%
  \BibitemOpen
  \bibfield  {author} {\bibinfo {author} {\bibfnamefont {A.~J.~S.}\
  \bibnamefont {Hamilton}},\ }\bibfield  {title} {\enquote {\bibinfo {title}
  {{Uncorrelated modes of the nonlinear power spectrum}},}\ }\href {\doibase
  10.1046/j.1365-8711.2000.03071.x} {\bibfield  {journal} {\bibinfo  {journal}
  {Mon. Not. Roy. Astron. Soc.}\ }\textbf {\bibinfo {volume} {312}},\ \bibinfo
  {pages} {257--284} (\bibinfo {year} {2000})},\ \Eprint
  {http://arxiv.org/abs/astro-ph/9905191} {arXiv:astro-ph/9905191} \BibitemShut
  {NoStop}%
\bibitem [{\citenamefont
  {Werthm{\"u}ller}(2020)}]{dieter_werthmuller_2020_3830476}%
  \BibitemOpen
  \bibfield  {author} {\bibinfo {author} {\bibfnamefont {Dieter}\ \bibnamefont
  {Werthm{\"u}ller}},\ }\href {\doibase 10.5281/zenodo.3830476} {\enquote
  {\bibinfo {title} {prisae/pyfftlog: First packaged release},}\ } (\bibinfo
  {year} {2020})\BibitemShut {NoStop}%
\bibitem [{\citenamefont {Waskom}\ \emph {et~al.}(2017)\citenamefont {Waskom}
  \emph {et~al.}}]{seaborn}%
  \BibitemOpen
  \bibfield  {author} {\bibinfo {author} {\bibfnamefont {Michael}\ \bibnamefont
  {Waskom}} \emph {et~al.},\ }\href {\doibase 10.5281/zenodo.883859} {\enquote
  {\bibinfo {title} {mwaskom/seaborn: v0.8.1 (september 2017)},}\ } (\bibinfo
  {year} {2017})\BibitemShut {NoStop}%
\bibitem [{\citenamefont {McKinney}(2010)}]{pandas:2010}%
  \BibitemOpen
  \bibfield  {author} {\bibinfo {author} {\bibfnamefont {Wes}\ \bibnamefont
  {McKinney}},\ }\bibfield  {title} {\enquote {\bibinfo {title} {Data
  structures for statistical computing in python},}\ }in\ \href@noop {} {\emph
  {\bibinfo {booktitle} {Proceedings of the 9th Python in Science
  Conference}}},\ \bibinfo {editor} {edited by\ \bibinfo {editor}
  {\bibfnamefont {St\'efan}\ \bibnamefont {van~der Walt}}\ and\ \bibinfo
  {editor} {\bibfnamefont {Jarrod}\ \bibnamefont {Millman}}}\ (\bibinfo {year}
  {2010})\ pp.\ \bibinfo {pages} {51 -- 56}\BibitemShut {NoStop}%
\bibitem [{\citenamefont {{Virtanen}}\ \emph {et~al.}(2020)\citenamefont
  {{Virtanen}} \emph {et~al.}}]{2020SciPy-NMeth}%
  \BibitemOpen
  \bibfield  {author} {\bibinfo {author} {\bibfnamefont {Pauli}\ \bibnamefont
  {{Virtanen}}} \emph {et~al.},\ }\bibfield  {title} {\enquote {\bibinfo
  {title} {{SciPy 1.0: Fundamental Algorithms for Scientific Computing in
  Python}},}\ }\href {\doibase https://doi.org/10.1038/s41592-019-0686-2}
  {\bibfield  {journal} {\bibinfo  {journal} {Nature Methods}\ } (\bibinfo
  {year} {2020}),\ https://doi.org/10.1038/s41592-019-0686-2}\BibitemShut
  {NoStop}%
\bibitem [{\citenamefont {da~Costa-Luis}(2019)}]{da2019tqdm}%
  \BibitemOpen
  \bibfield  {author} {\bibinfo {author} {\bibfnamefont {Casper~O}\
  \bibnamefont {da~Costa-Luis}},\ }\bibfield  {title} {\enquote {\bibinfo
  {title} {tqdm: A fast, extensible progress meter for python and cli},}\
  }\href@noop {} {\bibfield  {journal} {\bibinfo  {journal} {JOSS}\ }\textbf
  {\bibinfo {volume} {4}},\ \bibinfo {pages} {1277} (\bibinfo {year}
  {2019})}\BibitemShut {NoStop}%
\bibitem [{\citenamefont {{Field}}(1959)}]{1959ApJ...129..551F}%
  \BibitemOpen
  \bibfield  {author} {\bibinfo {author} {\bibfnamefont {George~B.}\
  \bibnamefont {{Field}}},\ }\bibfield  {title} {\enquote {\bibinfo {title}
  {{The Time Relaxation of a Resonance-Line Profile.}}}\ }\href {\doibase
  10.1086/146654} {\bibfield  {journal} {\bibinfo  {journal} {\apj}\ }\textbf
  {\bibinfo {volume} {129}},\ \bibinfo {pages} {551} (\bibinfo {year}
  {1959})}\BibitemShut {NoStop}%
\bibitem [{\citenamefont {Pritchard}\ and\ \citenamefont
  {Loeb}(2012)}]{Pritchard:2011xb}%
  \BibitemOpen
  \bibfield  {author} {\bibinfo {author} {\bibfnamefont {Jonathan~R.}\
  \bibnamefont {Pritchard}}\ and\ \bibinfo {author} {\bibfnamefont {Abraham}\
  \bibnamefont {Loeb}},\ }\bibfield  {title} {\enquote {\bibinfo {title}
  {{21-cm cosmology}},}\ }\href {\doibase 10.1088/0034-4885/75/8/086901}
  {\bibfield  {journal} {\bibinfo  {journal} {Rept. Prog. Phys.}\ }\textbf
  {\bibinfo {volume} {75}},\ \bibinfo {pages} {086901} (\bibinfo {year}
  {2012})},\ \Eprint {http://arxiv.org/abs/1109.6012} {arXiv:1109.6012
  [astro-ph.CO]} \BibitemShut {NoStop}%
\bibitem [{\citenamefont {{Zygelman}}(2005)}]{2005ApJ...622.1356Z}%
  \BibitemOpen
  \bibfield  {author} {\bibinfo {author} {\bibfnamefont {B.}~\bibnamefont
  {{Zygelman}}},\ }\bibfield  {title} {\enquote {\bibinfo {title} {{Hyperfine
  Level-changing Collisions of Hydrogen Atoms and Tomography of the Dark Age
  Universe}},}\ }\href {\doibase 10.1086/427682} {\bibfield  {journal}
  {\bibinfo  {journal} {\apj}\ }\textbf {\bibinfo {volume} {622}},\ \bibinfo
  {pages} {1356--1362} (\bibinfo {year} {2005})}\BibitemShut {NoStop}%
\bibitem [{\citenamefont {Furlanetto}\ and\ \citenamefont
  {Furlanetto}(2007{\natexlab{a}})}]{Furlanetto:2006su}%
  \BibitemOpen
  \bibfield  {author} {\bibinfo {author} {\bibfnamefont {Steven}\ \bibnamefont
  {Furlanetto}}\ and\ \bibinfo {author} {\bibfnamefont {Michael}\ \bibnamefont
  {Furlanetto}},\ }\bibfield  {title} {\enquote {\bibinfo {title} {{Spin
  Exchange Rates in Electron-Hydrogen Collisions}},}\ }\href {\doibase
  10.1111/j.1365-2966.2006.11169.x} {\bibfield  {journal} {\bibinfo  {journal}
  {Mon. Not. Roy. Astron. Soc.}\ }\textbf {\bibinfo {volume} {374}},\ \bibinfo
  {pages} {547--555} (\bibinfo {year} {2007}{\natexlab{a}})},\ \Eprint
  {http://arxiv.org/abs/astro-ph/0608067} {arXiv:astro-ph/0608067} \BibitemShut
  {NoStop}%
\bibitem [{\citenamefont {Furlanetto}\ and\ \citenamefont
  {Furlanetto}(2007{\natexlab{b}})}]{Furlanetto:2007te}%
  \BibitemOpen
  \bibfield  {author} {\bibinfo {author} {\bibfnamefont {Steven}\ \bibnamefont
  {Furlanetto}}\ and\ \bibinfo {author} {\bibfnamefont {Michael}\ \bibnamefont
  {Furlanetto}},\ }\bibfield  {title} {\enquote {\bibinfo {title} {{Spin
  Exchange Rates in Proton-Hydrogen Collisions}},}\ }\href {\doibase
  10.1111/j.1365-2966.2007.11921.x} {\bibfield  {journal} {\bibinfo  {journal}
  {Mon. Not. Roy. Astron. Soc.}\ }\textbf {\bibinfo {volume} {379}},\ \bibinfo
  {pages} {130--134} (\bibinfo {year} {2007}{\natexlab{b}})},\ \Eprint
  {http://arxiv.org/abs/astro-ph/0702487} {arXiv:astro-ph/0702487} \BibitemShut
  {NoStop}%
\bibitem [{\citenamefont {Fialkov}\ \emph
  {et~al.}(2014{\natexlab{b}})\citenamefont {Fialkov}, \citenamefont
  {Barkana},\ and\ \citenamefont {Visbal}}]{Fialkov:2014kta}%
  \BibitemOpen
  \bibfield  {author} {\bibinfo {author} {\bibfnamefont {Anastasia}\
  \bibnamefont {Fialkov}}, \bibinfo {author} {\bibfnamefont {Rennan}\
  \bibnamefont {Barkana}}, \ and\ \bibinfo {author} {\bibfnamefont {Eli}\
  \bibnamefont {Visbal}},\ }\bibfield  {title} {\enquote {\bibinfo {title}
  {{The Observable Signature of Late Heating of the Universe during Cosmic
  Reionization}},}\ }\href {\doibase 10.1038/nature12999} {\bibfield  {journal}
  {\bibinfo  {journal} {Nature}\ }\textbf {\bibinfo {volume} {506}},\ \bibinfo
  {pages} {197} (\bibinfo {year} {2014}{\natexlab{b}})},\ \Eprint
  {http://arxiv.org/abs/1402.0940} {arXiv:1402.0940 [astro-ph.CO]} \BibitemShut
  {NoStop}%
\bibitem [{\citenamefont {Fialkov}\ \emph {et~al.}(2013)\citenamefont
  {Fialkov}, \citenamefont {Barkana}, \citenamefont {Visbal}, \citenamefont
  {Tseliakhovich},\ and\ \citenamefont {Hirata}}]{Fialkov:2012su}%
  \BibitemOpen
  \bibfield  {author} {\bibinfo {author} {\bibfnamefont {Anastasia}\
  \bibnamefont {Fialkov}}, \bibinfo {author} {\bibfnamefont {Rennan}\
  \bibnamefont {Barkana}}, \bibinfo {author} {\bibfnamefont {Eli}\ \bibnamefont
  {Visbal}}, \bibinfo {author} {\bibfnamefont {Dmitriy}\ \bibnamefont
  {Tseliakhovich}}, \ and\ \bibinfo {author} {\bibfnamefont {Christopher~M.}\
  \bibnamefont {Hirata}},\ }\bibfield  {title} {\enquote {\bibinfo {title}
  {{The 21-cm signature of the first stars during the Lyman-Werner feedback
  era}},}\ }\href {\doibase 10.1093/mnras/stt650} {\bibfield  {journal}
  {\bibinfo  {journal} {Mon. Not. Roy. Astron. Soc.}\ }\textbf {\bibinfo
  {volume} {432}},\ \bibinfo {pages} {2909} (\bibinfo {year} {2013})},\ \Eprint
  {http://arxiv.org/abs/1212.0513} {arXiv:1212.0513 [astro-ph.CO]} \BibitemShut
  {NoStop}%
\bibitem [{\citenamefont {Barkana}\ and\ \citenamefont
  {Loeb}(2004)}]{Barkana:2003qk}%
  \BibitemOpen
  \bibfield  {author} {\bibinfo {author} {\bibfnamefont {Rennan}\ \bibnamefont
  {Barkana}}\ and\ \bibinfo {author} {\bibfnamefont {Abraham}\ \bibnamefont
  {Loeb}},\ }\bibfield  {title} {\enquote {\bibinfo {title} {{Unusually large
  fluctuations in the statistics of galaxy formation at high redshift}},}\
  }\href {\doibase 10.1086/421079} {\bibfield  {journal} {\bibinfo  {journal}
  {Astrophys. J.}\ }\textbf {\bibinfo {volume} {609}},\ \bibinfo {pages}
  {474--481} (\bibinfo {year} {2004})},\ \Eprint
  {http://arxiv.org/abs/astro-ph/0310338} {arXiv:astro-ph/0310338} \BibitemShut
  {NoStop}%
\bibitem [{\citenamefont {Press}\ and\ \citenamefont
  {Schechter}(1974)}]{Press:1973iz}%
  \BibitemOpen
  \bibfield  {author} {\bibinfo {author} {\bibfnamefont {William~H.}\
  \bibnamefont {Press}}\ and\ \bibinfo {author} {\bibfnamefont {Paul}\
  \bibnamefont {Schechter}},\ }\bibfield  {title} {\enquote {\bibinfo {title}
  {{Formation of galaxies and clusters of galaxies by selfsimilar gravitational
  condensation}},}\ }\href {\doibase 10.1086/152650} {\bibfield  {journal}
  {\bibinfo  {journal} {Astrophys. J.}\ }\textbf {\bibinfo {volume} {187}},\
  \bibinfo {pages} {425--438} (\bibinfo {year} {1974})}\BibitemShut {NoStop}%
\bibitem [{\citenamefont {{Bond}}\ \emph {et~al.}(1991)\citenamefont {{Bond}},
  \citenamefont {{Cole}}, \citenamefont {{Efstathiou}},\ and\ \citenamefont
  {{Kaiser}}}]{Bond}%
  \BibitemOpen
  \bibfield  {author} {\bibinfo {author} {\bibfnamefont {J.~R.}\ \bibnamefont
  {{Bond}}}, \bibinfo {author} {\bibfnamefont {S.}~\bibnamefont {{Cole}}},
  \bibinfo {author} {\bibfnamefont {G.}~\bibnamefont {{Efstathiou}}}, \ and\
  \bibinfo {author} {\bibfnamefont {N.}~\bibnamefont {{Kaiser}}},\ }\bibfield
  {title} {\enquote {\bibinfo {title} {{Excursion Set Mass Functions for
  Hierarchical Gaussian Fluctuations}},}\ }\href {\doibase 10.1086/170520}
  {\bibfield  {journal} {\bibinfo  {journal} {\apj}\ }\textbf {\bibinfo
  {volume} {379}},\ \bibinfo {pages} {440} (\bibinfo {year}
  {1991})}\BibitemShut {NoStop}%
\bibitem [{\citenamefont {Sheth}\ and\ \citenamefont
  {Tormen}(1999)}]{Sheth:1999mn}%
  \BibitemOpen
  \bibfield  {author} {\bibinfo {author} {\bibfnamefont {Ravi~K.}\ \bibnamefont
  {Sheth}}\ and\ \bibinfo {author} {\bibfnamefont {Giuseppe}\ \bibnamefont
  {Tormen}},\ }\bibfield  {title} {\enquote {\bibinfo {title} {{Large scale
  bias and the peak background split}},}\ }\href {\doibase
  10.1046/j.1365-8711.1999.02692.x} {\bibfield  {journal} {\bibinfo  {journal}
  {Mon. Not. Roy. Astron. Soc.}\ }\textbf {\bibinfo {volume} {308}},\ \bibinfo
  {pages} {119} (\bibinfo {year} {1999})},\ \Eprint
  {http://arxiv.org/abs/astro-ph/9901122} {arXiv:astro-ph/9901122} \BibitemShut
  {NoStop}%
\bibitem [{\citenamefont {Cohen}\ \emph {et~al.}(2019)\citenamefont {Cohen},
  \citenamefont {Fialkov}, \citenamefont {Barkana},\ and\ \citenamefont
  {Monsalve}}]{Cohen:2019vck}%
  \BibitemOpen
  \bibfield  {author} {\bibinfo {author} {\bibfnamefont {Aviad}\ \bibnamefont
  {Cohen}}, \bibinfo {author} {\bibfnamefont {Anastasia}\ \bibnamefont
  {Fialkov}}, \bibinfo {author} {\bibfnamefont {Rennan}\ \bibnamefont
  {Barkana}}, \ and\ \bibinfo {author} {\bibfnamefont {Raul}\ \bibnamefont
  {Monsalve}},\ }\bibfield  {title} {\enquote {\bibinfo {title} {{Emulating the
  Global 21-Cm Signal from Cosmic Dawn and Reionization}},}\ }\href {\doibase
  10.1093/mnras/staa1530} {\  (\bibinfo {year} {2019}),\
  10.1093/mnras/staa1530},\ \Eprint {http://arxiv.org/abs/1910.06274}
  {arXiv:1910.06274 [astro-ph.CO]} \BibitemShut {NoStop}%
\bibitem [{\citenamefont {Cohen}\ \emph {et~al.}(2018)\citenamefont {Cohen},
  \citenamefont {Fialkov},\ and\ \citenamefont {Barkana}}]{Cohen:2017xpx}%
  \BibitemOpen
  \bibfield  {author} {\bibinfo {author} {\bibfnamefont {Aviad}\ \bibnamefont
  {Cohen}}, \bibinfo {author} {\bibfnamefont {Anastasia}\ \bibnamefont
  {Fialkov}}, \ and\ \bibinfo {author} {\bibfnamefont {Rennan}\ \bibnamefont
  {Barkana}},\ }\bibfield  {title} {\enquote {\bibinfo {title} {{Charting the
  Parameter Space of the 21-cm Power Spectrum}},}\ }\href {\doibase
  10.1093/mnras/sty1094} {\bibfield  {journal} {\bibinfo  {journal} {Mon. Not.
  Roy. Astron. Soc.}\ }\textbf {\bibinfo {volume} {478}},\ \bibinfo {pages}
  {2193--2217} (\bibinfo {year} {2018})},\ \Eprint
  {http://arxiv.org/abs/1709.02122} {arXiv:1709.02122 [astro-ph.CO]}
  \BibitemShut {NoStop}%
\end{thebibliography}%

\clearpage
\onecolumngrid
\appendix

\onecolumngrid
\begin{center}
\textbf{\large Observable 21-cm Fluctuations from Millicharged Dark Matter} \\ 
\vspace{0.05in}
{\it \large Supplemental Material}\\ 
\vspace{0.05in}
{Rennan Barkana, Anastasia Fialkov, Hongwan Liu and Nadav Outmezguine}
\end{center}

In this supplemental material, we first review 21-cm cosmology for high-energy physicists that are new to this area. We next provide all of the details necessary for our power spectrum calculation, first writing down a general expression for the two-point correlation function, before examining the small-separation and large-separation limits of the function. We also provide some useful expressions for the bulk relative velocity correlation functions. Finally, we end with an elaboration of the simulations that we ran as part of our work in the main body. 

\section{21-cm Cosmology for High-Energy Physicists} 

In this section, we briefly review the physics of the 21-cm emission in a language familiar to high-energy physicists, and detail the full calculation for obtaining the brightness temperature $T_{21}$, paying special attention to deriving expressions that are valid even when the spin temperature is comparable to the hyperfine splitting, a scenario that is possible due to the significant cooling of baryons in the two-fluid dark sector model. For an extensive review, we refer the readers to Ref.~\cite{Barkana:2016nyr}. 

Consider a line-of-sight through a large region of space, parametrized by comoving coordinates $\vecb{x}$ and redshift $z$. Radiation from the cosmic microwave background (CMB) with temperature $T_\gamma(z)$ is incident on neutral hydrogen or HI gas at $\vecb{x}$ with number density $n_\text{HI}(\vecb{x}, z)$. Since the 21-cm line is very narrow (it has a decay width of $A_{10} = \SI{2.85e-15}{\per\second}$, compared to the energy level separation corresponding to a frequency of $\nu_{10} = \SI{1.42}{\giga\hertz}$), we can safely treat all interactions with neutral hydrogen as occurring only when the energy of the photon is exactly given by the hyperfine splitting $\omega_{10} \equiv 2 \pi \nu_{10}$ at each point in $z$. Photons approaching the point with frequency just above $\nu_{10}$ enter with the CMB blackbody intensity $I_{\nu,\text{BB}}(\nu = \nu_{10}, T_\gamma(z))$, with intensity being defined with respect to frequency $\nu$.\footnote{This assumes that the only 21-cm photons passing through this point originates only from the CMB, and that the CMB is a perfect blackbody with no other sources of distortion.} At this point, 21-cm photons can be absorbed or emitted by the gas, leading to a change in its intensity $\Delta I_{\nu_{10}}(z)$. This change in the intensity with respect to the blackbody is the observable in 21-cm cosmology. 

There are three processes at redshift $z$ that contribute to $\Delta I_{\nu_{10}}(z)$: spontaneous emission, absorption and stimulated emission. In the study of radiative transfer, it is conventional to treat absorption and stimulated emission together, while spontaneous emission is regarded as a source-term at each point. Under the narrow-width approximation, we can write the absorption cross section as $\sigma_{0 \to 1} \equiv S \delta(\omega - \omega_{10})$, where $S$ is a constant, so that the number of photons absorbed per volume per time is simply $n_0 (d n_\gamma/ d\omega)|_{\omega_{10}} S$, where $(d n_\gamma / d \omega |_{\omega_{10}})$ is the number density of photons per unit energy at frequency $\omega_{10}$, where $n_0$ is the number density of neutral hydrogen atoms in the ground state of the hyperfine splitting. Similarly, stimulated emission can be encapsulated in an effective cross section $\sigma_{1 \to 0} = R \delta(\omega - \omega_{10})$, and the number of photons emitted by stimulated emission per volume per time written as $n_1 (d n_\gamma / d \omega) |_{\omega_{10}} R$ for another constant $R$, where $n_1$ is the number density of neutral hydrogen atoms in the excited state of the hyperfine splitting. Then the usual detailed balance argument used in deriving the Einstein coefficients tells us that in equilibrium at any temperature $T$,
\begin{alignat*}{1}
    n_1^\text{eq} A_{10} + n_1^\text{eq} R \left. \frac{dn_{\gamma}^\text{eq}}{d\omega} \right|_{\omega_{10}}  = n_0^\text{eq} S \left. \frac{dn_\gamma^\text{eq}}{d\omega} \right|_{\omega_{10}} \,,
\end{alignat*}
where ``eq'' denotes equilibrium quantities. The various number densities are given by $n_1^\text{eq} / n_0^\text{eq} =  3e^{- \omega_{10} / T}$, and 
\begin{alignat*}{1}
    \left. \frac{dn_\gamma^\text{eq}}{d\omega} \right|_{\omega_{10}} = \frac{\omega_{10}^2}{\pi^2} \frac{1}{e^{\omega_{10}/T} - 1}
\end{alignat*}
Using the fact that detailed balance applies at any temperature, we find the following relationship: 
\begin{alignat}{1}
    R = \frac{S}{3} = \frac{\pi^2}{\omega_{10}^2} A_{10} \,.
\end{alignat}
We now define the optical depth of a photon passing through the point at redshift $z$ to be
\begin{alignat}{1}
    \tau(z) = \int dt \, [n_0(z) S - n_1(z) R] \delta(\omega - \omega_{10}) = \int dt \, n_0(z) \left[ 1 - e^{-\omega_{10}/T_\text{S}(z)} \right] S \, \delta(\omega - \omega_{10}) \,,
\end{alignat}
where we have used the definition of the spin temperature, 
\begin{alignat}{1}
    n_1 / n_0 \equiv 3 \exp(- \omega_{10} / T_\text{S}).
\end{alignat}
Notice that the optical depth is defined using the net rate of absorption and stimulated emission. As the photon travels through the point at redshift $z$, it comes into resonance with $\dot{\omega} = H(z) \omega$. We can therefore rewrite the integral in terms of $ d \omega = H(z) \omega \, dt$, and perform the integral to obtain
\begin{alignat*}{1}
    \tau(z) = \frac{3 \pi^2 A_{10} n_\text{HI}(z) }{H(z) \omega_{10}^3} \frac{1 - e^{-\omega_{10}/T_\text{S}(z)}}{1 + 3 e^{-\omega_{10}/T_\text{S}(z)}} \,,
\end{alignat*}
with $n_\text{HI} = n_0 + n_1$. In this expression, we have avoided the common approximation $\omega_{10} \ll T_\text{S}$, since baryonic cooling in our model can lead to very small values of $T_\text{b}$ and hence $T_\text{S}$. In the limit $\omega_{10} \ll T_\text{S}$, owing to the highly populated excited triplet state, the optical depth is small, and is numerically given by~\cite{Barkana:2016nyr}
\begin{alignat}{1}
    \tau(z) \simeq 9.85 \times 10^{-3} \left( \frac{T_\gamma(z)}{T_\text{S}(z)} \right) \left( \frac{\Omega_\text{b} h}{0.0327} \right) \left( \frac{\Omega_\text{m}}{0.307} \right)^{-1/2} \left( \frac{1+z}{10} \right)^{1/2} \,.
\end{alignat}

After passing through the gas, the 21-cm intensity changes because of a combination of absorption and stimulated emission---encapsulated by $\tau(z)$---and spontaneous emission. We can write the change in intensity as
\begin{alignat}{1}
    \Delta I_{\nu_{10}}(z) = -I_{\nu,\text{BB}}(\nu_{10}, T_\gamma(z)) \left[1 - e^{- \tau(z)} \right] + S(z) \,,
    \label{eq:radiative_transfer_intensity}
\end{alignat}
where $I_{\nu,\text{BB}}(\nu_{10}, z)$ is the incoming intensity of 21-cm CMB blackbody radiation, while $S(z)$ is the contribution from spontaneous emission, which only depends on the properties of neutral hydrogen atoms. We know that in thermal equilibrium, i.e.\ if the incoming intensity were a perfect blackbody with temperature $T_\text{S}$, then we must have $\Delta I_{\nu_{10}}(z) = 0$. From this, we can conclude that
\begin{alignat*}{1}
    S(z) = I_{\nu,\text{BB}} (\nu_{10}, T_\text{S}(z)) \left[ 1 - e^{-\tau(z)} \right] \,.
\end{alignat*}

In radio astronomy, the intensity $I_\nu$ at a particular frequency $\nu$ is often expressed as a brightness temperature $\Theta$ instead, with $ \Theta(\omega) \equiv 2 \pi^2 I_\nu / \omega^2$. 
For a blackbody, the relation between $\Theta$ and the thermodynamic temperature $T$ is
\begin{alignat}{1}
    \Theta(\omega) = T \times \frac{\xi}{e^{\xi}-1} \,,
\end{alignat}
where $\xi \equiv \omega/T$, with $\Theta = T$ in the limit $\xi \to 0$. The expression in Eq.~\eqref{eq:radiative_transfer_intensity} can therefore be written as
\begin{alignat*}{1}
    \Delta \Theta(z) = \left[\frac{\xi(z)}{e^{\xi(z)} - 1} T_\text{S}(z)  - T_\gamma (z)\right] \left[1 - e^{-\tau(z)} \right] \,,
\end{alignat*}
where $\xi(z) \equiv \omega_{10} / T_\text{S}$, and we have taken $\omega_{10} \ll T_\gamma(z)$. Finally, the observed 21-cm brightness temperature is precisely this absorption or emission relative to the background $T_\gamma(z)$, redshifted to the present day, i.e.\ 
\begin{alignat}{1}
    T_{21}(z) = \frac{1}{1+z} \left[\frac{\xi(z)}{e^{\xi(z)} - 1} T_\text{S}(z)  - T_\gamma (z)\right] \left[1 - e^{-\tau(z)} \right] \,.
\end{alignat}
Once again, we have not taken the usual approximation $\omega_{10} \ll T_\text{S}(z)$ or equivalently $\xi(z) \ll 1$, since this assumption can be violated with baryonic cooling. Adopting this limit allows one to drop the $\xi(z)/(e^{\xi(z)} - 1)$ term, recovering the more usual expression~\cite{Barkana:2016nyr}.  

Other than $\Lambda$CDM parameters, the only remaining unknown parameter that determines $T_{21}(z)$ is the spin temperature. $T_\text{S}$ is determined by a competition between \textit{1)} scattering of HI atoms with the CMB, which causes $T_\text{S} \to T_\gamma$; \textit{2)} collisions between HI atoms, which causes $T_\text{S} \to T_\text{b}$, and \textit{3)} Ly$\alpha$ scattering, which couples $T_\text{S}$ to the color temperature of the Ly$\alpha$ photons $T_\text{S} \to T_\text{C}$ through the Wouthuysen-Field (WF) effect~\cite{1952AJ.....57R..31W,1958PIRE...46..240F,1959ApJ...129..551F}. The spin temperature can be expressed as a weighted mean~\cite{1958PIRE...46..240F}
\begin{alignat}{1}
    T_\text{S}^{-1} = \frac{T_\gamma^{-1} + x_c T_\text{b}^{-1} + x_\alpha T_\text{C}^{-1}}{1 + x_c + x_\alpha} \,,
    \label{eq:spin_temperature_expr}
\end{alignat}
where $x_c$ and $x_\alpha$ represent coupling coefficients through collisions and Ly$\alpha$ scattering respectively. The collisional coupling coefficient can be written as~\cite{Pritchard:2011xb}
\begin{alignat}{1}
    x_c = \frac{\omega_{10}}{A_{10} T_\gamma} \left[ \kappa_{1-0}^\text{HH}(T_\text{b}) n_\text{HI} + \kappa_{1-0}^{e\text{H}}(T_\text{b}) n_e + \kappa_{1-0}^{p\text{H}}(T_\text{b}) n_p  \right] \,,
\end{alignat}
where $n_p$ and $n_e$ are the number densities of free protons and electrons; the rates $\kappa$ are calculated and tabulated in Refs.~\cite{2005ApJ...622.1356Z,Furlanetto:2006su,Furlanetto:2007te}. The Ly$\alpha$ coefficient is~\cite{Madau:1996cs,Barkana:2016nyr}
\begin{alignat}{1}
    x_\alpha = \frac{16 \pi^2 \alpha \omega_{10}}{27 A_{10} m_e T_\gamma} J_\alpha \,,
    \label{eq:x_alpha_high_temp}
\end{alignat}
where $\alpha \approx 1/137$ is the fine-structure constant, and $J_\alpha$ is defined as
\begin{alignat*}{1}
    J_\alpha = \int \frac{d \Omega}{4 \pi} \frac{I_{\nu_\alpha}}{ \omega_\alpha} \,,
\end{alignat*}
where $I_{\nu_\alpha}$ is the photon intensity at the Ly$\alpha$ frequency, and $\omega_\alpha \approx \SI{10.2}{\eV}$ is the Ly$\alpha$ transition energy. Including the effect of atomic recoil during Ly$\alpha$ scattering as well as the possibility of multiple scatterings allows us to write Eq.~\eqref{eq:spin_temperature_expr} as
\begin{alignat}{1}
    T_\text{S}^{-1} = \frac{T_\gamma^{-1} + x_\text{tot,eff} T_\text{b}^{-1}}{1 + x_\text{tot,eff}} \,,
\end{alignat}
where $x_\text{tot,eff} = x_c + x_{\alpha,\text{eff}}$, and
\begin{alignat}{1}
    x_{\alpha,\text{eff}} \equiv x_\alpha \left(1 + \frac{T_\text{se}}{T_\text{b}} \right)^{-1} \exp \left[-2.06 \left( \frac{\Omega_\text{b} h}{0.0327} \right)^{1/3} \left( \frac{\Omega_\text{m}}{0.307} \right)^{-1/6} \left( \frac{1+z}{10} \right)^{1/2} \left( \frac{T_\text{b}}{T_\text{se}} \right)^{-2/3} \right] \,.
    \label{eq:x_alpha_eff}
\end{alignat}
In this expression, $T_\text{se} \equiv m_\text{H} (\omega_{10}/ \omega_\alpha)^2 \approx \SI{0.402}{\kelvin}$, with $m_\text{H}$ being the mass of the hydrogen atom~\cite{Barkana:2016nyr}. For the results presented in the \textit{Letter}, we adopted 140 phenomenologically viable models of $x_\alpha(z)$ from Ref.~\cite{Reis:2021nqf}, chosen for particular large values of $x_\alpha(z)$ so that $T_\text{S}$ is coupled strongly to the $T_\text{b}$, leading to optimistically but realistically large values of $T_{21}$ that can be potentially probed in near-future 21-cm experiments.

\section{Details of the Power Spectrum Calculation} 
\label{sec:details_calculations}
As we discussed in the main \textit{Letter}, the statistics of the baryon-CDM bulk relative velocity $v_\text{bC}$ fully specifies the statistics of $T_{21}$ in the two-fluid dark sector model. $v_\text{bC}(\vecb{x})$ in this model is a Gaussian random field, as it is in $\Lambda$CDM cosmology; it is fully specified by its two-point function $P_v(k)$, defined as $\langle \tilde{v}_\text{bC} (\vecb{k}) \tilde{v}_\text{bC} (\vecb{k}') \rangle = (2 \pi)^3 \delta^{(3)} (\vecb{k} + \vecb{k}') P_v(k)$, where $\hat{k} \tilde{v}_\text{bC}(\vecb{k}) \equiv \tilde{\vecb{v}}_\text{bC}(\vecb{k})$, the Fourier transform of $\vecb{v}_\text{bC}(\vecb{x})$. In particular, the one-point probability density function (PDF) $f(\vecb{v}_\text{bC})$ is
\begin{alignat}{1}
    f(\vecb{v}_\text{bC}) = \left(\frac{3}{2 \pi v_\text{rms}^2} \right)^{3/2} \exp \left( - \frac{3 v_\text{bC}^2}{2 v_\text{rms}^2} \right) \,,
\end{alignat}
where $v_\text{rms} \equiv \langle v_\text{bC}^2(\vecb{x}) \rangle^{1/2} \approx \SI{29}{\kilo\meter\per\second}$ at $z = 1010$. With this expression, we can easily obtain the mean value of $T_{21}$ at redshift $z$, $\langle T_{21} \rangle (z)$, given any particular astrophysics model or new physics parameters, by integrating over $\vecb{v}_\text{bC}$, i.e.\
\begin{alignat}{1}
    \langle T_{21} \rangle (z) = \int d^3 \vecb{v}_\text{bC} f(\vecb{v}_\text{bC}) T_{21} (v_\text{bC}, z) \,. 
\end{alignat}

Similarly, the two-point correlation function (2PCF) is given by
\begin{alignat}{1}
    \xi_{T_{21}}(\vecb{x}) = \int d^3 \vecb{v}_{\text{bC}, a} \int d^3 \vecb{v}_{\text{bC}, b} \mathcal{P}(\vecb{v}_{\text{bC}, a}, \vecb{v}_{\text{bC}, b}; \vecb{x}) \delta_{T_{21}} (v_{\text{bC},a}) \delta_{T_{21}} (v_{\text{bC},b}) \,,
\end{alignat}
where $\delta_{T_{21}} \equiv (T_{21} - \langle T_{21} \rangle) / \langle T_{21} \rangle$ is the spatial fluctuation of $T_{21}$, and $\mathcal{P}$ is the two-point PDF, i.e.\ the joint probability density function of 3D velocities at two different points, separated by the vector $\vecb{x}$, given by
\begin{alignat}{1}
    \mathcal{P}(\vecb{v}_{\text{bC}, a}, \vecb{v}_{\text{bC}, b}; \vecb{x}) &= \frac{1}{(2\pi)^3 \sqrt{|\mathsf{C}(\vecb{x})|}} \exp \left[ - \frac{1}{2} \vec{U}^\mathsf{T} \mathsf{C}^{-1}(\vecb{x}) \vec{U} \right] \,,
\end{alignat}
where $\vec{U} \equiv (\vecb{v}_{\text{bC}, a}, \vecb{v}_{\text{bC}, b})$ is a 6D vector, and $\mathsf{C}$ is the covariance matrix for this multivariate Gaussian. To differentiate between different types of quantities, we use a boldface letters to represent 3D vectors, an arrow to indicate 6D vectors, \textsf{sans-serif font} for $6 \times 6$ matrices, and underscores for $3 \times 3$ matrices. Note that ultimately, after performing the velocity integrals, $\xi_{T_{21}}$ is only a function of $x \equiv |\vecb{x}|$ in a homogeneous and isotropic Universe. 

The covariance matrix of the 6D Gaussian is given by~\cite{Dalal:2010yt,Ali-Haimoud:2013hpa}
\begin{equation}
	\frac{\mathsf{C}(\vecb{x})}{\sigma_{v}^2}=\mathbb{1}_{6 \times 6}+\left(\begin{array}{cc}0 & \underline{c}(\vecb{x}) \\ \underline{c}(\vecb{x}) & 0\end{array}\right).
\end{equation}
Here, $\sigma_v^2\equiv v_{\rm rms}^2 / 3 =\int d\log k\, \Delta^2_v(k)$ is the 1D velocity dispersion, with $\Delta^2_v(k) \equiv k^3  P_v(k) / (2 \pi^2)$. The elements of the $3\times 3$ matrix $\underline{c}$ are given by
\begin{align}\label{eq:correlations}
	&\underline{c}(\vecb{x})_{i j}\equiv \frac{\langle v_\text{bC}^{i} v_\text{bC}^{j}\rangle(\vecb{x})}{\sigma_{v}^{2}}=c_{\|}(x) \hat{\vecb{x}}^{i} \hat{\vecb{x}}^{j}+c_{\perp}(x)\left(\delta^{i j}-\hat{\vecb{x}}^{i} \hat{\vecb{x}}^{j}\right)\,,\nonumber\\
	&c_{\|}(x)=\frac{1}{v_{\mathrm{rms}}^{2}} \int \frac{d k}{k} \Delta_{v}^2(k)\left[j_{0}(k x)-2 j_{2}(k x)\right]\,,\\
	&c_{\perp}(x)=\frac{1}{v_{\mathrm{rms}}^{2}} \int \frac{d k}{k} \Delta_v^2 (k)\left[j_{0}(k x)+j_{2}(k x)\right] \,,\nonumber
\end{align}
where $j_\ell$ are the spherical Bessel functions of order $\ell$, $i$ and $j$ denote spatial components, $\delta^{ij}$ is the Kronecker delta symbol, and $\hat{\vecb{x}}$ is a unit vector in the direction of $\vecb{x}$. $c_{\parallel}$ and $c_{\perp}$ give the correlation of the velocity component parallel and perpendicular to the separation vector $\vecb{x}$ respectively. For later uses, in a coordinate system where $\vecb{x}=x\hat{z}$, the 3D matrix takes the form
\begin{equation}\label{eq:c_mat_z_hat}
	\underline{c}(x\hat{z})=\left(\begin{array}{ccc}
	c_{\|} 	& 0 		& 0\\
	0 		& c_{\|}	& 0\\
	0 		& 0 		& c_{\perp}
	\end{array}\right) .
\end{equation}

Throughout this appendix, we find it convenient to use $\vecb{u}\sigma_v=\vecb{v}_\text{bC}$ as our integration variable. Using the Fourier transform of a 6D Gaussian, the 2PCF of any function $f(v)$ can be expressed as
\begin{equation}
	\xi_{f}(x)=\int d^3 \vecb{u}_a \int d^3 \vecb{u}_b \int \frac{d^3 \vecb{\omega}_a}{(2\pi)^3} \int \frac{d^3 \vecb{\omega}_b}{(2\pi)^3}e^{i \boldsymbol{\omega}_a\cdot\boldsymbol{u}_a}e^{i \boldsymbol{\omega}_b\cdot\boldsymbol{u}_b}\delta_{f}(u_a \sigma_v)\delta_{f}(u_b \sigma_v)\tilde{\cal P}(\boldsymbol{\omega}_a,\boldsymbol{\omega}_b; \vecb{x}),
	\label{eq:2PCF_Fourier_gaussian}
\end{equation}
where $\delta_f \equiv f/\thb{f}-1$, ($\vecb{\omega}_a$, $\vecb{\omega}_b$) are the Fourier transforms of ($\vecb{u}_a$, $\vecb{\omega}_b$), and
\begin{equation}\label{eq:gaussian_fourier}
		\tilde{\cal P}(\boldsymbol{\omega}_a,\boldsymbol{\omega}_b; \vecb{x})=\exp\left[-\frac{1}{2}\left(\omega_a^2+\omega_b^2\right)-\boldsymbol{\omega}_a^T\underline{c}(\vecb{x})\boldsymbol{\omega}_b\right].
\end{equation}
Since $T_{21}$ is an isotropic function of velocity, integrating over the angles of both $d^3u$ can be performed easily
\begin{equation}
	d^3 \vecb{u}_a \frac{d^3 \vecb{\omega}_a}{(2\pi)^3} d^3 \vecb{u}_b \frac{d^3 \vecb{\omega}_b}{(2\pi)^3} e^{i \boldsymbol{\omega}_a\cdot\boldsymbol{u}_a}e^{i \boldsymbol{\omega}_b\cdot\boldsymbol{u}_b}= u_a^2 du_a \frac{d^3 \vecb{\omega}_a}{2 \pi^2} u_b^2 du_b \frac{d^3 \vecb{\omega}_b}{2 \pi^2}\frac{\sin(u_a \omega_a)}{u_a \omega_a}\frac{\sin(u_b \omega_b)}{u_b \omega_b} \,.
\end{equation}
We can simplify the non-diagonal  part of the Gaussian in Eq.~\eqref{eq:gaussian_fourier} by noting that
\begin{equation}
	\boldsymbol{\omega}_a^T\underline{c}(\vecb{x}) \boldsymbol{\omega}_b=\omega_a\left|\underline{c}(\vecb{x}) \boldsymbol{\omega}_b\right| \cos \theta_a = \omega_a \omega_b \cos \theta_a \cdot R(x, \cos \theta_b) \,,
\end{equation}
where $\theta_a$ is the angle between $\vecb{\omega}_a$ and $\underline{c}\vecb{\omega}_b$, and $\theta_b$ is the angle between $\vecb{\omega}_b$ and $\vecb{x}$. Where through Eq.~\eqref{eq:c_mat_z_hat} we have identified
\begin{alignat}{1}
    R(x, \cos \theta_b) \equiv |\underline{c}(\vecb{x}) \vecb{\omega}_b| =  \sqrt{c_\perp^2(x) +\cos^2 \theta_b \left[c_\parallel^2(x) -c_\perp^2(x) \right]} \,. 
\end{alignat}
This result can be directly computed using the expressions in Eq.~\eqref{eq:correlations}. 

At this point, we can integrate over the two remaining azimuthal angles and $\theta_a$; after some algebra, we arrive at
\begin{equation}\label{eq:corrs_fin}
	\xi_{f}(x)=\frac{2}{\pi}\int d u_1 \, u_1^2 \int d u_2 \, u_2^2 \cdot \delta_{f}(u_1 \sigma_v)\delta_{f}(u_2 \sigma_v)W(u_1,u_2,x),
\end{equation}
with
\begin{equation}
	W(u_1,u_2,x)=\frac{1}{u_1 u_2}\int_{-1}^1\frac{d \cos \theta_b}{2} \frac{\exp\left[-\frac{1}{2}\frac{u_1^2+u_2^2}{1-R^2}\right]\sinh\left[\frac{R u_1u_2}{1-R^2}\right]}{R\sqrt{1-R^2}} \,.
\end{equation}
Eq.~\eqref{eq:corrs_fin} can now be integrated numerically for a given $\delta_{T_{21}}(v_\text{bC})$ to obtain the 2PCF, $\xi_{T_{21}}(x)$. Note that the numerical integration that needs to be performed is a 3D integral, which is numerically much simpler to evaluate than the equivalent 4D integral expressions found in Refs.~\cite{Dalal:2010yt,Ali-Haimoud:2013hpa}. 

Implementing Eq.~\eqref{eq:corrs_fin} numerically poses some mild numerical challenges. For separations $x$ much shorter than the sound horizon, which sets the typical scale of spatial fluctuations, $c_\perp, c_\parallel \to 1$ and $R \to 1$, and the 3D integral becomes extremely peaked and difficult to evaluate numerically with a regular mesh. On the other hand, for distances much larger than the sound horizon, the correlation between the two points becomes weak; a series expansion in the limit of large $x$ produces a simple expression for $\xi_{T_{21}}$ that is highly accurate, allowing us to avoid performing the relatively expensive numerical integration. We will now discuss the small $x$ and large $x$ limits in turn.

\subsection{The small separation limit} 
\label{sub:the_small_separation_limit_of_}

At distances much smaller than the sound horizon, we expect $c_{\parallel},c_{\perp}\to 1$. To understand what happens in this limit, we first write Eq.~\eqref{eq:2PCF_Fourier_gaussian} as
\begin{alignat}{1}
    \xi_f(x) = \int d^3 \vecb{u}_a \int d^3 \vecb{u}_b \int \frac{d^3 \vecb{\omega}_a}{(2\pi)^3} \int \frac{d^3 \vecb{\omega}_b}{(2\pi)^3} e^{i \vecb{\omega}_a \cdot \vecb{u}_a} e^{i \vecb{\omega}_b \cdot \vecb{u}_b} \exp \left[ - \frac{1}{2} (\vecb{\omega}_a + \vecb{\omega}_b)^2 + \vecb{\omega}_a^\mathsf{T} \underline{d}(x) \vecb{\omega}_b \right] \delta_f(u_a \sigma_v) \delta_f(u_b \sigma_v) \,,
\end{alignat}
with $\underline{d} \equiv 1 - \underline{c}$, which is a small parameter in this limit. We now rewrite $\exp(\vecb{\omega}_a^\intercal \underline{d}(x) \vecb{\omega}_b)$ as derivatives acting on $\exp(i \vecb{\omega}_a \cdot \vecb{u}_a) \exp(i \vecb{\omega}_b \cdot \vecb{u}_b)$, to obtain 
\begin{alignat}{1}
    \xi_f(x) &= \int d^3 \vecb{u}_a \int d^3 \vecb{u}_b \, \delta_f(u_a \sigma_v) \delta_f(u_b \sigma_v) \exp \left[ - \partial_a^i \partial_b^j \underline{d}_{ij} \right] \int \frac{d^3 \vecb{\omega}_a}{(2\pi)^3} \int \frac{d^3 \vecb{\omega}_b}{(2\pi)^3} e^{i \vecb{\omega}_a \cdot \vecb{u}_a}  e^{i \vecb{\omega}_b \cdot \vecb{u}_b} \exp \left[ - \frac{1}{2} ( \vecb{\omega}_a + \vecb{\omega_b})^2 \right] \,,
\end{alignat}
where we introduce the notation $\partial_a^i$ is the partial derivative with respect to the component $u_a^i$ (and likewise for $\partial_{b}^j$), and we adopt Einstein summation convention from here on. The last two integrals are now inverse Fourier transforms that we can evaluate, to obtain 
\begin{alignat}{1}
    \xi_f(x) &= \int d^3 \vecb{u}_a \int d^3 \vecb{u}_b \, \delta_f(u_a \sigma_v) \delta_f(u_b \sigma_v) \exp \left[ - \partial_a^i \partial_b^j \underline{d}_{ij} \right] \frac{e^{-u_b^2/2}}{(2\pi)^{3/2}} \delta_D^{(3)} (\vecb{u}_a - \vecb{u}_b) \nonumber \\
    &= \int d^3 \vecb{u}_a \int d^3 \vecb{u}_b \, \delta_D^{(3)}(\vecb{u}_a - \vecb{u}_b) \frac{e^{-u_b^2/2}}{(2\pi)^{3/2}} \exp \left[ - \partial_a^i \partial_b^j \underline{d}_{ij} \right] \delta_f(u_a \sigma_v) \delta_f(u_b \sigma_v) \,,
\end{alignat}
where in the last line we have performed an integration by parts to move the derivatives over. 

We are now ready to expand in terms of the small parameter $\underline{d}_{ij}$. Expanding the exponential to second order, we have
\begin{alignat}{1}
    \exp \left[ - \partial_a^i \partial_b^j \underline{d}_{ij} \right] \simeq 1 - \partial_a^i \partial_b^j \underline{d}_{ij} + \frac{1}{2} \partial_a^i \partial_b^j \partial_a^k \partial_b^l \underline{d}_{ij} \underline{d}_{kl} \,.
\end{alignat}
At leading order, we have
\begin{alignat}{1}
    \xi_f^{(0)}(x \to 0) = \int d^3 \vecb{u}_a \int d^3 \vecb{u}_b \, \delta_D^{(3)}(\vecb{u}_a - \vecb{u}_b) \frac{e^{-u_b^2/2}}{(2\pi)^{3/2}} \delta_f(u_a \sigma_v) \delta_f(u_b \sigma_v) = \langle \delta_f^2(v) \rangle \,,
\end{alignat}
since the PDF of 3D bulk relative velocities is $g(v) d^3 \vecb{v} = (2 \pi \sigma_v^2)^{-3/2} \exp[- v^2 / (2 \sigma_v^2)] d^3 \vecb{v}$, with $v = \sigma_v u$. 

At the next order, evaluating the derivatives, we find
\begin{alignat}{1}
    \xi_f^{(1)}(x \to 0) &= -d_{ij} \int d^3 \vecb{u}_a \int d^3 \vecb{u}_b \, \delta_D^{(3)}(\vecb{u}_a - \vecb{u}_b) \frac{e^{-u_b^2/2}}{(2\pi)^{3/2}} \frac{u_a^i u_b^j}{u_a u_b} \sigma_v^2 \delta_f'(u_a \sigma_v) \delta_f'(u_b \sigma_v) \nonumber \\
    &= -d_{ij} \int d^3 \vecb{u}_a \, \frac{e^{-u_a^2/2}}{(2\pi)^{3/2}} \frac{u_a^i u_a^j}{u_a^2} \sigma_v^2 \delta_f'^2(u_a \sigma_v) \,.
\end{alignat}
At this point, we make use of the identity
\begin{alignat}{1}
    \int d^3 \vecb{u} \, u^i u^j f(u) = \frac{\delta^{ij}}{3} \int d^3 \vecb{u} \, u^2 f(u) \,,
    \label{eq:symmetric_integral_rank_2}
\end{alignat}
which can be deduced from the fact that the symmetry of the integral makes it proportional to the rank-2 isotropic tensor, i.e.\ the Kronecker delta function. Putting everything together, we finally obtain
\begin{alignat}{1}
    \xi_f^{(1)}(x \to 0) = -\frac{1}{3} \underline{d}\indices{^i_i} \sigma_v^2 \langle \delta_f'^2 \rangle= -\frac{1}{3} {\rm Tr}\left[\underline{d}\right] \sigma_v^2 \langle \delta_f'^2 \rangle\,,
\end{alignat}
where $\rm Tr$ is the trace of a matrix.

Finally, at the second order, we first expand the derivatives carefully:
\begin{alignat}{1}
    \partial_a^i \partial_a^k = \frac{\delta^{ik}}{u_a} \partial_a + \frac{u_a^i u_a^k}{u_a^2} \left( \partial_a^2 - \frac{1}{u_a} \partial_a \right) \,,
\end{alignat}
which can be applied to the second order expression to give
\begin{alignat}{1}
    \xi_f^{(2)}(x \to 0) = \frac{1}{2} \underline{d}_{ij} \underline{d}_{kl} \int d^3 \vecb{u}_b \frac{e^{-u_b^2/2}}{(2\pi)^{3/2}} \left[ \left( \frac{\delta^{ik}}{u_b} - \frac{u_b^i u_b^k}{u_b^3}\right) \sigma_v \delta_f' (u_a \sigma_v) + \frac{u_b^i u_b^k}{u_b^2} \sigma_v^2 \delta_f''(u_a \sigma_v) \right] \times (i\to j, k \to l) \,.
\end{alignat}
To simplify the expression, we apply Eq.~\eqref{eq:symmetric_integral_rank_2}, as well as the analogous result at rank-4: 
\begin{alignat}{1}
    \int d^3 \vecb{u} u^i u^j u^k u^l f(u) = \frac{1}{15}(\delta_{ij} \delta_{kl} + \delta_{ik} \delta_{jl} + \delta_{il} \delta_{jk}) \int d^3 \vecb{u} \,  u^4 f(u) \equiv \frac{1}{15} S_{ijkl}\int d^3 \vecb{u} \,  u^4 f(u) \,,
\end{alignat}
to obtain
\begin{alignat}{1}
    \xi_f^{(2)}(x \to 0) &= \frac{1}{2} \underline{d}_{ij} \underline{d}_{kl} \int d^3 \vecb{u}_b \frac{e^{-u_b^2/2}}{(2\pi)^{3/2}} \left( \left[ \frac{\delta^{ik} \delta^{jl}}{3u_b^2} + \frac{S_{ijkl}}{15 u_b^2}\right] \sigma_v^2 \delta_f'^2 + 2\left[ \frac{\delta^{ik} \delta^{jl}}{3 u_b} - \frac{S_{ijkl}}{15u_b} \right] \sigma_v^3 \delta_f' \delta_f'' + \frac{S_{ijkl}}{15} \sigma_v^4 \delta_f''^2 \right) \nonumber \\
    &= \frac{1}{2} \underline{d}_{ij} \underline{d}_{kl} \int d^3 \vecb{u}_b \frac{e^{-u_b^2/2}}{(2\pi)^{3/2}} \left( \left[ \frac{1}{3} \delta^{ik}\delta^{jl} + \frac{2 S_{ijkl}}{15 u_b^2} - \frac{S_{ijkl}}{15} \right] \sigma_v^2 \delta_f'^2 + \frac{S_{ijkl}}{15} \sigma_v^4 \delta_f''^2 \right) \,,
\end{alignat}
where in the last line we have performed an integration by parts. Contracting the indices gives finally
\begin{alignat}{1}
\xi_f^{(2)}(x \to 0) = \left[ 3\underline{d}^{ik} \underline{d}_{ik} -\underline{d}\indices{^i_i} \underline{d}\indices{^k_k} \right] \frac{\sigma_v^2 \langle \delta_f'^2 \rangle}{30} + \frac{1}{30}\left(2 \underline{d}^{ik} \underline{d}_{ik} +\underline{d}\indices{^i_i} \underline{d}\indices{^k_k} \right) \left(2\sigma_v^4 \left\langle \frac{\delta_f'^2}{ v_\text{bC}^2}\right \rangle + \sigma_v^4 \langle \delta_f''^2 \rangle\right) \,.
\end{alignat}
The combined expression can then be written as
\begin{multline}
    \langle f \rangle^2 \xi_f(x \to 0) \simeq \langle f^2 \rangle - \langle f \rangle^2 \\
    - \frac{1}{9} \left\{ {\rm Tr}[\underline{d}] - \frac{1}{2} {\rm Tr}\left[\underline{d}^2\right]\right\}\langle (v_\text{rms} f')^2 \rangle
    + \frac{1}{270} \left\{ {\rm Tr}\left[\underline{d}\right]^2 + 2 {\rm Tr}\left[\underline{d}^2\right]\right\} \left< \left(2 \frac{v_\text{rms}^2}{v_\text{bC}^2} - 3 \right) (v_\text{rms} f')^2 + (v_\text{rms}^2 f'')^2 \right> \,.
    \label{eq:xi_large_exp}
\end{multline}

\subsection{Large separation limit} 
\label{sub:the_large_separation_limit_of_}
At large separations, both $c_{\parallel}$ and $c_{\perp}$ approach $0$. Once again, we can start from Eq.~\eqref{eq:2PCF_Fourier_gaussian} and write
\begin{alignat}{1}
    \xi_f (x) = \int d^3 \vecb{u}_a \int d^3 \vecb{u}_b \, \delta_f(u_a \sigma_v) \delta_f(u_b \sigma_v) \exp \left[\partial_a^i \partial_b^j \underline{c}_{ij} \right] \int \frac{d^3 \vecb{\omega}_a}{(2\pi)^3} \int \frac{d^3 \vecb{\omega}_b}{(2\pi)^3} e^{i \vecb{\omega}_a \cdot \vecb{u}_a} e^{i \vecb{\omega}_a \cdot \vecb{u}_b} \exp \left[-\frac{1}{2}(\omega_a^2 + \omega_b^2) \right] \,.
\end{alignat}
We can immediately take the Fourier transform of the Gaussian to find
\begin{alignat}{1}
    \xi_f(x) = \int d^3 \vecb{u}_a \int d^3 \vecb{u}_b \, \delta_f(u_a \sigma_v) \delta_f(u_b \sigma_v) \exp \left[\partial_a^i \partial_b^j \underline{c}_{ij} \right] \frac{e^{-u_a^2/2}}{(2\pi)^{3/2}} \frac{e^{-u_b^2/2}}{(2\pi)^{3/2}}.
\end{alignat}
At this point, we can expand the exponential in the limit of small $\underline{c}$, noting that any term with odd derivatives in $u_a$ is zero, since the resulting expression is odd under $\vecb{u}_a \to - \vecb{u}_a$. Because $\langle \delta_f \rangle$ is zero, the leading order result goes as two powers of $\underline{c}$: 
\begin{alignat}{1}
    \xi_f^{(1)}(x \to \infty) = \frac{1}{2} \underline{c}_{ij} \underline{c}_{kl} \left( \int d^3 \vecb{u}_a \, \delta_f(u_a \sigma_v) \partial_a^i \partial_a^k \frac{e^{-u_a^2/2}}{(2\pi)^{3/2}} \right) \times (a \to b, i \to j, k \to l).
\end{alignat}
As in the small separation limit, we can exploit the symmetry of the integral to rewrite $\partial_a^i \partial_a^k \to \partial_a^2 \delta^{ik} / 3$, giving
\begin{alignat}{1}
    \xi_f^{(1)}(x \to \infty) &= \frac{1}{18} \underline{c}^{ik} \underline{c}_{ik} \left( \int d^3 \vecb{u}_a \, \delta_f(u_a \sigma_v) \partial_a^2 \frac{e^{-u_a^2/2}}{(2\pi)^{3/2}} \right)^2 \,,
\end{alignat}
which we can evaluate to obtain 
\begin{alignat}{1}
    \xi_f^{(1)}(x \to \infty) = \frac{1}{2} {\rm Tr}\left[\underline{c}^2\right] \frac{\langle v^2 \delta_f \rangle^2}{v_\text{rms}^4} \,.
\end{alignat}

At the next order, we have
\begin{alignat}{1}
    \xi_f^{(2)}(x \to \infty) = \frac{1}{24} \underline{c}_{ij} \underline{c}_{kl} \underline{c}_{mn} \underline{c}_{pq} \left( \int d^3 \vecb{u}_a \, \delta_f(u_a \sigma_v) \partial_a^i \partial_a^k \partial_a^m \partial_a^p \frac{e^{-u_a^2/2}}{(2\pi)^{3/2}} \right) \times (a \to b, i \to j, k \to l, m \to n, p \to q) \,.
\end{alignat}
Once again, we can replace $\partial_a^i \partial_a^k \partial_a^m \partial_a^p \to S^{ikmp} \vecb{\nabla}_a^2 \vecb{\nabla}_a^2 / 15$, where $\vecb{\nabla}_a^2 \equiv \partial_a^i \partial_{a,i}$, leading to
\begin{alignat}{1}
    \xi_f^{(2)}(x \to \infty) = \frac{S^{ikmp}S^{jlnq}}{24(15)^2}  \underline{c}_{ij} \underline{c}_{kl} \underline{c}_{mn} \underline{c}_{pq} \left( \int d^3 \vecb{u}_a \, \delta_f(u_a \sigma_v) \vecb{\nabla}_a^2 \vecb{\nabla}_a^2 \frac{e^{-u_a^2/2}}{(2\pi)^{3/2}} \right)^2 \,.
\end{alignat}
Contracting the tensor indices gives
\begin{alignat}{1}
    S^{ikmp} S^{jlnq} \underline{c}_{ij} \underline{c}_{kl} \underline{c}_{mn} \underline{c}_{pq} &= S^{ikmp} \left( \underline{c}^2_{ik} \underline{c}^2_{mp} + \underline{c}^2_{im} \underline{c}^2_{kp} + \underline{c}^2_{ip} \underline{c}^2_{km} \right) \nonumber \\
    &= 3 \left[ (\underline{c}^{ik} \underline{c}_{ik})^2 + 2 (\underline{c}^2)^{ik} (\underline{c}^2)_{ik} \right]=3\left\{{\rm Tr}\left[\underline{c}^2\right]^2+2{\rm Tr}\left[\underline{c}^4\right]\right\} \,.
\end{alignat}
Evaluating the integral and simplifying leads us to the final expression, 
\begin{alignat}{1}
    \xi_f^{(2)} (x \to \infty) = \frac{1}{8} \left\{{\rm Tr}\left[\underline{c}^2\right]^2+2{\rm Tr}\left[\underline{c}^4\right]\right\} \left( \frac{3}{5} \frac{\langle v^4 \delta_f \rangle}{v_\text{rms}^4} - 2 \frac{\langle v^2 \delta_f \rangle}{v_\text{rms}^2} \right)^2 \,.
\end{alignat}

The final combined result is
\begin{alignat}{1}
    \xi_f(x \to \infty) \simeq \frac{1}{2} {\rm Tr}\left[\underline{c}^2\right]\frac{\langle v^2 \delta_f \rangle^2}{v_\text{rms}^4} + \frac{1}{8} \left\{{\rm Tr}\left[\underline{c}^2\right]^2+2{\rm Tr}\left[\underline{c}^4\right]\right\} \left( \frac{3}{5} \frac{\langle v^4 \delta_f \rangle}{v_\text{rms}^4} - 2 \frac{\langle v^2 \delta_f \rangle}{v_\text{rms}^2} \right)^2 \,,
    \label{eq:xi_small}
\end{alignat}
which to leading order is in agreement with Ref.~\cite{Ali-Haimoud:2013hpa}.

\subsection{Velocity correlation functions} 
\label{sub:v2v4_correlation_function}
Finally, in this section, we compute the correlation functions for powers of $v_\text{bC}$, which we denote as $v$ for simplicity in this section. We also write $\xi_f$ in terms of various bias factors multiplying velocity correlation functions, which is a common approximation scheme used in the literature.  

The correlation function for $v^2$ and $v^4$ are both simple Gaussian integrals, resulting with
\begin{equation}
	\xi_{v^2}(x)=\frac{2}{9} {\rm Tr}\left[\underline{c}^2\right]\,, \qquad \xi_{v^4}(x)=4\xi_{v^2}+\frac{8}{225} \left\{ {\rm Tr}\left[\underline{c}^2\right]^2 + 2{\rm Tr}\left[\underline{c}^4\right]\right\}.
\end{equation}
Alternatively, these expressions could have been obtained by noting that the large separation expansion is exact to first order for $\xi_{v^2}$ and to second order for $\xi_{v^4}$. At large separation, we see that $\xi_{v^4}\simeq 4\xi_{v^2}$ to leading order, in agreement with Eq.~\eqref{eq:xi_small}. It is common to write the correlation function $\xi_f$ as a bias parameter multiplied by $\xi_{v^2}$ and $\xi_{v^4}$; comparing our expressions here and  Eq.~\eqref{eq:xi_small}, we find in the large separation limit
\begin{equation}
	\xi_{f}(x \to \infty)\simeq\frac{9}{4}\left(\frac{\la v^2 \delta_{f}\ra}{\la v^2\ra}\right)^2\xi_{v^2}(x)-\frac{225}{64}\left(2\frac{\la v^2 \delta_{f}\ra}{\la v^2\ra}-\frac{\la v^4 \delta_{f}\ra}{\la v^4\ra}\right)^2\left[4\xi_{v^2}(x)-\xi_{v^4}(x)\right].
\end{equation}
In the small separation limit, we write $\underline{c}=1 - \underline{d}$ as before, and noting that $\underline{c}^2=1-2 \underline{d} + \underline{d}^2$, we obtain
\begin{equation}
	\xi_{v^2}(x)=\frac{2}{3}-\frac{4}{9} {\rm Tr}\left[\underline{d}\right] + \frac{2}{9} {\rm Tr}\left[\underline{d}^2\right],
\end{equation}
which is an exact expression, while keeping terms up to order $\underline{d}^2$, we find
\begin{equation}
	\xi_{v^4}(x \to 0)\simeq\frac{16}{5}-\frac{112}{45}\left( {\rm Tr}\left[\underline{d}\right] -\frac{1}{2} {\rm Tr}\left[\underline{d}^2\right] \right) + \frac{32}{225}\left( {\rm Tr}\left[\underline{d}\right]^2 + 2 {\rm Tr}\left[\underline{d}^2\right] \right) \,.
\end{equation}
The above two expressions are in agreement with Eq.~\eqref{eq:xi_large_exp}, which we can now rewrite as
\begin{align}
	\bar{\xi}_{f}(x\to 0) &\simeq \left(\frac{A}{4}-\frac{7B}{48}\right)\bar{\xi}_{v^2}(x)+\frac{5B}{192}\bar{\xi}_{v^4}(x)\,, \nonumber\\
	\bar{\xi}(x)=\xi(x)-\xi(0) \,, \qquad A &=\la\left[v_{\rm rms}f'(v)\right]^2\ra \,, \qquad B=\la\left(2\frac{v_{\rm rms}^2}{v^2}-3\right)\left[v_{\rm rms}f'(v)\right]^2+\left[v_{\rm rms}^2f''(v)\right]^2\ra \,.
\end{align}

\section{Details of the Simulations} 
\label{sec:details_of_the_simulations}
We rely on a large-scale, semi-numerical 21-cm code based on Refs.~\cite{Visbal:2012aw,Fialkov:2013uwm,Reis:2021nqf,Fialkov:2014kta}. Driven by the specifications of radio telescopes such as the Square Kilometer Array, the simulation models large cosmic volumes ($384^3$ comoving \SI{}{\mega\parsec\cubed}) with a resolution of 3 comoving \SI{}{\mega\parsec}. The initial conditions for density fields, bulk relative velocities between dark matter and baryons~\cite{Tseliakhovich:2010bj,Fialkov:2012su,Visbal:2012aw}, and IGM temperature are generated at $z = 60$. The halo abundance is calculated within each resolution element using the approach of Ref.~\cite{Barkana:2003qk}, which is based on Refs.~\cite{Press:1973iz,Bond,Sheth:1999mn}. The resulting number of halos is biased by the local values of the large-scale density and velocity fields. Subsequently, star formation is derived assuming that every halo with a mass higher than the star-formation threshold will form stars at a given star-formation efficiency~\cite{Cohen:2019vck}, which is a function of halo mass as well as the local value of the Lyman-Werner (LW) radiative background, which suppresses star formation via the molecular-cooling channel~\cite{Fialkov:2012su}. 

Radiation produced by stars and stellar remnants is propagated taking into account redshifting and absorption in the IGM. We follow the evolution and spatial fluctuations in several key radiative backgrounds: X-rays heat up and mildly ionize the IGM, Ly$\alpha$ photons are responsible for the WF coupling and contribute to IGM heating~\cite{Reis:2021nqf}, LW photons affect the efficiency of star formation, and ionizing photons drive the process of reionization. We then compute the 21-cm signal of neutral hydrogen affected by all the sources of light within the light-cone. The simulation produces three dimensional cubes of the fluctuating 21-cm signal at a selection of redshifts, which can be used to calculate $\langle T_{21} \rangle$ and $P_{T_{21}}$.

Since the Universe at the time of primordial star formation is practically observationally unconstrained, 
we perform multiple simulations, varying several free astrophysical parameters within their allowed ranges~\cite{Cohen:2017xpx,Reis:2021nqf}. The relevant astrophysical parameters include the minimum circular velocity of star forming halos $V_c$, the star formation efficiency $f_*$, the  X-ray spectral energy distribution (SED), X-ray heating efficiency $f_X$, the ionizing efficiency of sources, and the mean free path of ionizing photons. In this \textit{Letter}, we aim to highlight the discovery potential of the two-fluid dark sector model in the 21-cm power spectrum; we therefore take an ensemble of simulations of 140 astrophysical models from this realistic parameter range, chosen to minimize astrophysical heating and maximize Ly$\alpha$ coupling, both of which result in a stronger absorption for $T_{21}$. Specifically, we choose models where $V_c > \SI{16.5}{\kilo\meter\per\second}$, $f_X < 0.001$ and $f_\text{radio} < 1$~\cite{Reis:2020arr}. Although it is included in the model, the process of reionization has a subdominant impact on the high redshift signals discussed in this paper.

\end{document}